\documentclass[useAMS,usenatbib,usegraphicx]{mn2e}
\usepackage{amsmath, amssymb}
\usepackage{bm}
\usepackage{color}
\usepackage{enumitem}
\usepackage{hyperref}

\def\gsim{ \lower .75ex \hbox{$\sim$} \llap{\raise .27ex \hbox{$>$}} }
\def\lsim{ \lower .75ex\hbox{$\sim$} \llap{\raise .27ex \hbox{$<$}} }
\def\lesssim{ \lower .75ex \hbox{$\sim$} \llap{\raise .27ex \hbox{$<$}} }
\def\gtrsim{ \lower .75ex \hbox{$\sim$} \llap{\raise .27ex \hbox{$>$}} }

\def\half{{\textstyle{\frac{1}{2}}}}
\def\gothg{\,\mathfrak G}

\def\ffrac#1#2{{\textstyle\frac{#1}{#2}}}

\newcommand{\pc}{\,{\rm pc}}

\newcommand{\yr}{\,\mathrm{yr}}

\newcommand{\D}{\mathrm{d}}
\newcommand{\Msun}{\,\mathrm{M}_{\odot}}
\newcommand{\E}{\mathrm{RR}}

\newcommand{\J}{\mathcal{J}}
\newcommand{\Ln}{\bm{\hat{\L}}}
\renewcommand{\L}{\bm{L}}
\newcommand{\T}{\bm{T}}
\renewcommand{\O}{\mathbf{O}}
\newcommand{\G}{\gothg}
\newcommand{\In}{\rm in}
\newcommand{\Out}{\rm out}
\newcommand{\rms}{\rm RMS}
\renewcommand{\r}{\bm{r}}
\newcommand{\sat}{\mathrm{sat}}
\newcommand{\vrr}{\mathrm{vrr}}

\title{A numerical study of vector resonant relaxation}
\author[B. Kocsis \& S. Tremaine]{
Bence Kocsis$^{1}$\thanks{bkocsis@ias.edu}
and Scott Tremaine$^{1}$\thanks{tremaine@ias.edu}\\
$^{1}$ Institute for Advanced Study, Princeton, NJ 08540, USA \\
}
\begin{document}

\date{Received ---}
\maketitle

\begin{abstract}
  Stars bound to a supermassive black hole interact
  gravitationally. Persistent torques acting between stellar orbits
  lead to the rapid resonant relaxation of the orbital orientation
  vectors (``vector'' resonant relaxation) and slower relaxation of
  the eccentricities (``scalar'' resonant relaxation), both at rates
  much faster than two-body or non-resonant relaxation. We
  describe a new parallel symplectic integrator, \textsc{n-ring},
  which follows the dynamical evolution of a cluster of $N$ stars
  through vector resonant relaxation, by averaging the pairwise
  interactions over the orbital period and periapsis-precession
  timescale. We use \textsc{n-ring} to follow the evolution of
  clusters containing over $10^4$ stars for tens of relaxation
    times. Among other results, we find that the evolution is
  dominated by torques among stars with radially overlapping orbits,
  and that resonant relaxation can be modelled as a random walk of the
  orbit normals on the sphere, with angular step size ranging from
  $\sim0.5$--1 radian. The relaxation rate in a cluster with a
    fixed number of stars is proportional to the RMS mass of the stars.
  The RMS torque generated by the cluster stars
  is reduced below the torque between Kepler orbits due to
  apsidal precession and declines weakly with the eccentricity
    of the perturbed orbit. However since the angular momentum of an
  orbit also decreases with eccentricity, the relaxation rate is
  approximately eccentricity-independent for $e\,\lesssim\, 0.7$ and
  grows rapidly with eccentricity for $e\,\gtrsim\, 0.8$. We quantify
  the relaxation using the autocorrelation function of the spherical
  multipole moments; this decays exponentially and the $e$-folding
  time may be identified with the vector resonant relaxation timescale. 
\end{abstract}

\section{Introduction}
\label{s:intro}
\noindent
Most galaxies harbour a supermassive black hole (SMBH) of mass
$10^6$--$10^{10}\Msun$ at their centres.  The SMBH is typically
surrounded by a dense stellar system, which is sometimes a distinct
cluster and sometimes a smooth inward continuation from larger radii
of the galaxy's stellar distribution.

We focus in this paper on the near-Keplerian region where the
gravitational force is dominated by the SMBH. The dynamical behavior
of the stars in this region involves the following processes
\citep[e.g.,][hereafter KT11]{2011MNRAS.412..187K}. (i) To a first approximation,
the stars follow eccentric Keplerian orbits with orbital periods
 $P=1$--$10^4\yr$ (for the sake of
concreteness, all numerical estimates are for the near-Keplerian
region of the Milky Way between $0.001\pc$ and $\sim 1\pc$ of
the central black hole at Sgr A*). (ii) On longer timescales,
$10^3$--$10^5\yr$, the spherical component of the
gravitational field from the stellar system and relativistic effects
lead to apsidal precession (retrograde and prograde, respectively) of
the stellar orbits. (iii) Non-spherical components of the
gravitational field from the stellar system lead to diffusion in the
orientation of the orbits on even longer timescales,
$10^5$--$10^7\yr$. (iv) Non-axisymmetric torques between individual
stellar orbits lead to diffusion of the eccentricities of the orbits
on timescales of $10^7$--$10^{10}\yr$. Processes (iii) and (iv) are
called vector and scalar resonant relaxation, respectively
\citep{1996NewA....1..149R}.  (v) Finally, the semimajor axes diffuse
due to two-body encounters and dynamical friction on timescales
$\gtrsim 10^9\yr$. A review of these and other dynamical
  processes in galactic nuclei is given in \cite{2013degn.book.....M}.

A rough guide to the relevant timescales is obtained by
  considering a cluster of $N\gg1$ stars of mass $m$ surrounding a
  central mass $M_\bullet$, with $Nm\ll M_\bullet$.  If the typical
   orbital radius is $a$ and the corresponding orbital period is
   $P=2\pi(a^3/GM_\bullet)^{1/2}$, then
   \newpage

\begin{itemize}[leftmargin=0.5cm,itemsep=1ex]

\item the apsidal precession time is $\sim P\,M_{\bullet}/(Nm)$;

\item the orbital planes are re-oriented on the vector resonant relaxation
 timescale, $\sim P\, M_{\bullet}/(m\sqrt{N})$;

\item the eccentricities are re-distributed on the scalar resonant
  relaxation timescale, $\sim P\, M_{\bullet}/m$;

\item the semimajor axes diffuse on the two-body or non-resonant
  relaxation timescale, $\sim P\, M_{\bullet}^2/(m^2N)$.

\end{itemize}

The large number of stars ($\sim 10^7$) and vast range of spatial and
temporal scales ($10^{-6}$--$1\pc$ and 10--$10^{10}\yr$), as well as
the long-range spatial and temporal correlations of the forces
involved in resonant relaxation, prohibit the accurate dynamical
modeling of these environments with the tools used for stellar
clusters, namely Fokker--Planck calculations and direct N-body
integrations.  However, the hierarchy of timescales in near-Keplerian
stellar systems leads to adiabatic invariants, and algorithms that
enforce their conservation can increase numerical accuracy and
decrease computational demands.  For example, by averaging over
timescales long compared to the orbital period but short compared to
the apsidal precession timescale, we obtain Gauss's method for secular
dynamics \citep{2009MNRAS.394.1085T}, in which each body on an
eccentric orbit is replaced by a ``wire'' on which the linear density
is proportional to the corresponding residence time, i.e., inversely
proportional to the velocity.  On even longer timescales, we can
average the wires over the apsidal precession timescale and thereby
represent them with annuli.  Since these structures are stationary and
axisymmetric, the energy and magnitude of the angular momentum of a
stellar orbit are conserved but the direction of the angular momentum
is not; in other words the geometry of the annulus (periapsis,
apoapsis, and surface density) is  fixed, but its orientation is
not. Vector resonant relaxation (hereafter VRR) is the stochastic process arising from
the gravitational interaction of these annuli, leading to relaxation
of their orientations.

Here we describe a new symplectic integrator, \textsc{n-ring}, which
follows VRR in near-Keplerian stellar systems.
First, we derive the surface density of the annulus describing an
eccentric stellar orbit by averaging over orbital phase and apsidal
angle. Next we derive the corresponding secular Hamiltonian describing
the interaction between a pair of stars.  The resulting equations of
motion for a pair of stars can be solved analytically.  We construct a
symplectic integrator by combining the effects of the pairwise
interactions.  We parallelize, refine, and optimize the algorithm by
evaluating independent pairs in parallel, and by evaluating the
strongest interactions with a smaller timestep than the weaker ones.

We use \textsc{n-ring} to study VRR in spherical
near-Keplerian stellar systems containing up to 16k stars. We
measure the temporal correlation function of the orbit normals and
determine the timescales for relaxation and complete mixing as a
function of the semimajor axis, eccentricity, and stellar mass
distributions. We construct a simple model of the relaxation process
as a Markovian random walk on a sphere and show that this provides a good
representation of the numerical results. We also provide empirical
formulae that can be used to estimate the VRR timescale in spherical systems.

\section{Secular evolution}

\subsection{Hamiltonian for vector resonant relaxation}\label{s:Hamiltonian}

We consider a system of $N$ stars, of masses $m_i$ with $ i \in
\{1,2,\dots,N\}$, orbiting an SMBH of mass $M_\bullet$ located at the
origin. We denote the Keplerian orbit by $\bm{r}_{i}(t)$ and the
semimajor axis, eccentricity, and period by $a_i$, $e_i$, and
$P_i\equiv 2\pi/\Omega_i$ with $\Omega_i \equiv (G
M_{\bullet}/a_i^3)^{1/2}$.  We make the following assumptions:

\begin{enumerate}[itemsep=1.5ex,leftmargin=0.5cm]

\item \label{i:mass} the mass in stars is much less than the mass of
the SMBH, $\sum_im_i\ll M_\bullet$, although the number of stars $N\gg1$;

\item \label{i:binary} there are no binaries (although binaries with semimajor
    axes much less than the system size can be treated as single
    stars over the timescales considered here);

\item\label{i:GR} the stellar system is sufficiently far from the SMBH
  that each star follows an approximately Keplerian orbit around the
  SMBH;

\item\label{i:apsidal} the apsidal precession time of each orbit is
  much longer than the longest orbital period in the stellar system;

\item\label{i:apse} the apsidal precession time of each orbit is 
much shorter than the shortest orbital plane re-orientation time\footnote{\label{foot:prec}
This assumption fails for a small fraction of stars with eccentricity very close
  to unity, since the angular momentum goes to zero as $e\to 1$ so
  even a tiny torque will rapidly re-orient the orbit. More precisely, the
  apsidal precession rates due to the mean mass distribution and due
  to general relativity vary as $(1-e^2)^{1/2}$ and $(1-e^2)^{-1}$
  respectively, while the re-orientation rates due to VRR and due to
  Lense--Thirring precession vary as $(1-e^2)^{-1/2}$ and
  $(1-e^2)^{-3/2}$.};

\item all orbital and apsidal precession periods are incommensurate, 
so mean-motion and apsidal secular resonances do not play a role;

\item\label{i:nodal} the Newtonian potential of the stellar cluster is
  the main driver of the re-orientation of orbital planes, as opposed
  to either Lense--Thirring precession or a massive perturber (e.g., a second black hole, a galactic
  bar, or a molecular torus).

\end{enumerate}
These assumptions may be satisfied for most stars and compact objects
between $\sim 0.001$ and $\sim0.2\pc$ in the Galactic center on timescales
$10^5$--$10^7\,$yr (see KT11). In particular, assumption~\ref{i:mass}
requires that the apoapsides $r_{a,i}=a_i(1+e_i)$ are much smaller
than the radius $1.8\pc$ where the SMBH mass equals the enclosed
stellar mass. The expected binary fraction in galactic nuclei,
assumption~\ref{i:binary}, is quite 
uncertain \citep{2008MNRAS.389.1655A,2009ApJ...700.1933H}, but a recent study
suggests that $30^{+34}_{-21}\%$ of massive young stars in the
Galactic centre may be in binaries \citep{2014ApJ...782..101P}. 
Assumption~\ref{i:GR} requires that the periapsides
$r_{p,i}=a_i(1-e_i)$ are much larger than the gravitational radius
$r_g = GM_{\bullet}/c^2=2\times10^{-7}\pc$.
Assumption~\ref{i:apsidal} is valid for stars with semimajor axes
$\lesssim 1\pc$ (see Fig.\ 1 of KT11). Assumption \ref{i:apse} is
generally valid for stars with semimajor axes $\lesssim 1\pc$, except
in a narrow range of radii where the prograde general-relativistic
apsidal precession cancels the retrograde Newtonian precession, in
particular $a (1-e^2)^{0.54}\simeq 7\,\rm mpc$ (see
\citealt{2010PhRvD..81f2002M,2014CQGra..31x4003B}, KT11, and Eq.\ \ref{e:appprec} with
$s=1$).  Orbits outside this narrow range of radii approximately conserve their 
eccentricity; during one VRR timescale $\Delta e \sim (t_\vrr/t_{\rm rr})^{1/2} \sim N^{-1/4}$.
As for assumption \ref{i:nodal}, Lense--Thirring precession
is negligible if $r_{p,i}$ is much larger than the rotational
influence radius $r_{r} = [4 \chi M_{\bullet} / (m_{\rms} \sqrt{N})
]^{2/3} r_g \sim 1\,\chi^{2/3}\,$mpc where $0<\chi<1$ is the
dimensionless spin parameter of the SMBH (see
\citealt{2010PhRvD..81f2002M}, Fig.\ 1 of KT11, and
\citealt{2012PhRvD..86j2002M}).  The most prominent known massive
perturber in the Galactic Centre is the molecular torus at radii
$1.5$--$7\,$pc, whose influence is significant outside of $\sim
0.2\,$pc (see KT11 and references therein).

The Keplerian orbits evolve slowly due to the gravitational forces
from the other stars. To follow this evolution we first average the
gravitational interaction potential between stars $i$ and $j$ over the
orbital periods of both stars\footnote{Note that because of this orbit
  averaging the net force on the SMBH is zero, so it remains at rest at
  the origin in this approximation.}. This average is
\begin{align}\label{e:Hellipse}
H^{( i  j )}_{\E}
&\equiv \left\langle- \frac{G m_i m_j}{\|\r_i(t) - \r_j(t')\|}
\right\rangle_{t,t'}\nonumber \\
&=-\frac{1}{P_i P_j}
\oint\D \bm{r}_ i  \oint \D\bm{r}_j\frac{G m_i m_j}{v_i v_j \|\r_i  - \r_j \|}
\end{align}
where the subscript ``RR'' stands for ``resonant relaxation'' and
\begin{equation}
v=\|\dot{\r}\|=\sqrt{GM_{\bullet} \left(\frac{2}{\|\r\|} -
  \frac{1}{a}\right)}
\end{equation}
is the speed. The integrations run over the Keplerian elliptical
trajectories. The interaction energy is that of two elliptical wires
with linear density $m/(P v)$.

We assume that the stellar system is approximately spherical. Then its
dominant effect on the orbit of an individual star is apsidal
precession. The characteristic precession time is approximately
$t_{\rm prec}= 2\pi \|\bm{\Omega}_{\rm prec}\|^{-1}\approx\Omega /
[G \rho(a) ]$, where $\rho(a)$ is the average stellar mass
density in the vicinity of the orbit (see
Appendix~\ref{app:precession}).  We next average the interaction
Hamiltonian $H^{( i j )}_{\E}$ over the apsidal precession period
$t_{\rm prec}$, so the eccentric wires are replaced by axisymmetric
rings or annuli. For each star the mass between radii $r$ and $r+\D r$
is $\D m = 2 m \,\D r/(P|v_{r}|)$ where $v_{r}$ is the radial
component of the Keplerian velocity.  Using $|v_{r}| = (v^2
- v_{\theta}^2)^{1/2}$ and the conservation of angular momentum $L =
m r v_{\theta} = m\sqrt{G M_{\bullet} a (1-e^2)}$, the surface
density becomes
\begin{equation}\label{e:sigma}
\sigma(r)= \frac{\D m}{2\pi r \D r}=\frac{m}{ 2\pi^2 a \sqrt{(r_{a}-r)(r-r_{p})}}
\end{equation}
if $r_{p}\leq r\leq r_{a}$ and $\sigma(r)=0$ otherwise; here
$r_{a}=a(1+e)$, $r_{p}=a(1-e)$ are the apoapsis and periapsis of
the orbit. Thus,
\begin{equation}
H^{( i  j )}_{\E}
=-\int \D \r  \int \D\r'
\frac{G \sigma_i(r) \sigma_j(r')}{\|\r - \r'\|}\,,
\label{e:Hannuli}
\end{equation}
where the integration is over the annular surfaces swept out by the rotating ellipses
in the range $r_{p,i}\leq r\leq r_{a, i}$ and
$r_{p, j}\leq r'\leq r_{a, j}$.

We evaluate the integral using a multipole expansion in
Appendix~\ref{app:interactionenergy} to find (Eqs.\
\ref{e:Hintdefinion}, \ref{e:Phi_ell}, \ref{e:Rdef}, and \ref{e:sijl-def})
\begin{equation}\label{e:Hresult}
H^{( i  j )}_{\E} = -\frac{ G m_i m_j}{a_{\Out}}
 \sum_{\ell=0}^{\infty}
P_{\ell}(0)^2\, s_{ i  j \ell}\, \alpha_{ij}^{\ell}
\,P_{\ell}(\cos I_{ i  j })\,.
\end{equation}
where $I_{ij}$ is the inclination angle between the orbital planes of
star $i$ and $j$, $P_{\ell}(x)$ is a Legendre polynomial, and in
particular for integer $n\ge0$
\begin{equation} \label{e:pnzero}
P_{2n}(0)=(-1)^n\frac{(2n)!}{2^{2n}(n!)^2}\,,\quad P_{2n+1}(0)=0\,.
\end{equation}
Furthermore (Eqs.\ \ref{e:sijl-def}, \ref{e:www})
\begin{align}\label{e:s_ijl}
 s_{ij\ell} &=
\frac{1}{\pi^2}\int_0^{\pi} \D \phi \int_0^{\pi} \D \phi'
\\&\quad\times\nonumber
\frac{\min\left[\; (1 + e_{\In}\cos\phi),\; \alpha_{ij}^{-1}(1 + e_{\Out} \cos\phi')\;\right]^{\ell+1}
}{ \max\left[\; \alpha_{ij}(1 + e_{\In} \cos\phi),\; (1 + e_{\Out}  \cos\phi')\;\right]^{\ell}}
\end{align}
where ``${\Out}$'' and ``${\In}$'' label the index $i$ or $j$ with the
larger and the smaller semimajor axis, respectively, and
$\alpha_{ij}=a_{\In} / a_{\Out} < 1$.  In
Appendix~\ref{app:interactionenergy} we show that one of the two
integrals in Eq.~(\ref{e:s_ijl}) can be evaluated analytically and we
use this result to derive a generating function of $s_{ij \ell}$.  Analytic
closed expressions are available in special cases: for example, for circular,
non-overlapping orbits $s_{i j \ell} = 1$ for all $\ell$, and for
eccentric radially non-overlapping orbits we have (Eq.~\ref{e:S-nonoverlapping})
\begin{equation}
s_{ij\ell} = \frac{\chi_{ \Out }^{\ell}}{\chi_{ \In }^{\ell+1}}
 P_{\ell+1}(\chi_{ \In })P_{\ell-1}(\chi_{ \Out })\quad {\rm
   if}~r_{a,\In}<r_{p,\Out}\,,
\label{e:xxxyyy}
\end{equation}
for $\ell > 0$, where $\chi_i$ is the aspect ratio of the elliptical
orbit of star $i$, i.e., $\chi_i=a_i/b_i=1/\sqrt{1-e_i^2}$, where
$b_i=a_i \sqrt{1-e_i^2}$ is the semiminor axis.  The integral
$s_{ij\ell}$ in Eq.~(\ref{e:s_ijl}) depends on the four parameters
$\alpha_{ij}$, $e_{\In}$, $e_{\Out}$, and $\ell$, and can be tabulated
on a four-dimensional grid.  The integral for all stellar pairs may
then be obtained by interpolation on the grid\footnote{The grid must
  be sufficiently dense to resolve the resonance peaks shown in
  Figure~\ref{f:energy-alpha} below.}.

\begin{figure*}
\centering
\mbox{\includegraphics{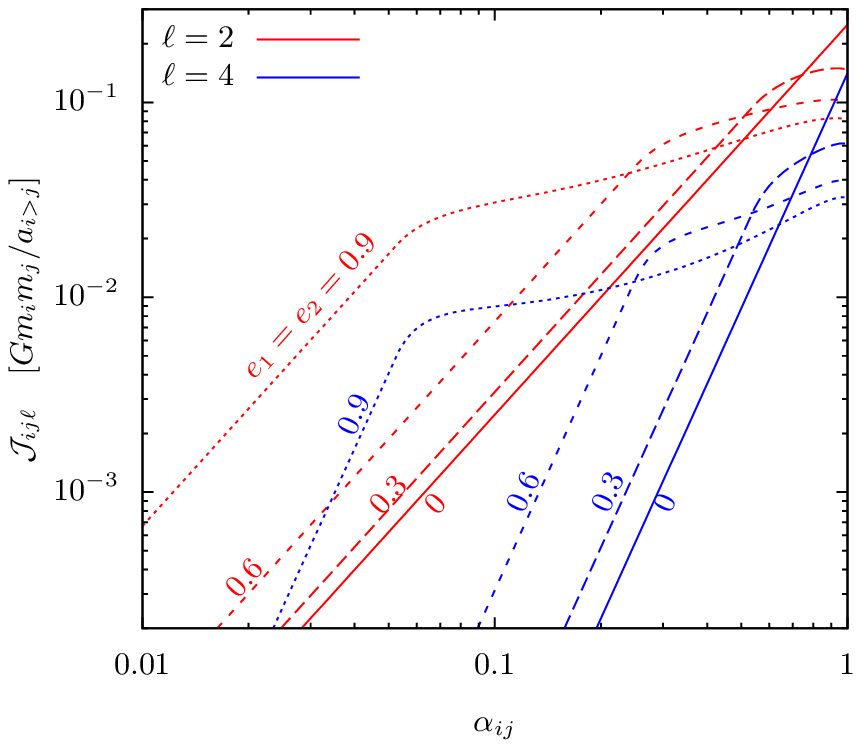}
\quad\includegraphics{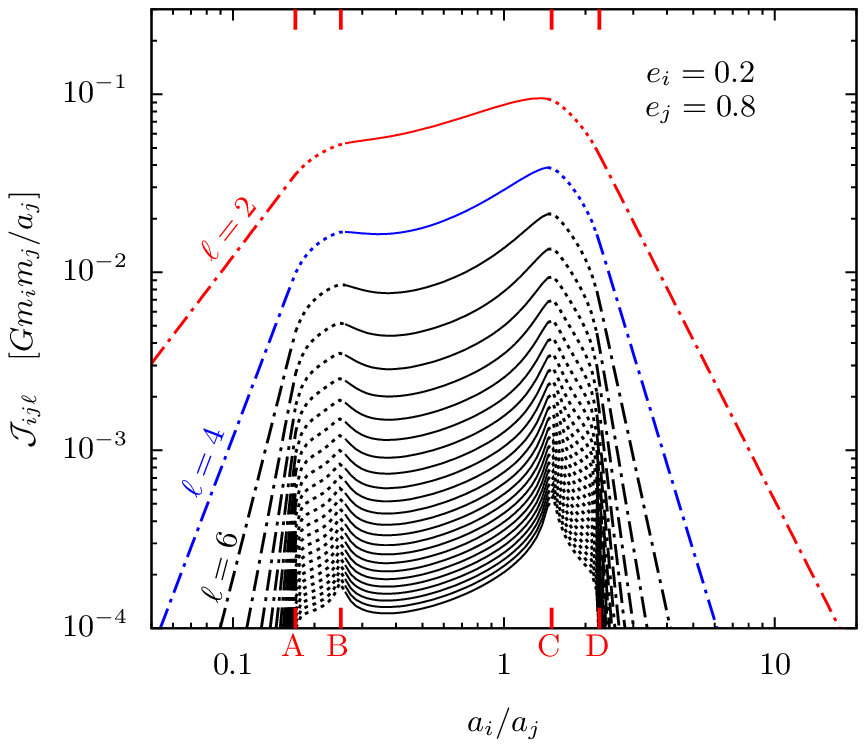}}\\
\mbox{\includegraphics{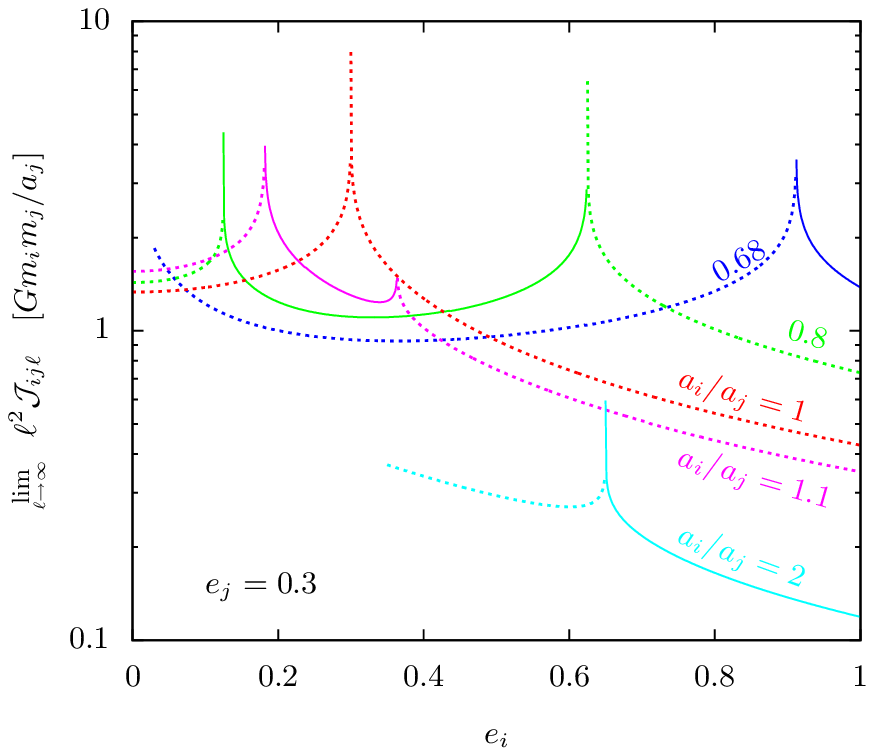}\quad
\includegraphics{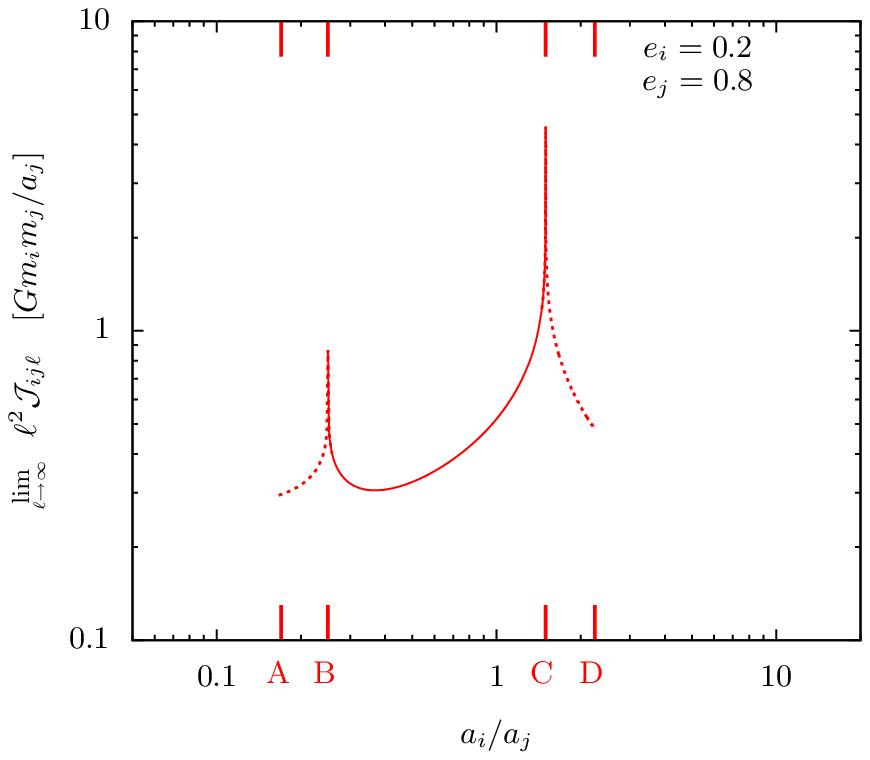}
}
\caption{\label{f:energy-alpha} VRR coupling coefficients
  $\J_{ij\ell}$ (Eqs.\ \ref{e:HRR} and \ref{e:Jijell}). The subscripts
  $i$ and $j$ label the stars and $\ell$ labels the (even) multipole
  order.  {\it Top left panel:} Eccentricities $e_i = e_j=0$ (solid
  line), 0.3 (long-dashed), 0.6 (short-dashed), and 0.9 (dotted).  The
  red and blue curves show $\J_{ij\ell}$ for the multipoles $\ell = 2$
  and $4$, respectively, as a function of the semimajor axis ratio
  $\alpha_{ij}=\min(a_i,a_j)/\max(a_i,a_j)$.  Circular orbits are
  coupled more strongly than eccentric orbits for comparable semimajor
  axes ($\alpha\sim 1$), but the coupling falls off more slowly for
  eccentric orbits in the range $1\geq \alpha_{ij} \geq (1-e)/(1+e)$
  where there is radial overlap.  {\it Top right panel:} $e_i=0.2$ and
  $e_j=0.8$. Here additional multipoles up to $\ell=50$ are shown as a
  function of the semimajor axis ratio $a_i/a_j$. Different line
  styles show different radial regimes, as defined in Appendix
  \ref{app:interactionenergy}: non-overlapping orbits (dash--dotted),
  overlapping (dotted), and embedded (solid).  The boundaries between
  these regions are marked with $A$, $B$, $C$, and $D$ which satisfy
  $a_{i}/a_{j}=(1\pm e_j)/(1\pm e_i)$.  {\it Bottom panels:} The
  limiting behavior of $\ell^2 \J_{ij\ell}$ for asymptotically large
  $\ell$, as a function of eccentricity and semimajor axis.  In the
  bottom left panel, $e_j=0.3$ and $a_i/a_j=0.68$, 0.8, 1, 1.1, and 2
  for different curves, as labeled. In the bottom right panel
  $e_i=0.2$ and $e_j=0.8$ and $a_i/a_j$ is varied. The limit of
  $\ell^2 \J_{ij\ell}$ is zero for non-overlapping orbits, finite and
  non-zero for overlapping (dotted lines) or embedded orbits (solid
  lines), and divergent if the periapsides or the apoapsides coincide
  (see Appendix \ref{app:convergence}). }
\end{figure*}

The sum over $\ell$ in Eq.~(\ref{e:Hresult}) converges very quickly
for radially non-overlapping orbits with $\alpha_{ij}\ll 1$.  The
convergence is slower for $\alpha_{ij}\sim 1$ or for radially
overlapping orbits, but even so the terms in the sum decrease
asymptotically as $\ell^{\,-2}$--$\ell^{\,-2.5}$ except for a set of
measure zero (see Appendix~\ref{app:convergence} for a thorough
discussion of convergence).  The first 10 even multipoles are
typically sufficient for at least $\sim1\%$ accuracy.  The series
converges more slowly if the periapsides or the apoapsides of the two
orbits coincide and the orbits are coplanar ($\sim\ell^{-2}\ln\ell$),
especially if one of the orbits is circular ($\sim \ell^{-1.5}$), or
if the orbits are circular with the same radii but not coplanar ($\sim
\ell^{-1.5}$).  The sum diverges (terms $\sim \ell^{-1}$) only if the
two orbits are circular with the same radii and coplanar
($\alpha_{ij}=1$ and $e_i=e_j=I_{ij}=0$).

Since the averaged surface density representing each star is
stationary and axisymmetric on the orbital timescale $P$, and the
precession timescale $t_{\rm prec}\gg P$, the orbits conserve their
Keplerian energy and their scalar angular momentum $L=\|\L\|$ as they
interact. Thus, the semimajor axes and eccentricities are conserved
during the evolution. In summary,
\begin{equation}\label{e:HRR}
H_{\E} =
 -\sum_{ij \ell}^{i<j}\J_{ij\ell}\,P_{\ell}\big( \Ln_{i}\cdot \Ln_{j} \big)\,,
\end{equation}
where the dynamical variables are the unit vectors normal to the
orbits, $ \Ln_{i} \equiv \L _{i}/L_{i}$,
and $\J_{ij\ell}$ are constant coupling coefficients
\begin{equation}\label{e:Jijell}
\J_{ij\ell} = \frac{ G m_i m_j}{a_{\Out}}
 P_{\ell}(0)^2\, s_{ i  j \ell}\, \alpha_{ij}^{\ell}\,.
\end{equation}
The top panels of Figure~\ref{f:energy-alpha} show $\J_{ij\ell}$ for
$\ell=2$--$4$ (top left panel) and 2--50 (top right panel), for a
range of semimajor axis ratios $a_i/a_j$ and selected values of the
eccentricities $e_i$ and $e_j$.  At all semimajor axes and
eccentricities, the interaction energy is dominated by the $\ell=2$
quadrupolar term and decreases monotonically with $\ell$.  The
coupling declines rapidly with $\ell$, as
$\alpha_{ij}^{\ell}(1+e_{\In})^{\ell}/(1-e_{\Out})^{\ell}$, for radially
non-overlapping orbits, i.e., for $\alpha_{ij}<
(1-e_{\Out})/(1+e_{\In})$.  The coupling coefficients exhibit peaks
when the periapsides or apoapsides coincide, which become increasingly
prominent as $\ell$ increases.  The bottom panels show the limit of
$\ell^2 \J_{ij\ell}$ for large $\ell$, as a function of $e_i$ and
$a_i/a_j$, respectively.  This quantity is relevant for the torque
exerted between inclined orbits as we show below.  The limit is zero
for non-overlapping orbits, but finite positive for overlapping or
embedded orbits (see Appendix \ref{app:interactionenergy} for precise
definitions of these terms). Thus, a larger number of multipoles is needed to
calculate accurately the torques between overlapping or embedded
orbits.

\subsection{Equations of motion}

We have argued that only the directions of the angular momenta of the
stellar orbits change due to the averaged star-star interactions,
while the scalar angular momenta $L=\|\L\|$ are conserved. The equations
of motion for the angular momenta can be derived using Poisson
brackets.

We shall use Greek subscripts to denote Cartesian coordinates
$(x,y,z)$. The Poisson brackets of the angular-momentum vectors
satisfy $\{L_{i \alpha},L_{j
  \beta}\}=\sum_{\gamma}\delta_{ij}\epsilon_{\alpha\beta\gamma}L_{i
  \gamma}$;  here $i$ and $j$
label the stars, $\delta_{ij}=1$ if $i=j$ and zero otherwise, and
$\epsilon_{\alpha\beta\gamma}$ is the Levi-Civita or antisymmetric
tensor.  For any complete set of phase space variables $\{X_s\}$ and a
function $f$ of phase-space variables, we have
\begin{equation}
\frac{\D f}{\D t}=\{f,H\} = \sum_s \{f, X_s \} \frac{\partial
  H}{\partial X_s}
\end{equation}
where $H$ is the Hamiltonian. Using Eqs.\ (\ref{e:HRR}) and
(\ref{e:Jijell}) the equations of motion become
\begin{align}\label{e:EOM1}
\frac{\D L_{i \alpha}}{\D t} &= \{L_{i \alpha}, H\} =
\sum_{j=1}^N\sum_{\beta=1}^3\{L_{i \alpha}, L_{j \beta}\}
\frac{\partial H}{\partial L_{j\beta}} \nonumber \\
&= -\sum_{j\ell\beta\gamma} \epsilon_{\alpha\beta\gamma} \frac{L_{i \gamma} L_{j\beta}}{L_iL_j}\J_{ij\ell}
P'_{\ell}\big(\Ln_i\cdot \Ln_j\big),
\end{align}
where $P'_\ell(x)$ is the derivative of the Legendre polynomial\footnote{Note that
$P'_n(x)=n[P_{n-1}(x)-xP_{n}(x)]/(1-x^2)$.
},
and $L=\|\L\|$. This can be expressed more simply as
\begin{align}
\dot{\L }_i &=   \bm{\Omega}_i \times \L _i, \nonumber \\
\bm{\Omega}_i &= -\sum_{j\ell} \frac{\J_{ij\ell}}{L_i L_j}
P'_{\ell}\big(\Ln_i\cdot \Ln_j\big)\, \L _j .
\label{e:EOM2}
\end{align}
The vector $\bm{\Omega}_i$ is the  angular velocity of the
precession of the angular-momentum vector of a star $i$ due to its
averaged interactions with the other stars.

Using the $\L _i$ as phase-space variables, the phase space has $3N$
dimensions. There are $N+2$ conserved quantities:
\begin{align}\label{e:Ltot}
&\frac{\D}{\D t} E_{\E} = -\frac{\D}{\D t}\sum_{ij\ell} \J_{ij\ell}
P_{\ell}\big(\Ln_i\cdot \Ln_j\big)=0, \nonumber \\
&\frac{\D}{\D t} \sum_i \L _i = -\sum_{ij\ell} \frac{\J_{ij\ell}}{L_i
  L_j} P'_{\ell}\big(\Ln_i\cdot \Ln_j\big)\,\L _j\times  \L _i=0,
\nonumber \\
&\frac{\D}{\D t}(\L _i\cdot \L _i)=0\quad{\rm for~all~}i\in \{1,\dots,N\}.
\end{align}
The first is the conservation of total energy, which follows because
the Hamiltonian $H_{\E}$ (Eq.\ \ref{e:HRR}) is independent of time.
The second is the conservation of the total angular-momentum vector, which follows from
the double sum over $i$ and $j$ of products of symmetric
($\J_{ij\ell}=\J_{ji\ell}$) and antisymmetric terms ($\Ln_j\times\Ln_i$).
The third is the conservation of the scalar angular momentum of each
star, $L_i=m_i \sqrt{G M_{\bullet} a_i(1-e_i^2)}$, due to the orthogonality of
$\L _i$ and $\dot{\L }_i$ in Eq.~(\ref{e:EOM2}). The first two
conservation laws are valid for the original N-body system, but the
third holds only after we average over the orbital period $P$ and
apsidal precession time $t_{\rm prec}$.

\section{Numerical integrator}

\subsection{Pairwise evolution}\label{s:pairwise}

Since the Hamiltonian $H_{\E}$ is a sum of pairwise inter\-action
terms it is useful to first examine the evolution under a single such
term and then superimpose the effects of all the pairs.

The interaction between a single pair of stars leads to uniform
precession of their angular momenta around their common total
angular-momentum vector. Because of this simple behavior, the
equations of motion can be integrated analytically, as we now
show. Eq.~(\ref{e:EOM2}) implies that
\begin{align}
\frac{\D \L _i}{\D t} &= -\sum_{\ell=2}^{\infty}
\frac{\J_{ij\ell}}{L_i L_j} P'_\ell\big(\Ln_i\cdot
\Ln_j\big)\, \L _j \times \L _i, \nonumber \\
\frac{\D \L _j}{\D t} &= -\frac{\D \L _i}{\D t}.
\end{align}
Introduce new variables  $\bm{J}_{ij} = (\L _i + \L _j)/2$
and $\bm{K}_{ij} = (\L _i - \L _j)/2$. Then the equations become
\begin{align}
\frac{\D \bm{J}_{ij}}{\D t} = 0\quad{\rm and}\quad
\frac{\D \bm{K}_{ij}}{\D t} = \bm{\Omega}_{ij} \times \bm{K}_{ij},
\end{align}
where
\begin{equation}\label{e:omjk}
 \bm{\Omega}_{ij}= -\sum_{\ell=2}^{\infty} \frac{2 \J_{ij\ell}}{L_i L_j}
P'_\ell\bigg(\frac{J_{ij}^2 - K_{ij}^2 }{L_i L_j}\bigg)\, \bm{J}_{ij}={\rm const}\,.\\
\end{equation}
The magnitudes of $\bm{J}_{ij}$ and $\bm{K}_{ij}$ are both
conserved. Thus $\bm{\Omega}_{ij}$ is conserved, so $\bm{K}_{ij}$
rotates uniformly with angular velocity $\bm{\Omega}_{ij}$, and we
have
\begin{align}
\bm{J}_{ij}(t) &= \bm{J}_{ij0}  \nonumber \\
\bm{K}_{ij}(t) &=  \cos\left[\Omega_{ij} (t -t_0) \right] \bm{K}_{ij0} \nonumber\\
&\quad +  \sin\left[\Omega_{ij} (t -t_0) \right]\bm{\hat{\Omega}}_{ij}\times \bm{K}_{ij0}\\\nonumber
&\quad+ \left\{1 - \cos\left[\Omega_{ij} (t-t_0)\right]\right\}\big(\bm{K}_{ij0}\cdot \bm{\hat{\Omega}}_{ij}\big)
\bm{\hat{\Omega}}_{ij}
\end{align}
where $\bm{K}_{ij0} = \bm{K}_{ij}(t_0)$ and $\bm{J}_{ij0} =
\bm{J}_{ij}(t_0)$ denote the initial conditions.

The angular momenta are fixed if $\L_i$ and $\L_j$ are parallel,
antiparallel, or perpendicular.  Nearly perpendicular angular momenta
precess with nearly zero angular velocity, but nearly parallel angular
momenta with mutual inclination $I_{ij}\ll 1$ precess with a nonzero
angular speed $\Omega_{ij} \approx \sum_{\ell\mbox{ \scriptsize even}}\ell
J_1(\ell I_{ij}) \J_{ij\ell}(L_i+L_j)/(I_{ij}L_i L_j)$ in a retrograde
direction relative to $\bm{L_{i}}+\bm{L_{j}}$; here $J_1$ is a Bessel
function (see Eq.~\ref{e:P'cosI1}).  For overlapping or embedded
orbits, $\ell^2 \J_{ij\ell}$ approaches a finite limit
(Eq.~\ref{e:Jasymptotic-overlap}) shown in
Figure~\ref{f:energy-alpha}, thus the angular velocity tends
asymptotically to
\begin{align}\label{e:Omegaasymptotics}
\bm{\Omega}_{ij} &\approx -\lim_{\ell\rightarrow \infty} (\ell^2
\J_{ij\ell})\sum_{\ell\mbox{ \scriptsize even}} \frac{J_1(\ell I_{ij})}{\ell
  I_{ij}}\frac{\L_i+\L_j}{L_i L_j} \nonumber \\
&\approx -\lim_{\ell\rightarrow \infty} (\ell^2
\J_{ij\ell})\frac{\L_i+\L_j}{2I_{ij} L_i L_j}\,,
\end{align}
where the sum has been approximated by an integral in the last
equation. Thus the precession speed $\|\dot{\L}_i\|=\|\bm{\Omega}_{ij}
\times\L_i\|$ approaches a finite non-zero limit for
$I_{ij}\rightarrow 0$ for overlapping or embedded orbits.  The bottom
panels of Figure~\ref{f:energy-alpha} show that $\lim_{\ell\rightarrow
  \infty} \ell^2 \J_{ij\ell}$ is singular when the periapsides or
apoapsides of the two orbits coincide, so the precession speed is
singular in this case.  Furthermore, since the torque is non-zero when
either eccentricity tends to unity, $\bm{\Omega}_{ij}$ tends to
infinity as ${\Ln}_{j} I_{ij}^{-1} (1-e_i^2)^{-1/2}$ when
$e_i\rightarrow 1$; thus very eccentric orbits precess very
rapidly. Similar remarks apply for nearly antiparallel angular
momenta.  We derive the asymptotic angular velocity for arbitrary
inclinations in Appendix~\ref{s:asymptotics}
(Eq.~\ref{e:Omegaasymptotics4}).

\subsection{Symplectic integrator}

A system of $N$ stars has $\frac{1}{2}N(N-1)$ pairwise
interactions. Clearly, we can integrate this system
numerically by advancing the angular
momentum of each pair of stars in turn using the results of the
previous subsection. However, there is some advantage to deriving this
result in a more systematic and general way.

The evolution is governed by the first-order differential
equations~(\ref{e:EOM2}).  We may write these as
\begin{equation}
 \dot{\L} = \gothg \L
\label{e:eqmot}
\end{equation}
where $\L\equiv (\L_{1},\ldots,\L_N)$ and $\gothg$ is the operator
defined by
\begin{equation}
\gothg\L=(\bm{\Omega}_{1}\times\L_{1},\ldots,\bm{\Omega}_N\times\L_N).
\end{equation}
The operator $\gothg$ can be written as a sum over pairs,
\begin{equation}
\gothg=\sum_{i=1}^N\sum_{j>i} \gothg_{ij}
\end{equation}
where $\gothg_{ij}$ operates only on the pair of angular momenta
$\L_i,\L_j$ as described in Section \ref{s:pairwise}. Thus the commutator
$[\gothg_{ij},\gothg_{mn}]$ is zero if and only if the pairs $ij$ and
$mn$ have no member in common. Since $\bm{\Omega}_i$
depends explicitly on $\L$, $\G_{ij}$ is a nonlinear operator.

The solution to the equations of motion (\ref{e:eqmot})
is formally
\begin{equation}
 {\L}(t) = \exp(\Delta t\, \gothg) \L(t_0)
 = \sum_{n=0}^{\infty} \frac{\Delta t^n}{n!} \gothg^n \L(t_{0}), \quad \Delta
 t\equiv t-t_0\,.
\end{equation}
Since $\gothg$ is a sum of operators $\gothg_{ij}$ that do not all
commute, the exponential of $\gothg$ is not simply the product of the
exponentials $\gothg_{ij}$.  The Zassenhaus formula shows that to
second order in $\Delta t$ \citep[see][and references
therein]{2012CoPhC.183.2386C}
\begin{align}\label{e:exponential}
 \exp(\Delta t\,\G) &=
\bigg(\prod_J\exp(\Delta t\, \G_J)\bigg)
\\&\quad\times
 \bigg(\prod_{J<K}\exp\big(- \half\Delta t^2
   [\G_J,\G_K]\big)\bigg). \nonumber
\end{align}
Here $J,K=1,2,\ldots,N(N-1)/2$ are indices labeling all of the
particle pairs in an arbitrary order. Assuming that the first product
of exponentials is evaluated in this order [$\exp(\Delta
t\G_{1})\exp(\Delta t\G_{2})\cdots$], the second product can be
evaluated in any order so long as $J<K$ in each commutator
$[\G_J,\G_K]$.

In the following we keep only the first product which corresponds to a
composition of the actions of independent pairwise interactions
generated by Hamiltonians $H^{(ij)}_\E$. The state vector of the
system $\L = (\L_{1}, \L_{2}, \dots, \L_N)$ then follows as
\begin{equation}\label{e:L(t)}
 \L(t) = \prod_{i,j}^{i>j}\O_{ij}(\Delta t) \L(t_0),\  {\rm where }\  \O_{ij}(\Delta t)=\exp(\Delta t\, \G_{ij})\,.
\end{equation}
In Section \ref{s:pairwise} we have derived the analytic solution to
the pairwise interaction: $\O_{ij}(\Delta t)$ rotates $\L_i$ and
$\L_j$ around their common total angular-momentum vector by a finite
angle $\Omega_{ij}\,\Delta t$, keeping all other $\L_k$ fixed.

The integrator given by Eq.~(\ref{e:L(t)}) is symplectic since each
component operator $\O_{ij}(\Delta t)$ is the exact solution of the
equations of motion for the Hamiltonian $H_{\E}^{(ij)}$. However it is
only first-order accurate, i.e., the truncation error after a fixed
integration time $\Delta T=n\Delta t$ varies as $\Delta t$ or as
$n^{-1}$.  Errors arise due to the non-commutativity of different
interaction pairs and the effects of higher order interactions in
Eq.~(\ref{e:exponential}).  Convergence may be improved either by
using a higher order integrator or by choosing a particular ordering
of the evaluation of the $\O_{ij}$.  We discuss these and other
improvements to the numerical algorithm in the following subsections.

\subsection{Higher order accuracy}\label{s:s:second order}

\begin{figure*}
\centering
\mbox{\includegraphics[height=7cm]{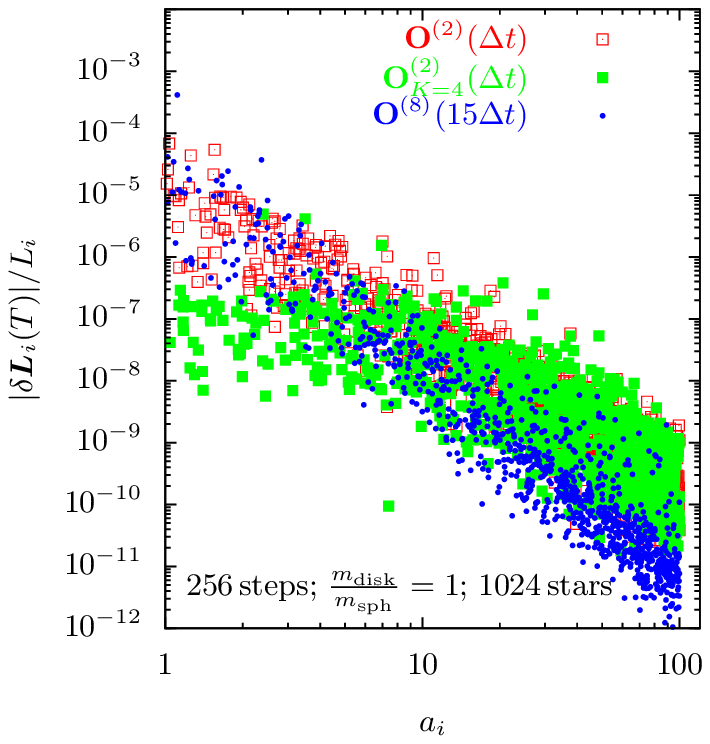}
\includegraphics[height=7cm]{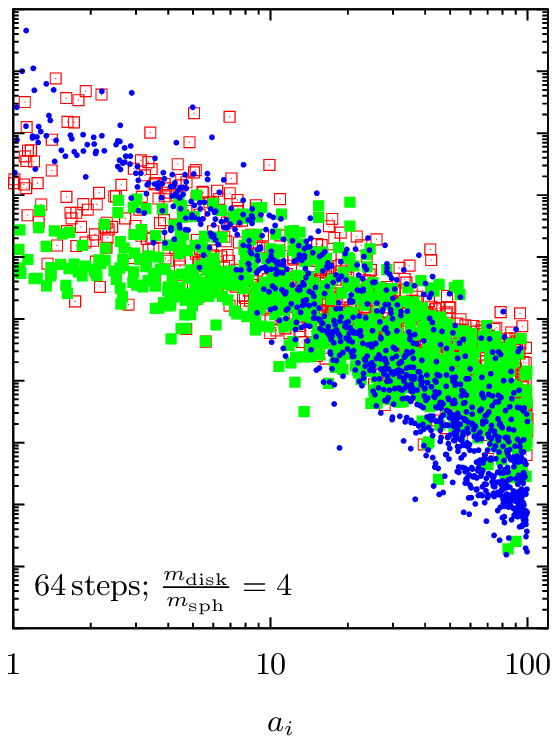}
\includegraphics[height=7cm]{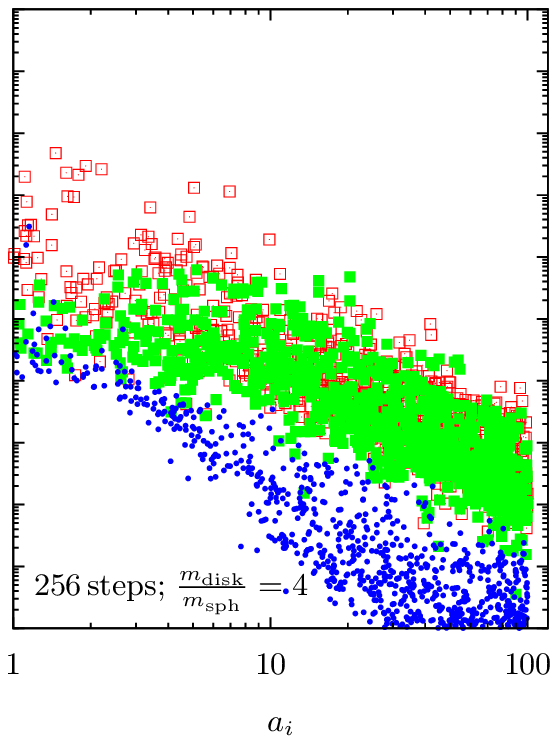}}
\mbox{\includegraphics[height=7cm]{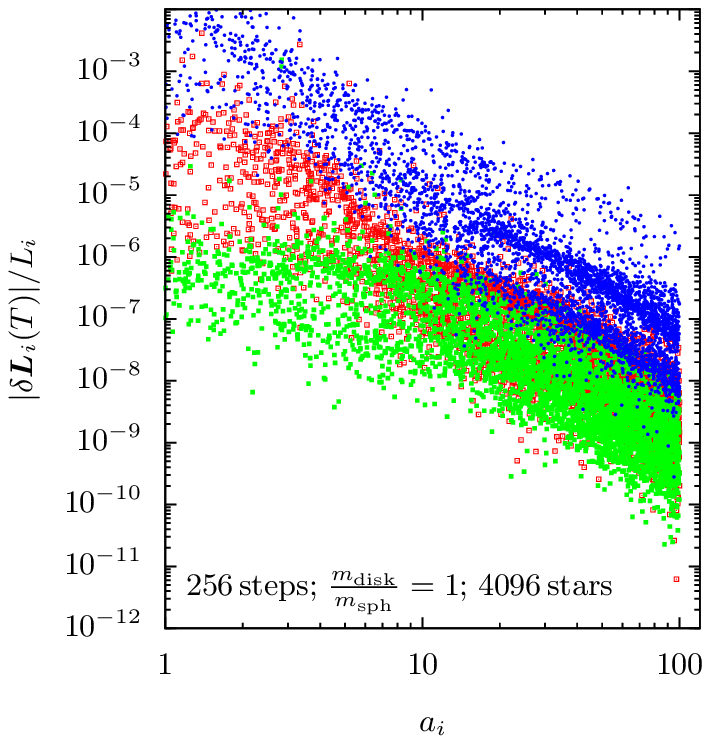}
\includegraphics[height=7cm]{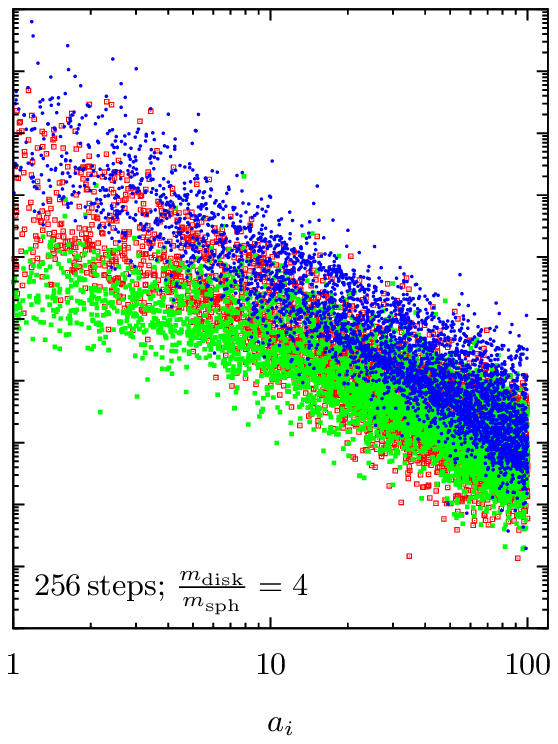}
\includegraphics[height=7cm]{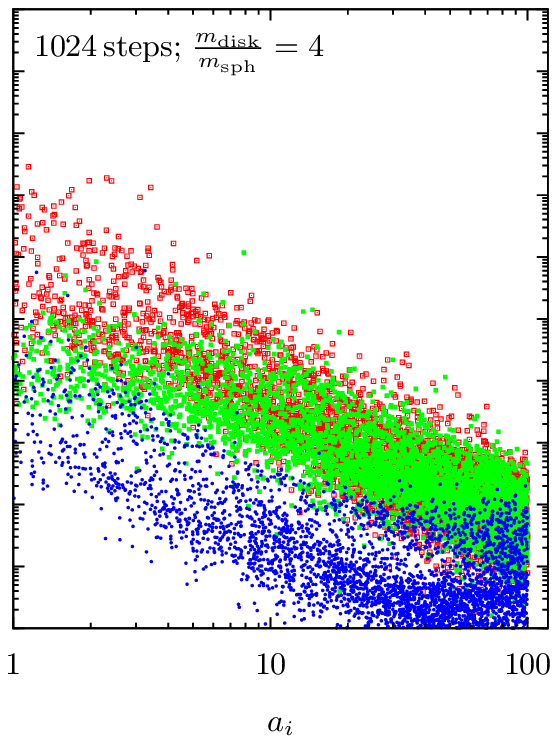}}

\caption{\label{f:error-a}
Angular-momentum convergence errors
  $\|\delta\L\|/L\equiv \|\L -\L^{\rm true}\|/L$ after a fixed time $T$,
  as a function of semimajor axis $a$. The cluster has $N=1024$ and 4096
  stars in the top and bottom rows, respectively.
  The number of simulation steps varies across the panels as marked;
  the reference angular momentum $\L^{\rm true}$ is determined by
  integrating with 4096 timesteps. Three different integrators are
  shown: (i) the open red squares show the second-order integrator
  $\O^{(2)}(\Delta t)$ (Eq.\ \ref{e:L(t)2}); (ii) the filled green
  squares show the same integrator, but with the timestep for the
  innermost $N/4$ stars reduced by a factor of 16; (iii) the small
  blue circles show the eighth-order integrator
  $\O^{(8)}(15\Delta t)$ (Eq.\ \ref{e:8th-order}; the factor 15 is
  chosen so that the second-order and eighth-order integrators have
  the same number of function evaluations per unit time).
  All simulations neglect multipoles beyond $\ell_{\max}=20$.
  The stars are initially distributed spherically and in a disk with root mean square (RMS) inclination 0.1;
  the two components have the same total mass in all panels.
  In the left panel, the sphere and disk stars have equal mass,
  in the middle and right panels the disk stars are 4 times as massive
  as the stars in the spherical component.
  The total simulated time interval corresponds to a VRR timescale of the inner edge of the cluster
$t_{\vrr}=M_{\bullet}/(m_{\rms} \sqrt{N}) P_{\min}$, $m_{\rms}=\langle m^2\rangle^{1/2}$,
  $P_{\min}=2\pi a_{\min}^{3/2}/(GM_{\bullet})^{1/2}$ (Eq.~\ref{e:vrr}).  For both the
  disk and the sphere the initial conditions are
  $n(a)\propto a^{-2.4}$, $a_{\max}/ a_{\min}=100$, $dN/de\propto e$ for $e\leq0.9$.
}
\end{figure*}

A simple way to improve the integrator to second-order (error of order
$(\Delta t)^2$ after a fixed integration time) is to choose a
time-reversible ordering of terms, e.g.,
\begin{equation}\label{e:L(t)2}
 \L(t) = \prod_{i=2}^{N}\prod_{j=1}^{i-1}\O_{ij}(\Delta t/2)\times\prod_{i=N}^{2}\prod_{j=i-1}^{1}\O_{ij}(\Delta t/2)\,
\L(t_0)
\end{equation}
Products are ordered from the initial to final values (shown on the
bottom and top of the product symbols) here and below if not stated
otherwise.  Since each term is time-reversible, i.e., $\O_{ij}(\Delta
t)\O_{ij}(-\Delta t)=\bm{I}$ is the identity operator for arbitrary
$\Delta t$, their reversible composition is time-reversible. Hence,
the truncation error after a fixed time interval must be even in the
timestep $\Delta t$ and so must be at least of order $(\Delta
t)^2$.

Higher order algorithms can be constructed by varying $\Delta t$ in
successive iteration steps
\citep{1990PhLA..150..262Y,1990PhLA..146..319S}.  For example, if we
label the second-order operator on the right side of
Eq.~(\ref{e:L(t)2}) $\O^{(2)}(\Delta t)$, an eighth-order integrator
is
\begin{equation}\label{e:8th-order}
\O^{(8)}(\Delta t) = \prod_{s=0}^{14} \O^{(2)}(r_s \Delta t)
\end{equation}
where \citep{1994PhyA..205...65S}
\begin{align}
r_0 &= r_{14} = 0.74167036435061295344822780\nonumber \\
r_1 &= r_{13} = -0.4091008258000315939973001\nonumber \\
r_2 &= r_{12} = 0.19075471029623837995387626\nonumber \\
r_3 &= r_{11} = -0.57386247111608226665638773\nonumber \\
r_4 &= r_{10} =  0.29906418130365592384446354\nonumber \\
r_5 &= r_{9} =  0.33462491824529818378495798\nonumber \\
r_6 &= r_{8} = 0.31529309239676659663205666\nonumber \\
r_7 &= -0.79688793935291635401978884\,.	
\label{e:8th-order-coefficients}			
\end{align}
Note that here 15 evaluations are required for each $\Delta t$, i.e.
the execution time of $\O^{(8)}(15\Delta t)$ is equivalent to
that of $\O^{(2)}(\Delta t)$ repeated 15 times. The truncation error of $\O^{(8)}(15\Delta t)$
is much smaller than that of $\O^{(2)}(\Delta t)$ for sufficiently small
$\Delta t$ as shown in the right panels of Figure~\ref{f:error-a}.

\subsection{Timestep refinement}\label{s:s:timerefinement}

As seen in Figure \ref{f:error-a}, the integration errors of the
innermost stars in a cluster typically greatly exceed those of the
outer stars.  This is not surprising, since the coupling coefficients
satisfy $\J_{ij\ell} \propto 1/a_{\Out}$ (Eq.~\ref{e:Jijell}), and from
Eq.\ (\ref{e:EOM1}) the characteristic timescale for changes in the
angular momentum of star $i$ is $\Delta t_{\rm int} \approx L_i/(a^3
n(a) \J_{ij \ell}) \propto a^{{\gamma}-1.5}$ where $n(a)\propto a^{-{\gamma}}$
is the number density of stars in the cluster\footnote{The interaction
  is often strongest for a stellar disk component even if it is
  subdominant in mass. The observed disk of young stars in the
  Galactic Centre has ${\gamma}=2.4$--$2.9$ and the spherical component of
  old stars has ${\gamma}=1.2$--$1.75$
  \citep{2009ApJ...697.1741B,2010ApJ...708..834B}.}.  Thus, stars at
smaller semimajor axes require a smaller timestep $\Delta t$ for the
same integration accuracy. The errors may be efficiently reduced by
implementing a block timestep procedure that preserves the symplectic
and time-reversible properties
\citep{1992JChPh..97.1990T,1994AJ....108.1962S}.  We reduce the
timestep to $\Delta t/k$ for a block containing the innermost $N/K$
stars, and calculate the mutual interactions of the stars within the
block $k$ times before calculating their interactions with the rest of
the stars.  Thus, the integrator can be written as
\begin{equation}\label{e:refinement}
\O_{\rm in, in}(\Delta t,k) \O_{\rm in, out}(\Delta t) \O_{\rm out, out}(\Delta t)
\end{equation}
where
\begin{align}\label{e:refinement-inin}
 \O_{\rm in, in}(\Delta t,k) &= \bigg[\prod_{i,j}^{j<i\leq N/K}\O_{ij}(\Delta t/k)\bigg]^{k}\,,\\
\O_{\rm in, out}(\Delta t) & = \prod_{i,j}^{j\leq N/K<i}\O_{ij}(\Delta t)\,,\\
\O_{\rm out, out}(\Delta t) & = \prod_{i,j}^{N/K<j<i}\O_{ij}(\Delta t)\,.
\end{align}

The two-level timestep refinement procedure reduces the truncation
errors of the stars in the inner block by a factor $\sim k^n$ for a
method that converges as $\mathcal{O}(\Delta t^n)$.  If the algorithm
execution time is proportional to $N^2$, the calculation of the inner
block is approximately the same cost as the calculation of the rest of
the system when $k=K^2$.

Figure~\ref{f:error-a} shows the effects of the two-level timestep
refinement procedure for a cluster with $a_{\max}/a_{\min}=100$. The
red squares show the errors when a single timestep is used, and the
green squares show the errors when using the two-level timestep
procedure (with $K=4$ and $k=16$). The errors are indeed improved by
close to $K^4=256$ at the smallest semimajor axes.  The
optimal value of $K$ may be set according to the radial range of the
simulated cluster and the number density exponent $\gamma$.

The errors may be further decreased using a Trotter decomposition in which
the combined action of the operators $e^A$ and $e^B$ is represented as
$e^{A/2} e^B e^{A/2}$ \citep{Trotter,1992JChPh..97.1990T}.
For $e^A\equiv \O_{\rm in,\In}(\Delta t,k)$ and
$e^B \equiv \O_{\rm in, out}(\Delta t) \O_{\rm out, out}(\Delta t)$,
Eq.~(\ref{e:refinement}) becomes
\begin{equation}\label{e:refinement2}
 \O_{\rm in, in}\left(\half\Delta t,\half k\right) \O_{\rm in, out}(\Delta t) \O_{\rm out, out}(\Delta t)
\O_{\rm in, in}\left(\half\Delta t,\half k\right)\,.
\end{equation}
The algorithm may be made time-reversible and hence second-order
accurate as discussed in Section~\ref{s:s:second order} by evaluating
all operators in the reverse order in successive timesteps. An
improved variant with even smaller errors is obtained by making each
$\O_{\rm in, in}\left(\half \Delta t,\half k\right)$ term in
Eq.~(\ref{e:refinement2}) time-reversible by choosing the reverse
order of the pairwise operators $\O_{ij}$ for steps
$2,4,\ldots,\half k$.

The operators $\O_{\rm in, out} \O_{\rm out, out} $ may be further
Trotter decomposed or time-symmetrized but we find that this
does not improve convergence  significantly.
The left and middle panels of Figure~\ref{f:error-a-refinement} show how the errors change for
various implementations of the two-level timestep refinement.

\begin{figure}
\centering
\mbox{\includegraphics[width=8.5cm]{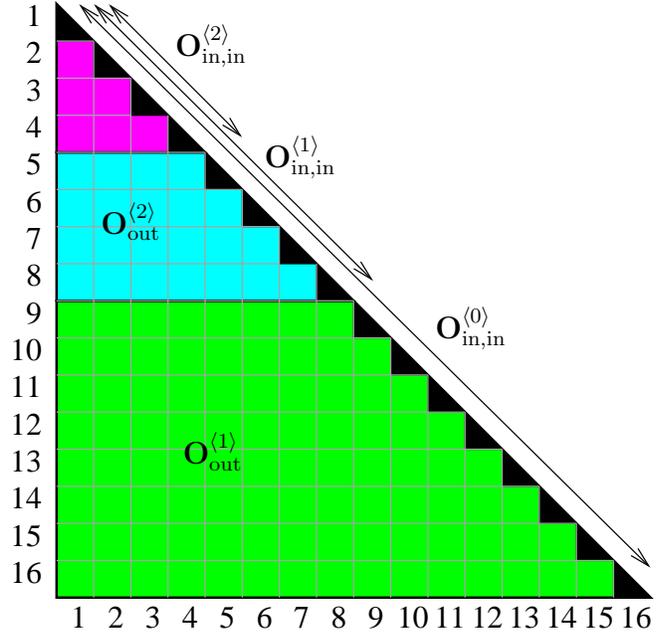}}
\caption{\label{f:refinement} Timestep refinement scheme of the
  symplectic integrator, shown for a three-level refinement with $K_1
  = K_2 = 2$ for a cluster of $N=16$ stars.  We depict the operators
  as elements of a lower triangular matrix as shown.  The algorithm
  for an arbitrary number of refinement levels runs recursively as
  follows.  In each refinement level $n<n_{\max}$ a block of $N_n$
  stars is grouped in two sets based on their specific angular
  momentum: the $(n+1)^{\rm st}$ ``inner block'' of $N_{n+1} \equiv
  N_n/K_{n+1}$ stars and the $(n+1)^{\rm st}$ ``outer block''of
  $N_n(K_{n+1}-1)/K_{n+1}$ stars.  For each refinement level, the
  inner block is further refined and the refined operators are
  executed $2k_{n+1}$ times with timestep $\Delta t_{n+1} \equiv
  \Delta t_{n}/(2k_{n+1})$ each, while the interactions among the
  outer stars and the interactions of the inner stars with the outer
  stars are executed only twice with timestep $\half\Delta t_{n}$.
  The algorithm starts with $O_{\rm in,\rm in}^{\langle 0 \rangle}$
  for $N_0=N$, which includes all stars in the inner block.
}
\end{figure}
Figure~\ref{f:error-a} shows that even after the two-level
  timestep refinement, the convergence errors vary systematically by
  three orders of magnitude over a factor 100 in semimajor axis. To
  obtain more uniform convergence, we may choose a larger inner block
(i.e., smaller $K$) and implement a multilevel refinement by
recursively refining the innermost block of stars.  To start, set the
$0^{\rm th}$ refinement level to be the whole cluster of stars
$N_0\equiv N$.  Then set the stars in the $n^{\rm th}$ refinement
level to be the innermost $N_{n} \equiv N_{n-1}/K_{n}$ stars, where
$K_{n}$ is an integer. In each refinement step, we execute the
operators corresponding to interactions among these $N_{n}$ stars with
a reduced timestep $ \Delta t_{n} \equiv \Delta t_{n-1}/(2k_{n})$;
each such operator is applied $2k_n$ times, as follows.  In the
$n^{\rm th}$ level refinement, we define the operators within the
inner block recursively as
\begin{align}\label{e:refinement3}
 \O_{\rm in,in}^{\langle n \rangle}\left( \Delta t_{n} \right) &=
 \left[ \O_{\rm in,in}^{\langle n+1 \rangle}\left(\frac{\Delta t_{n}}{2 k_{n+1}}\right)\right]^{\frac12 k_{n+1}}
\nonumber\\&\quad\times
\O_{\rm out}^{\langle n+1 \rangle}\left( \frac{\Delta t_{n}}{2} \right)
 \left[ \O_{\rm in,in}^{\langle n+1 \rangle}\left(\frac{\Delta t_{n}}{2 k_{n+1}}\right)\right]^{k_{n+1}}
\nonumber\\&\quad\times
 {\O'}_{\rm out}^{\langle n+1 \rangle}\left( \frac{\Delta t_{n}}{2} \right)
 \left[ \O_{\rm in,in}^{\langle n+1 \rangle}\left(\frac{\Delta t_{n}}{2 k_{n+1}}\right)\right]^{\frac12 k_{n+1}}\!\!.
 \end{align}
where
\begin{align}
{\O}_{\rm out}^{\langle n+1 \rangle}\left( \frac{\Delta t_{n}}{2} \right)
&=\O_{\rm in, out}^{\langle n+1 \rangle}\left( \frac{\Delta t_{n}}{2} \right)
\O_{\rm out, out}^{\langle n+1 \rangle}\left( \frac{\Delta t_{n}}{2} \right)\,,\\
 \O_{\rm in, out}^{\langle n+1 \rangle}\left( \frac{\Delta t_{n}}{2} \right)  & =
\prod_{i,j}^{j\leq N_{n+1}<i\leq N_{n}}\O_{ij}\left( \frac{\Delta t_{n}}{2} \right) \,,\\
\O_{\rm out, out}^{\langle n+1 \rangle}\left( \frac{\Delta t_{n}}{2} \right)  & =
\prod_{i,j}^{N_{n+1}<j<i\leq N_n}\O_{ij}\left( \frac{\Delta t_{n}}{2} \right) \,.
\label{e:refinement3c}
\end{align}
Here the index inside the angle brackets $\langle \cdot \rangle$
labels the refinement level, and  primed operators use the
 reverse-order composition of the unprimed operator (as in the
   operators on either side of the $\times$ in Eq.\ \ref{e:L(t)2}).
The recursion ends at the final level of refinement $n_{\max}$ for which
\begin{align}
\O_{\rm in, in}^{\langle n_{\max} \rangle}\left( \Delta t_{n_{\max}} \right)&=
\prod_{i,j}^{j<i\leq N_{n_{\max}}}\O_{ij}\left( \frac{\Delta t_{n_{\max}}}{2}\right)
\nonumber\\
&\times
\left[\prod_{i,j}^{j<i\leq N_{n_{\max}}}{\O}_{ij}\left(\frac{\Delta t_{n_{\max}}}{2}\right)\right]'
\label{e:refinement3-finalin}
\end{align}

In practice, the simulation is advanced  by $\Delta t$ by running
\begin{align}\label{e:refinement3-simulation}
 \O_{\rm simulation}(\Delta t) =  \O_{\rm in,in}^{\langle 0 \rangle}\left( \Delta t \right)
 \end{align}
 where $\O_{\rm in,in}^{\langle \cdot \rangle} (\cdot)$
 is defined by Eq.~(\ref{e:refinement3}). It is
instructive to verify that $\O_{\rm simulation}(\Delta t)$ executes
each $\O_{ij}$ operator for a total interval of $\Delta t$.  To see
this, note that Eqs.~(\ref{e:refinement3})--(\ref{e:refinement3c})
imply that $\O_{\rm in,in}^{\langle 0 \rangle}\left( \Delta t
\right)$ executes the interactions among the outer stars of the first
refinement level ($N_1=N/K_1<j\le N$) for a total time $\Delta t$,
via two operations of timestep $\half\Delta t$. These operators
will not be executed any more during this simulation step.
Furthermore $\O_{\rm in,in}^{\langle 0 \rangle}\left( \Delta t
\right)$ executes $\O_{\rm in,in}^{\langle 1 \rangle}\left(
  \half\Delta t/k_1 \right)$ for $2k_1$ times. When doing so
Eq.~(\ref{e:refinement3}) is invoked again, each time executing the
interactions among the outer stars of the second refinement level
twice with timestep $\frac{1}{4}\Delta t/k_1$ each, thus in total
for $4k_1$ times.  Thus every outer operator of the second refinement
level is run for a total time of $\Delta t$; these operators are not
executed any more during this simulation step.  The recursion
continues until the maximum refinement level is reached; at this stage
each of the inner operators is applied twice with timestep $\half
\Delta t_{n_{\max}}$. The maximum refinement level has $\Delta
t_{n_{\max}} = \Delta t / (2^{n_{\max}} k_1 k_2\cdots k_{n_{\max}})$.
Figure~\ref{f:refinement} shows the subdivisions of the operators for
a three-level refinement with $K_0=1$, $K_1=2$, and $K_2=2$.

\begin{figure*}
\centering
\mbox{\includegraphics[width=0.38\textwidth]{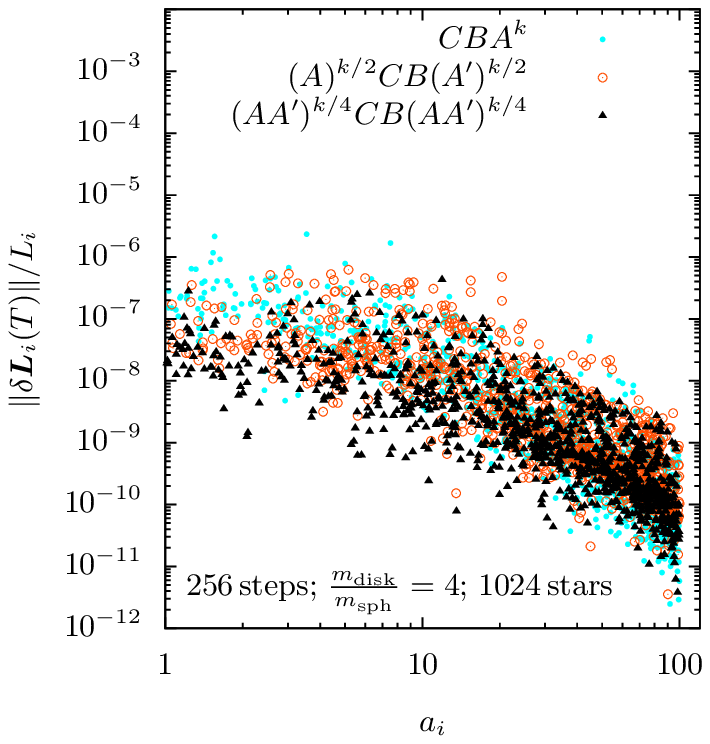}
\includegraphics[width=0.295\textwidth]{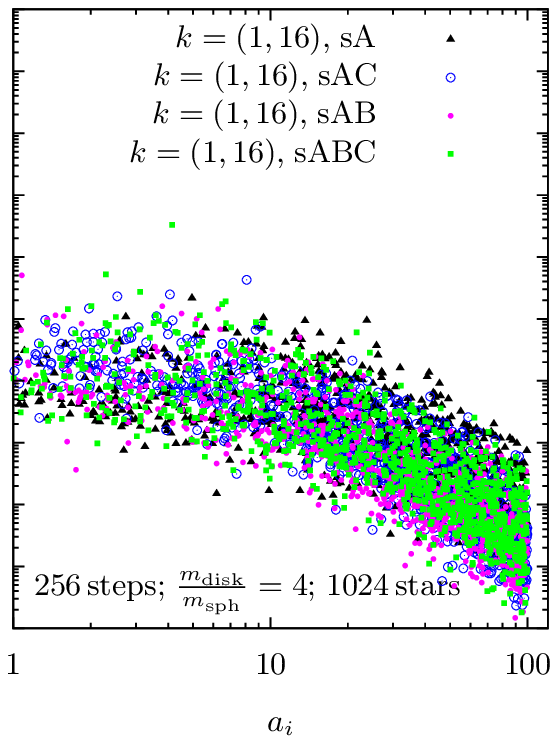}
\includegraphics[width=0.295\textwidth]{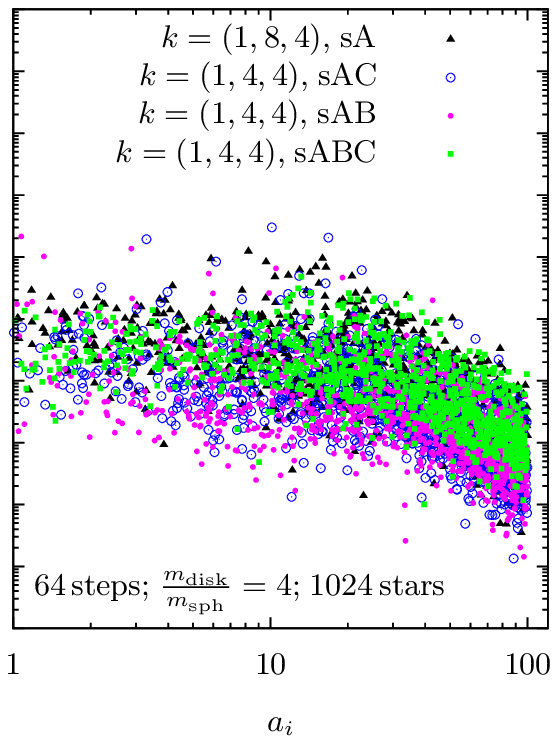}}
\caption{\label{f:error-a-refinement} Angular-momentum convergence
  errors for simulations with different refinement methods.  The left
  and middle panels show different algorithms with a two-level
  timestep refinement, the right panel shows a three-level timestep
  refinement. {\it Left panel:} The operators are labeled as follows:
  $A$ represents the interactions among the members of the inner block
  of $N/K$ stars (with semi-latus rectum $a_i(1-e_i^2)\leq 8$),
  followed with timestep $\Delta t/k$ where $K=4$ and $k=16$; $B$ is
  the mutual interaction between the members of the inner and outer
  blocks followed with timestep $\Delta t$, $B$ is the interactions
  among the members of the outer block of $N-(N/K)$ stars (with
  $a_i(1-e_i^2)> 8$) followed with timestep $\Delta t$.  Primed
  operators use the reverse-order composition of the operators in the
  corresponding unprimed operators. The simulation parameters are the
  same as in the top right panel of Figure~\ref{f:error-a}. The legend
  shows the order in which the operators are evaluated for a single
  simulation step from right to left. All refinement schemes employ
  the reverse order of operators for every second simulation step.
  The simplest refinement method $C B A^k$ improves the errors by a
  factor $\sim 40$ relative to an integrator with no refinement
  (cf. open red squares in Figure~\ref{f:error-a}).  The Trotter
  decomposition $A^{k/2} C B (A')^{k/2}$ helps to decrease errors
  further by a factor $\sim 4$--$5$. The inner-symmetric Trotter
  decomposition $(AA')^{k/4} C B (A A')^{k/4}$ method is even better,
  by another factor $\sim 2$--$3$.  {\it Middle panel:} Different
  variants of the inner-symmetric Trotter decomposition given by
  Eqs.~(\ref{e:refinement3-sA})--(\ref{e:refinement3-sABC}) labeled
  sA, sAB, sAC, and sABC, respectively.  All variants show comparable
  errors.  {\it Right panel:} Three-level $K=2$ refinement using the
  same algorithms and timestep as in the middle panel.  The $sABC$
  method produces the most uniform errors, and smallest maximum
  errors.  The two-level timestep-refined simulations execute 256
  steps in $\sim50\%$ more time than the three-level timestep-refined
  algorithms with 64 simulation steps.  }
\end{figure*}

Note that the reverse-order composition of operators, denoted by
primes, has been invoked in Eqs.~(\ref{e:refinement3}) and
(\ref{e:refinement3-finalin}) to make the algorithm time-reversible.
For an overview, suppressing the arguments, the refinement scheme may
be summarized as
\begin{align}\label{e:refinement3-sA}
\O_{\rm in,in}^{\langle n \rangle} = &
 \big(\O_{\rm in,in}^{\langle n+1 \rangle}\big)^{\frac12 k_{n+1}} \O_{\rm in,out}^{\langle n+1 \rangle} \O_{\rm out,out}^{\langle n+1 \rangle}
\big(\O_{\rm in,in}^{\langle n+1 \rangle}\big)^{k_{n+1}}\nonumber\\
&\;\times {\O'}_{\rm out,out}^{\langle n+1 \rangle} {\O'}_{\rm in,out}^{\langle n+1 \rangle}
\big(\O_{\rm in,in}^{\langle n+1 \rangle}\big)^{\frac12 k_{n+1}}.
\end{align}
With this algorithm $\O_{\rm in,in}^{\langle n\rangle}=\O_{\rm
  in,in}^{\prime\langle n\rangle}$ at all refinement levels $n$.
Alternatively, we may time-symmetrize according to any of the
following schemes,
\begin{align}\label{e:refinement3-sAB}
&(\O_{\rm in,in}^{\langle n+1 \rangle})^{k_{n+1}} \O_{\rm in,out}^{\langle n+1 \rangle}\O_{\rm out,out}^{\langle n+1 \rangle}
{\O'}_{\rm out,out}^{\langle n+1 \rangle}{\O'}_{\rm in,out}^{\langle n+1 \rangle} (\O_{\rm in,in}^{\langle n+1 \rangle})^{k_{n+1}},\\
\label{e:refinement3-sAC}
&(\O_{\rm in,in}^{\langle n+1 \rangle})^{k_{n+1}} \O_{\rm out,out}^{\langle n+1 \rangle}\O_{\rm in,out}^{\langle n+1 \rangle}
{\O'}_{\rm in,out}^{\langle n+1 \rangle}{\O'}_{\rm out,out}^{\langle n+1 \rangle} (\O_{\rm in,in}^{\langle n+1 \rangle})^{k_{n+1}},\\
\label{e:refinement3-sABC}
&(\O_{\rm in,in}^{\langle n+1 \rangle})^{k_{n+1}} \O_{\rm in,out}^{\langle n+1 \rangle}{\O'}_{\rm in,out}^{\langle n+1 \rangle}
\O_{\rm out,out}^{\langle n+1 \rangle}{\O'}_{\rm out,out}^{\langle n+1
  \rangle} (\O_{\rm in,in}^{\langle n+1 \rangle})^{k_{n+1}}\,.
\end{align}
Figure~\ref{f:error-a-refinement} shows the convergence
errors for Eqs.~(\ref{e:refinement3-sA})--(\ref{e:refinement3-sABC})
labeled by sA, sAB, sAC, and sABC, respectively. All four methods
employ a three-level timestep refinement with $K=(1,2,2)$.  The
repetition factors are $k=(1,8,4)$ for sA and $k=(1,4,4)$ for the
other three methods.  The execution times are comparable for each
algorithm with 64 simulation steps and for the two-level timestep
algorithms with 256 steps in Figure~\ref{f:error-a-refinement}.
The sABC method
(Eq.~\ref{e:refinement3-sABC}) has the most homogeneous errors and
smallest maximum errors.  This algorithm has the most number of
time-reversible factors, including the inner and outer blocks of stars
and the mutual interactions between the two.

\subsection{Grouping terms in blocks}\label{s:s:blocks}

The accuracy of the integrator can be significantly improved by
choosing a particular order in which the interactions in
Eq.~(\ref{e:L(t)2}) are calculated.  One way to achieve this is by
grouping the stars into blocks such that the most strongly coupled
stars are mostly in the same block. Since the interactions are much
weaker if the semimajor axes are widely separated, $\alpha_{ij}\ll 1$,
and the precession rate is slower for less eccentric orbits,
it is natural to define the blocks using criteria based on the
semimajor axes or specific angular momenta $L_i/m_i \propto
\sqrt{a_i(1-e_i^{2})}$ of the stars. A specific assignment procedure
is described in the next subsection. After defining the blocks, we
evaluate the interactions block-by-block, first evaluating all the
interactions within each block then the interactions between blocks,
\begin{equation}\label{e:blocksprod}
 \prod_{a=1}^{B}\prod_{b=1}^{a} \O^{a,b} \times {\rm reverse~order}
\end{equation}
where $\O^{a,b}$ denotes the product of all pairwise interaction terms
between blocks $a$ and $b$, and ``$\rm reverse~order$'' denotes the
time-reversed composition of operators.

\subsection{Parallelization}

The main bottleneck of the symplectic integrator outlined above is the
steep scaling with the number of stars, at least $\mathcal{O}(N^2)$.
Each timestep requires the calculation of $N(N-1)/2$ interactions.
Furthermore, errors arise due to the noncommutativity of different
terms which further increase with $N$.  The steep scaling with $N$
makes it unfeasible to simulate clusters with a realistic number of
stars on a single processor. Here we show how to parallelize the
algorithm to reduce the execution time.

Since the symplectic algorithm outlined above uses a composition of
operators in a particular order, it is not immediately obvious whether
it is possible to run the algorithm on parallel threads.  Fortunately,
we may realize that each operator $\O_{ij}$ affects only $\L_i$ and
$\L_j$ and the strict sequential ordering of $\O_{ij}$ and $\O_{kl}$
is not necessary if $i$ and $j$ are different from $k$ and $l$. In
particular if we split the stars into two disjoint blocks, the
self-interactions of the blocks may be calculated in parallel by two
threads, followed by a sequential calculation of the mutual
interaction between blocks. More generally, we may split the operators
into many segments of the form $\O_{i_1,i_2} \O_{i_3,i_4}\dots
\O_{i_{N-1},i_N}$ where $(i_1,i_2,\dots, i_N)$ is a permutation of
$(1,2,\dots,N)$. Then all of these $N/2$ operators commute within this
sequence, and can be evaluated independently on parallel
threads (we show how to do this below).

\begin{figure}
\centering
\mbox{\includegraphics[width=8.5cm]{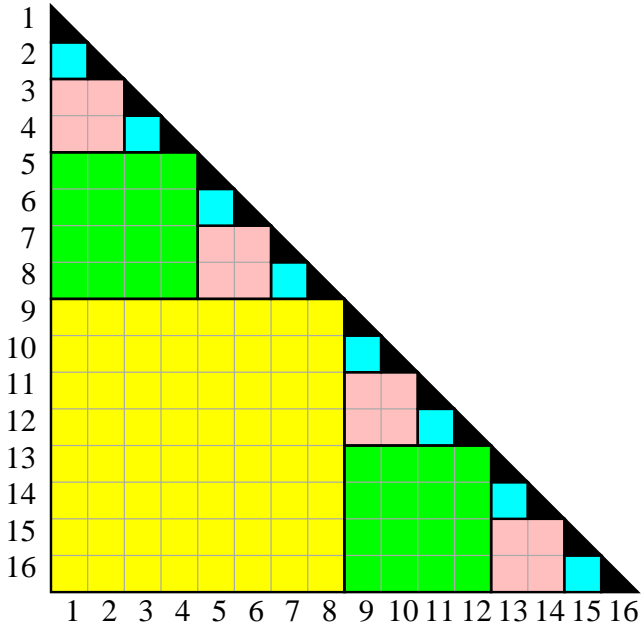}}
\caption{\label{f:parallel}
Parallelization scheme of the symplectic integrator. The interaction is
calculated as the composition of the effects of pairwise interaction terms. We
depict the interaction terms between stars $i$ and $j$ as elements of a lower
triangular matrix and group them in tiles of size $2^t$ as shown.
Tiles of the same size commute, and can be executed in parallel.
Further, interactions within a diagonal of a given tile also commute,
but different diagonals within a given tile do not, nor do different size tiles.
Thus, synchronization is necessary between executing the interactions of different
diagonals within a given tile and between different size tiles.
For an unlimited number of processors, the algorithm execution time is
$\mathcal{O}(N)$.
If the number of available processors $P$ is less than $N/2$,
the parallel algorithm run-time scales as $\mathcal{O}[(N(N-1)/2P]$
and requires exactly $2P$ synchronizations independent of
$N$.
}
\end{figure}

With this background in mind, we construct a parallel method for
$N=2^n$ stars as shown in Figure~\ref{f:parallel}.  We depict the
operators as elements of a lower triangular matrix, and group them
into tiles of size $2^t$ with $t=0,1,\dots,n-1$ as shown for $N=16$.
We construct the tiling by recursively removing square tiles of size
$2^t\times 2^t$ starting with the largest, $t=n-1$. Removing this
submatrix leaves two lower triangular matrices, half the size of the
original.  Next we remove the $2^{t-1}\times 2^{t-1}$ square matrices
from the two triangular matrices, leaving two smaller triangular
matrices each. We repeat this iteration down to $t=0$, thereby
covering the matrix completely. This gives $2^{n-t-1}$ square tiles of
size $2^t$. The elements of tile $k=0,1,\ldots,2^{n-t-1}-1$ of
size $2^t$ are $\O_{ij}$ where $1\leq i - (2k+1) 2^t \leq 2^t$ and
$1\leq j - (2k) 2^{t}\leq 2^t$. All tiles of a given size represent
interactions between distinct groups of stars (i.e., the tiles of a
given color in Figure \ref{f:parallel} do not overlap horizontally or
vertically). Thus, different tiles of the same size commute.

Next we discuss the commutativity of operators within a given tile.
Note that the operators in any diagonal within a tile
commute. This leads to a
parallelization scheme based on diagonals, which is best described by
an example. In the top green square in Figure~\ref{f:parallel},
$(n,t,k)=(4,2,0)$, we may choose the following ordering
\begin{align}
&(\O_{51}\O_{62}\O_{73}\O_{84})(\O_{52}\O_{63}\O_{74}\O_{81})\nonumber\\
&\quad\times(\O_{53}\O_{64}\O_{71}\O_{82})
(\O_{54}\O_{61}\O_{72}\O_{83})\,.
\end{align}
The terms in each parenthesis commute and can be evaluated in parallel,
but synchronization is required between the parentheses.
In summary, we may evaluate the action of all the $\O_{ij}$ as follows
\begin{align}\label{e:parallel1}
 \prod_{t=0}^{n-1}
 \prod_{d=1}^{2^{t}}\bigg(
 \prod_{i=1}^{2^{t}}
 \prod_{k=0}^{2^{n-t-1}-1}
\O_{(2k+1) 2^t + i,\,(2k) 2^t + [ (i+d)\, {\rm mod}\, 2^t]}\bigg)
\end{align}
where the terms in the large parentheses commute and can be run on independent threads.

More generally, instead of diagonals, we may choose any $2^t$
long cycle of permutations of $(1\dots 2^t)$, labelled $Z_{(2^t)}$,
to cover all elements of a tile
\begin{align}\label{e:parallel2}
 \prod_{t=0}^{n-1}
 \prod_{\sigma\in Z_{(2^t)}}\bigg(
 \prod_{i=1}^{2^{t}}
 \prod_{k=0}^{2^{n-t-1}-1}
\O_{(2k+1) 2^t + i,\,(2k) 2^t + \sigma_i}\bigg)\,.
\end{align}
Choosing random instead of fixed permutations for different simulation steps
helps to decrease systematic errors that arise due to the noncommutativity
of terms.

With at least $N/2$ processors, each parenthesis in
Eqs.~(\ref{e:parallel1})--(\ref{e:parallel2}) can be evaluated in a
time $\tau_e$, where $\tau_e$ denotes the execution time corresponding
to a single $\O_{ij}$ operator.  Different threads need to synchronize
data between evaluations of non-commuting operators, and we denote the
corresponding time overhead by $\tau_s$. The execution time of one
timestep of the simulation is then $\sum_{t=0}^{n-1}\sum_{d=1}^{2^t}
(\tau_e+\tau_s)= (N-1)(\tau_e + \tau_s)$, so the parallelized
simulation time scales as $N$.  For a limited number of processors
$P=2^p \leq N/2$, the time for evaluating the operators in one
timestep without synchronizations is $N(N-1)\tau_e/(2P)$.  In this
case the optimal processor allocation that provides the minimum number
of synchronizations is determined as follows. First split the stellar
system into $B=2P$ blocks of stars, and calculate all of the
interactions within a block on the same processor. Next, to calculate
the $B(B-1)/2$ mutual interactions between blocks, we tile the blocks
according to the same binary tree scheme as shown in
Figure~\ref{f:parallel}.  The interactions of different tiles of the
same size commute.  Therefore we can evaluate the mutual interactions
between blocks in the order given by
Eqs.~(\ref{e:parallel1})--(\ref{e:parallel2}). The calculation
requires synchronization after each diagonal of the tiles and after
calculating the self-interactions of blocks: $2P$ synchronizations in
total, independent of $N$. Thus, the execution time for $P< N/2$ is
$N(N-1)\tau_e/(2P) + 2 P \tau_s $.

The parallelization scheme outlined above applies for an arbitrary
indexing of stars.  In practice we may also employ all of the
improvements discussed in Sections~\ref{s:s:second
  order}--\ref{s:s:timerefinement} to further speed up the
calculation. The multilevel refinement outlined
in Section~\ref{s:s:timerefinement} is commensurate with
this parallelization scheme as long as the $K$ refinement levels are
powers of 2. When the timestep is decreased by a factor $k$,
the execution time increases by the same factor for the corresponding
$(N/K)^2$ operators. However, the number of synchronization steps
increases significantly for each refinement level since
that is independent of $N$.

\subsection{Summary}
First we summarize the algorithm for the eighth-order integrator
(\ref{e:8th-order}), but without timestep refinement; the description
for the second-order integrator is an obvious simplification of this one:

\begin{enumerate}[leftmargin=0.5cm,itemsep=1.5ex,label=\arabic*.]
 \item Calculate and store the coupling coefficients $\J_{ij\ell}$ for all $i$ and $j$
and for $\ell=2,4,\dots, \ell_{\max}$.

 \item Order stars according to semimajor axis or specific angular
   momentum and divide into tiles as illustrated in Figure
   \ref{f:parallel}.

 \item Choose a random permutation for each tile.  Set the timestep to
   $\Delta t_s = r_s\Delta t$ for the eighth-order integrator
   (Eq.~\ref{e:8th-order}). Repeat the following for $s=0,\ldots,14$
   to advance all pairs of stars $i$ and $j$ by substeps $\Delta t_s$:

\begin{enumerate}[itemsep=1.5ex]

 \item\label{tile} Starting with the smallest tile size ($t=0$ in Eq.\
   \ref{e:parallel2}), use parallel processors to operate on the
   elements of a given permutation within a tile and the different
   tiles of the same size [the products over $k$ and $i$ in
   Eq.~(\ref{e:parallel2})].

 \item\label{permutation} Repeat this process for the different
   permutations of a given tilesize [the product over $\sigma$ in
   Eq.\ (\ref{e:parallel2})].

 \item\label{tilesize} Repeat this for the different size tiles
   ($t=1,\ldots,n$).

 \item\label{reverse} Repeat the previous three steps in reverse order.

\end{enumerate}

 \end{enumerate}

 In the algorithm with a two-level timestep timestep refinement and
 second-order integrator, iterations \ref{tile}--\ref{tilesize} go as
 follows:

\begin{enumerate}[leftmargin=0.5cm,itemsep=1.5ex]

\item\label{i:inner1} Advance the innermost $N/K$ stars (those with
  the smallest indices) with a reduced timestep $\Delta t_s/{k}$ for a
  total time interval $\Delta t_s/2$, by repeating iterations
  \ref{tile}--\ref{tilesize} $k/2$ times.  In every second iteration
  we reverse the ordering of the operators.

\item\label{i:outer} Evolve the rest of the interactions among the
  outer $N(K-1)/K$ stars and the mutual interactions between the inner
  and outer stars with a timestep $\Delta t_s/2$ and then in the reverse order
  for $\Delta t_s/2$.

 \item\label{i:inner2} Repeat step \ref{i:inner1} to evolve the inner block again
for a total time interval $\Delta t_s/2$.

\end{enumerate}

Note that each operator is evaluated for a total $\Delta t_s$ after each iteration
\ref{i:inner1}--\ref{i:inner2}. Methods with higher order refinements decompose
the inner cluster further and repeat steps \ref{i:inner1}--\ref{i:inner2} for
each level of refinement.

The cluster composition (in particular the mass and radius
distribution of the stars) and the error tolerance determine the
optimal $K$ and repetition factors $k$ and the most efficient order
for the integrator (see Figures~\ref{f:error-a} and \ref{f:error-a-refinement}).
The value of $\ell_{\max}$ is chosen such
that $\ell_{\max}=\pi/(2I_{\min})$ where $I_{\min}$ is the minimum
inclination that must be resolved by the simulation (see
Appendix~\ref{app:convergence}).

\section{Vector Resonant Relaxation as a Stochastic Process}

\label{s:random}
As an application of these results, we examine VRR of a spherical stellar cluster around a
SMBH \citep{1996NewA....1..149R, 2006ApJ...645.1152H,
  2007MNRAS.379.1083G,2009ApJ...698..641E,2011MNRAS.412..187K,
  2011MNRAS.411L..56G,2011ApJ...738...99M,2011ApJ...726...61M}.  As
discussed in Section \ref{s:intro}, VRR is the stochastic process
arising from the torques between the annuli that represent stellar
orbits that have been averaged over the orbital period and apsidal
precession time. The adjective ``vector'' refers to the fact that such
torques change the orientation of the angular-momentum vector but not
the scalar angular momentum (Eq.~\ref{e:Ltot}).

In the standard (Chandrasekhar) model of two-body relaxation in
stellar systems \citep[e.g.,][]{bt08}, each star undergoes a random
walk in Cartesian velocity space due to encounters with stars passing
nearby. In the incoherent phase of VRR, each star undergoes a
random walk in $\Ln$ on the unit sphere due to torques from other
stars. Two-body relaxation can be approximated as Brownian motion,
that is, most of the relaxation is due to a large number of encounters
of short duration. In contrast, in VRR the stochastic motion of the
orbit normals cannot be divided into discrete steps occurring at a
fixed and very short time interval $\Delta t$. In other words, VRR is unlike Brownian motion or
diffusion in that the angular momenta move in a coherent, spatially
correlated manner until their directions change substantially and they
exibit incoherent, stochastic evolution only over much longer times.  For this
reason, the correlation function of angular momentum vector directions
\emph{cannot} be expressed as
$\|\L_i(t_0+\tau)-\L_i(t_0)\|/\|\L_i(t_0)\| = (\tau/{t_\vrr})^{1/2}$
in the incoherent evolutionary phase, and the definition of the vector resonant
relaxation timescale $t_\vrr$ must be revised.

In Section~\ref{s:random-theory}, we introduce a simple stochastic
model to describe incoherent VRR in a spherical stellar cluster,
in which the angular momentum vector directions undergo an isotropic
random walk on a spherical surface with a step size which is not
infinitesimal and which is drawn from a probability distribution
function (PDF). For any given PDF, we show that the stochastic
evolution may be solved analytically and that the multipole moments of
the correlation function with $\ell>0$ decay exponentially
(Eq.~\ref{e:randomwalk-mean}).  We use this property to define the VRR
timescale (Eq.~\ref{e:V-incoherent}) and construct moments of the
stellar distribution (Eq.~\ref{e:V}) that evolve linearly in time
(Eq.~\ref{e:Vell-theory2}). In Section~\ref{s:random-application}, we
analyse the results of our numerical simulations in this framework,
and in Section~\ref{s:comparison}, we compare results in the
literature for the coherent evolutionary phase of VRR with those in
this study.

\subsection{Random walk on the sphere -- general theory}
\label{s:random-theory}

In general, a random walk on a sphere can be described as follows
(\citealt{Roberts_Ursell60}; see also \citealt{1929pomo.book.....D,2012leas.book.....C}).
Suppose that the probability distribution
for the initial position of a point $\r_0$ on the spherical surface of
unit radius, $S_2$, is $\rho_0(\r)$. At step $n$, $\r$ moves an angle
$\alpha_n=\cos^{-1}\mu_n$ on the sphere in a random direction with
probability $p(\mu_n)\D\mu_n$. Therefore, the probability density after
the $n^{\rm th}$ step is set by the probability density of the
preceding step as\footnote{
We define the distribution function of $\r$ as a random field
$\rho_n(\r) \equiv
\rho_n[\r ; \rho_{n-1}(\r')_{\r'\in S_2}, \mu_n]\equiv
 \rho_n[\r ; \rho_0(\r')_{\r'\in S_2}, \mu_1,\dots, \mu_n]$ using Eq.~(\ref{e:randomwalkdef}).
Here the $\mu_i$ are independent random variables for all $i$ and
$\rho_0(\r')$ is a given initial distribution for $\r'\in S_2$.
}
\begin{equation}
\rho_n(\r) =
\frac{1}{2\pi}\int_{S_2} \D \r'
\delta(\r \cdot \r'- \mu_n)\,\rho_{n-1}(\r')\,.
\label{e:randomwalkdef}
\end{equation}
This equation is linear in $\rho$ and can be solved using the
eigenbasis of the corresponding linear operator.  In
Appendix~\ref{s:app:randomwalk}, we show that the eigenfunctions are
the spherical harmonics\footnote{See definition in Eq.~(\ref{e:Y}).}
$Y_{\ell m}(\r)$ with eigenvalues $P_{\ell}(\mu_n)$.  Expanding the
initial distribution in this basis as
\begin{equation}\label{e:randomwalk-rho0}
 \rho_0(\r) = \sum_{\ell, m}a_{\ell m,0}Y_{\ell m}(\r)\,,
\end{equation}
the distribution after a single step is
\begin{equation}\label{e:randomwalk1}
 \rho_1(\r) = \sum_{\ell,m}
 P_{\ell}(\mu_1) \, a_{\ell m,0} Y_{\ell m}(\r)\,,
\end{equation}
and after the $n^{\rm th}$ step it is
\begin{equation}\label{e:randomwalk-rho}
 \rho_{n}(\r) = \sum_{\ell, m} a_{\ell m, n}Y_{\ell m}(\r)
\end{equation}
where
\begin{equation}\label{e:randomwalk-a}
 a_{\ell m, n} = \prod_{k=1}^n P_{\ell}(\mu_k) \, a_{\ell m,0}\,.
\end{equation}
The expectation value of the $(\ell,m)$ spherical multipole moment in the $n^{\rm th}$ step
is
\begin{equation}\label{e:randomwalk-mean}
\langle a_{\ell m, n} \rangle = \langle P_{\ell}(\mu) \rangle^n a_{\ell m,0}
\end{equation}
where $\langle F(\mu_k) \rangle = \int_{-1}^1 F(\mu_k) p(\mu_k) \D \mu_k$ for any function $F(\mu_k)$.
The RMS fluctuations around the mean are given by
\begin{equation}\label{e:randomwalk-sigma}
\sigma^2\equiv \langle a_{\ell m,n }^2\rangle -\langle a_{\ell m, n}\rangle^2
=  \left\{ \left\langle [P_{\ell}(\mu)]^2 \right\rangle^n - \left\langle P_{\ell}(\mu) \right\rangle^{2n}  \right\}\, a_{\ell m,0}^2
\end{equation}
and the cross-correlation of $a_{\ell m, n}$ and $a_{\ell' m', n}$
\begin{align}\label{e:randomwalk-correlation}
C^{\ell m}_{\ell' m'} &\equiv \langle a_{\ell m,n } a_{\ell' m',n } \rangle -\langle a_{\ell m, n}\rangle\langle a_{\ell' m', n}\rangle\\
&=  \left\{ \left\langle P_{\ell}(\mu) P_{\ell'}(\mu) \right\rangle^n
- \left\langle P_{\ell}(\mu)\right\rangle^n \left\langle P_{\ell'}(\mu) \right\rangle^n  \right\}a_{\ell m,0}a_{\ell' m',0}.
\end{align}

Since $|\langle P_\ell(\mu) \rangle | \leq 1$ for $\ell>0$, each
multipole moment with $\ell>0$ decays exponentially in the number of
steps as $|\langle a_{\ell m, n} \rangle|/|a_{\ell m, 0} | = \exp[n
\ln |\langle P_\ell(\mu) \rangle|]$; the system ``isotropizes'' with a
decay time of $-\Delta t / \ln |\langle P_{\ell}(\mu) \rangle |$ where
$\Delta t$ is the timestep.

Since $x\equiv |a_{\ell m,n}/a_{\ell m,0}|$ is an
$n$-element product of independent and identically distributed positive random variables for any
$\ell$ and $m$, the distribution of $\ln x$ for $n\gg 1$
follows from the central limit theorem, and we find that the
probability density function of $x$ is approximately
\begin{equation}\label{e:distribution1}
 \varphi(x) \approx \frac{1}{\sqrt{2\pi n} \,x \sigma_0 }
\exp\hspace{-1pt}\left[ - \frac{(\ln x -
n\nu)^2}{2 n \sigma_0^2} \right]
\end{equation}
where
\begin{equation}
\nu\equiv \langle \ln |P_\ell(\mu)|\rangle, \quad \sigma_0^2\equiv\langle
[\ln|P_\ell(\mu)|]^2\rangle -\langle \ln |P_\ell(\mu)|\rangle^2\,.
\label{e:jjjppp}
\end{equation}
Note that the mean and RMS of $a_{\ell m,n}$ are given generally by
Eqs.~(\ref{e:randomwalk-mean})--(\ref{e:randomwalk-sigma}), while
Eqs.~(\ref{e:distribution1})--(\ref{e:jjjppp}) are approximate
statements valid only when $n\gg1$.

The Green's function corresponding to an initial density
$\rho_0$ that is concentrated at the $\theta=0$ pole corresponds to
$a_{\ell m,0} =\sqrt{(2\ell + 1) /(4\pi)}\delta_{m,0}$. Thus the
probability distribution function for the angle $\theta$
between the initial and final position after $n$ steps is given by
 \begin{equation}
p_n(\theta)= 2\pi\rho_n(\r)\sin\theta =\sum_{\ell=0}^\infty
\frac{2\ell +1}{2}\prod_{k=1}^n P_{\ell}(\mu_k)\,P_{\ell}(\cos\theta)\sin\theta,
\end{equation}
which implies that\footnote{This quantity is related to the
  autocorrelation function of the spherical multipole moments since
  $\overline{P_\ell(\cos\theta)}= {4\pi\,(2\ell+1)^{-1}
  \sum_{m=-\ell}^{\ell} a_{\ell m,n} a^*_{\ell m,0}}$. }
\begin{equation}\label{e:ppp}
\overline{ P_\ell(\cos\theta)}
\equiv \int_0^\pi P_\ell(\cos\theta) p_n(\theta)\,\D\theta = \prod_{k=1}^n P_{\ell}(\mu_k)
 \end{equation}
where overbar denotes the average over $p_n(\theta)$. Thus after averaging over all $\mu_k$
and $p_n(\theta)$ we get
\begin{equation}\label{e:probability-average}
 \overline{ \langle  P_\ell(\cos\theta) \rangle}   = \langle P_{\ell}(\mu)\rangle^{n}\,.
\end{equation}

In a planar random walk with step $\alpha$, the RMS distance traveled
after $n$ steps is $\sqrt{n}\alpha$. This formula does not apply to
the random walk on a sphere unless $\sqrt{n}\alpha\ll1$, since the
geometry is not planar (for example, the maximum angular distance
 between any two points on a sphere is $\pi$).  To generalize some of
 the concepts of planar random walks to the sphere, we first consider the limiting case
 of Brownian motion, in which the angular step $\alpha=\cos^{-1}\mu$ and
 the timestep $\Delta t$ both approach zero with
 $\alpha^2\sim \Delta t$. In this limit $P_{\ell}(\mu)
 \approx \exp[-\frac14 \ell (\ell+1) \alpha^2]$, and so Eqs.\
 (\ref{e:randomwalk-rho})--(\ref{e:randomwalk-a}) become\footnote{
Brownian motion on the sphere also satisfies the diffusion equation \citep{1929pomo.book.....D}
\begin{equation}
 \frac{d\rho}{dt} = \frac14 \nabla\cdot \frac{\langle\alpha^2\rangle}{\Delta t}\nabla\rho\,.
\end{equation}
where $\nabla$ is the gradient operator on the unit sphere.
}
\begin{equation}\label{e:randomwalk-brownian}
 \rho_{n}(\r) = \sum_{\ell,m} a_{\ell m,0} Y_{\ell m}(\r) e^{-\frac14 \ell (\ell+1) v_n}
\end{equation}
where $v_n = \sum_{k=1}^n\alpha_k^2$,  so that
$\langle v_n\rangle = n \langle \alpha^2\rangle = \langle\alpha^2\rangle t
/\Delta t $ is the variance of the corresponding planar
Brownian motion.
The analog of Eq.~(\ref{e:ppp}) is
\begin{equation}
\overline{P_\ell(\cos\theta)} = e^{-\frac14 \ell(\ell+1) v_n}.
\end{equation}

Motivated by the results above, we define the quantity
\begin{equation}
\label{e:V}
 V_{\ell}(t) \equiv - \frac{4}{\ell(\ell+1)} \ln \bigg| \frac1{N'} \sum_{i=1}^{N'}\frac1{T} \int_0^{T} \D t_0 P_{\ell}[ \cos \alpha_i(t,t_0) ] \bigg|
\end{equation}
which we call the angular variance; here $\alpha_i(t,t_0)$ is the angular distance traversed
by the orbit normal $\Ln_{i}$ between time $t_0$ and time $t_0+t$, i.e.,
\begin{equation}
  \cos \alpha_i(t,t_0) \equiv {\Ln}_{i}(t+t_0) \cdot {\Ln}_{i}(t_0)\,.
\end{equation}
In Eq.~(\ref{e:V}), we have averaged $P_{\ell}(\cos \alpha_i)$ over
both the cluster index and the reference time to reduce statistical noise.
 The ensemble average is either
over the full population ($N'=N$) or over a subset of the stars
($N'<N$, e.g., over stars within a restricted range of mass,
eccentricity, and semimajor axis).  For Brownian motion
$V_{\ell}(t_n)$ is an estimator of the variance $v_n$ for all $\ell$
so long as $v_n\ll 1$, and for a general random walk it estimates
$-4\ell^{-1}(\ell+1)^{-1}n\ln|\langle P_{\ell}(\mu)\rangle|$.  In
either case $V_{\ell}(t)$ grows linearly with time over timescales
long compared to the timestep $\Delta t$ until the $\ell^{\rm th}$
multipole becomes completely mixed.  Complete mixing occurs when the
level of anisotropy becomes less than the stochastic variations which
arise due to the finite number of stars. Thus for a single component
cluster, complete mixing occurs when $V_\ell\approx V_{\ell,\sat}$
with $\ell\geq 1$ and
\begin{align}
 \exp\left[ -\ffrac{1}{4}\ell (\ell+1) V_{\ell,\sat}\right]
&\equiv \frac{1}{\sqrt{N}}\langle [P_{\ell}(\cos\alpha)]^2\rangle^{1/2}\nonumber\\
&= \frac{1}{\sqrt{N(2\ell+1)}}\,;
\end{align}
 in the last line we assumed that $\alpha$ is drawn from an isotropic
 distribution.  Solving for $V_{\ell, \sat}$ gives
\begin{align}\label{e:Vell-sat}
V_{\ell,\sat} = \frac{2\ln\left[(2\ell+1)N\right]}{\ell(\ell+1)}\,.
\end{align}

In summary, for Brownian motion, the angular variance is expected to follow
\begin{align}\label{e:Vell-theory1}
 V_{\ell}(t)
=\left\{
\begin{array}{l}
\displaystyle{ \frac{t}{\Delta t} \langle \alpha^2 \rangle  {~~\rm if~~}\langle \alpha^2\rangle^{1/2} \ll 1 {~~\rm and ~~} V_{\ell}< V_{\ell,\sat}}\,,\\[2ex]
\text{stochastic variations around $V_{\ell,\sat}$ otherwise\,,}
\end{array}
\right.
\end{align}
and for a general random walk
\begin{align}\label{e:Vell-theory2}
 V_{\ell}(t)
=\left\{
\begin{array}{l}
\displaystyle{ -\frac{4}{\ell(\ell+1)} \frac{t}{\Delta t} \ln|\langle P_{\ell}(\mu)\rangle|  {~~\rm if~~}V_{\ell}< V_{\ell,\sat}}\,,\\[2ex]
\text{stochastic variations around $V_{\ell,\sat}$ otherwise\,.}
\end{array}
\right.
\end{align}
Complete mixing occurs when all multipole moments are completely
mixed. We find below that in general the dipole moment is the
  slowest to mix, so complete mixing occurs after approximately
  $n_{\sat} = -\ln 3N/(2\ln
|\langle \mu \rangle|) $ timesteps. For small angular steps $n_{\sat} =
\ln 3N/\langle\alpha^2 \rangle$.

\subsection{Application to resonant relaxation}
\label{s:random-application}

\begin{figure*}
\centering
\mbox{
\includegraphics[width=0.48\textwidth]{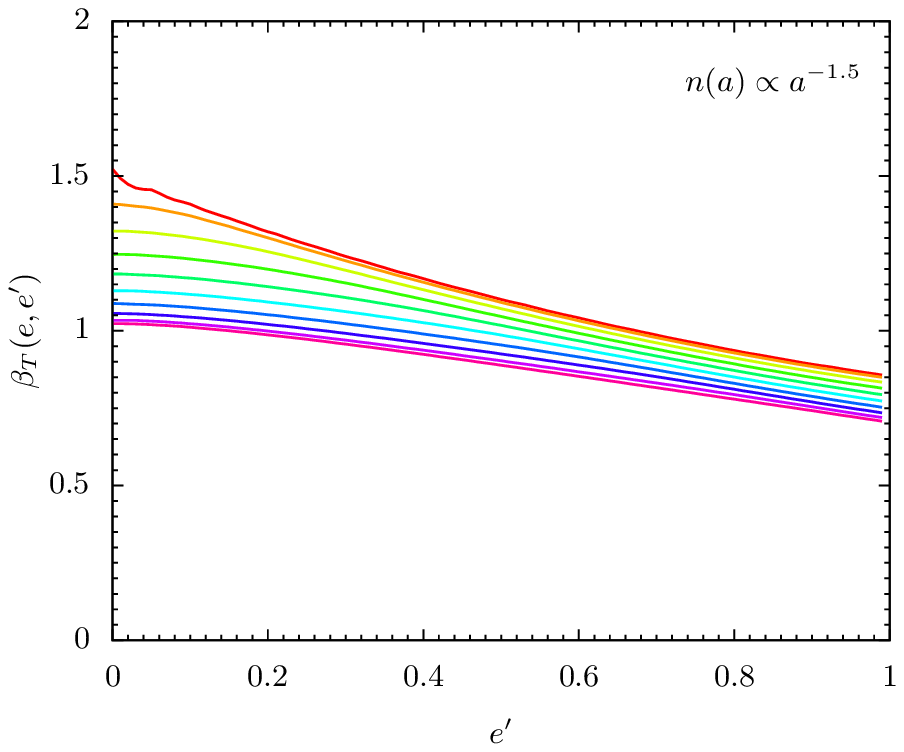}\hfill
\includegraphics[width=0.48\textwidth]{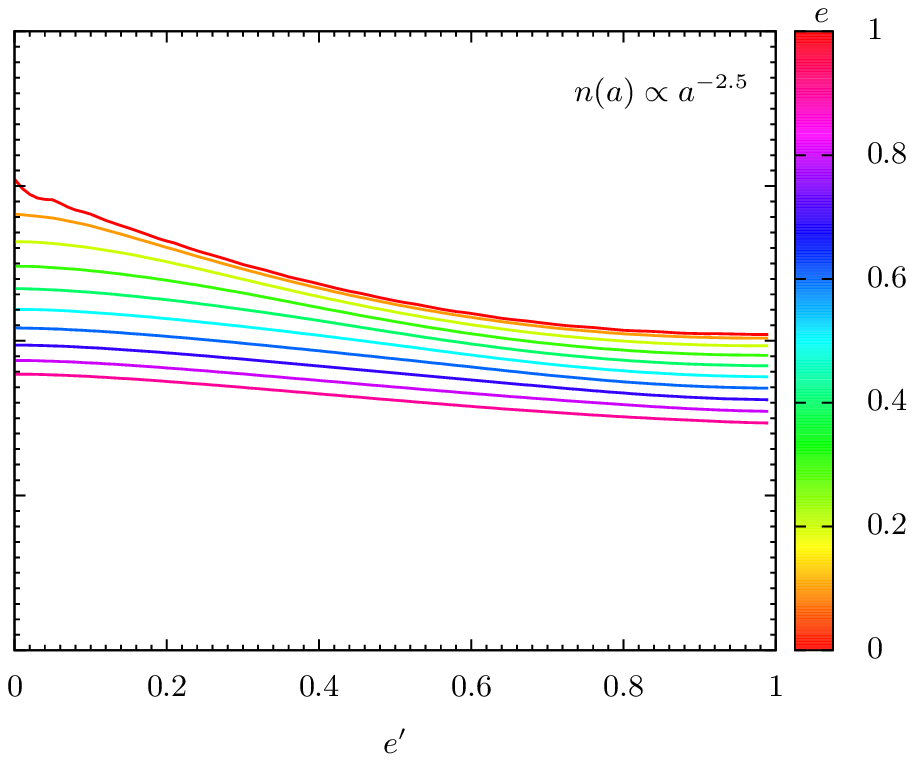}
}
\caption{\label{f:betaT}
  The dimensionless coherent torque parameter $\beta_T$
  (Eqs.~\ref{e:torque-coherent-powerlaw} and \ref{e:betaT3}) for a star with
  eccentricity $e$, orbiting in a spherical population of stars with a
  fixed eccentricity $e'$ and a distribution of semimajor axes
  $n(a)\propto a^{-1.5}$ (left panel) and $\propto a^{-2.5}$ (right
  panel).  The colored curves have $e=0$, 0.1, \dots, 0.9 from top to
  bottom, and $e'$ is varied on the horizontal axis.  }
\end{figure*}

We now apply these results to VRR.  In the incoherent phase of VRR,
each star undergoes a random walk in $\r\equiv \Ln$ on the unit sphere
due to torques from other stars.

We introduce a decoherence time $t_{\phi}$: over time intervals much
less than the decoherence time the stochastic torque on a star is
temporally correlated\footnote{In practice we identify the decoherence
  time with the time over which the the torque is approximately
  constant.}  (``coherent evolution''), while the torques at times
separated by much more than the decoherence time are temporally
uncorrelated (``incoherent evolution'').  Of course, the decoherence
time will depend on the eccentricity and semimajor axis of the star
and the properties of the stellar cluster of which it is a member.  We
first determine the RMS torque that characterizes the coherent
evolutionary phase, then we use the stochastic model of the previous
section to characterize the incoherent evolution. We analyse our
numerical simulations in this framework and determine how the model
parameters depend on the physical parameters of the stellar orbits in
the two regimes.

A second parameter that characterizes the evolution of a star $i$ during
VRR is related to the RMS torque that it experiences.
For a cluster composed of stars of similar semimajor
axes $a$, and a distribution of eccentricities and masses,
\begin{align}\label{e:torque-coherent}
T_{\rms,i} = \langle\T_{i}^2\rangle^{1/2}
&\simeq \frac{\beta_T}{2\pi}\frac{ G \sqrt{N} m_{\rms} m_i }{a}
\nonumber\\
 &= \beta_T \frac{\sqrt{N} m_{\rms}}{M_{\bullet}}
  \frac{m_i \sqrt{G M_{\bullet} a}}{P}\,,
\end{align}
where $P=2\pi(a^3/G M_\bullet)^{1/2}$ is the orbital period,
$m_{\rms} = (N^{-1}\sum_i m_i^2)^{1/2}$, $\beta_T $ is a dimensionless constant of
order unity, and averaging is over the distribution of the other stars
in a spherical cluster, $\Ln_{j\neq i}$.
Similarly, the RMS rate of change of the orbit normal for star $i$ is
\begin{equation}\label{e:om-coherent}
\Omega_{\rms,i} = \bigg\langle\bigg(\frac{\D\Ln_{i}}{\D t}\bigg)^2\bigg\rangle^{1/2}
=\left\langle \frac{\T_i^2}{L_i^2}\right\rangle^{1/2} \simeq
\beta_\Omega \frac{\sqrt{N}m_{\rms}}{M_\bullet P}.
\end{equation}
Using the notation of Eqs.~(\ref{e:HRR}) and (\ref{e:EOM2}),
\begin{align}\label{e:betaT0}
  \beta_T &= \frac{2\pi a}{G m_i m_{\rms}} \bigg[
    \frac{1}{N}\sum_{j,k=1}^N\sum_{\ell,n}\J_{ij\ell}\J_{ikn}P_\ell'(\cos I_{ij})
    \bigg.\nonumber \\
&\quad\times P_n'(\cos I_{ik}) \,(\cos I_{jk}-\cos I_{ij}\cos I_{ik})\bigg]^{1/2},
\end{align}
and $\beta_\Omega=\beta_T(1-e_i^2)^{-1/2}$. We simplify this
expression in Appendix~\ref{s:beta}.  We find that the series
in $\ell$ converges very quickly, and so the coherent torques
in a spherical cluster are predominantly quadrupolar.  The torque is a
Gaussian random variable with zero mean and dispersion set by
$\beta_T$.

\begin{figure}
\centering
\mbox{
\includegraphics{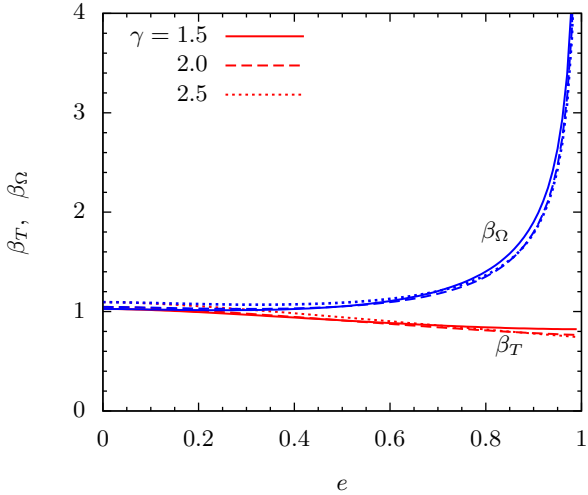}
}
\caption{\label{f:beta}
  The dimensionless parameters $\beta_T$ and $\beta_{\Omega}$
  (Eq.~\ref{e:torque-coherent-powerlaw}) describing
  the RMS coherent torque and precession rate for a star with
  eccentricity $e$ due to a spherical population of stars with a
  thermal distribution of eccentricity $\D N = 2 e \D e$ and a
  distribution of semimajor axes $n(a)\propto a^{-\gamma}$, where
  $1.5\leq \gamma \leq 2.5$ as labeled. The evaluation is done using
  Eq.~(\ref{e:betaT3}).  }
\end{figure}

More generally, if there is a range of semimajor axes with $\D N=4\pi
a^2n(a)\,\D a$ stars in the semimajor axis interval $a\to a+da$, we
can replace $N$ by $\D N /\D \ln a = 4\pi a^3n(a)$ in all these
equations where $a\equiv a_i$. For example, Eqs.\
(\ref{e:torque-coherent}) and (\ref{e:om-coherent}) become
\begin{align}\label{e:torque-coherent-powerlaw}
T_{\rms,i} &\simeq \beta_T \frac{\sqrt{\D N/\D\ln a}\, m_{\rms}}{M_{\bullet}}
  \frac{m_i \sqrt{G M_{\bullet} a}}{P}\,, \nonumber \\
\Omega_{\rms,i} &\simeq \beta_\Omega \frac{\sqrt{\D N/\D\ln a}\, m_{\rms}}{M_\bullet P}.
\end{align}
In Appendix~\ref{s:beta}, we show that with this definition $\beta_T$
is independent of $a$ if the distribution of $a$ is a power law, and
independent of the distribution of stellar masses. We evaluate the
average in Eq.~(\ref{e:betaT0}) as integrals over orientation,
eccentricity, and semimajor axis (Eq.~\ref{e:betaT3}) for $n(a)\propto
a^{-1.5}$ and $\propto a^{-2.5}$.  Figure~\ref{f:betaT} shows
$\beta_T$ for an orbit with eccentricity $e$, assuming that all stars
in the cluster have a fixed eccentricity $e'$.  The Figure shows that
$0.7\lesssim \beta_T\lesssim 1.5$ and that $\beta_T$ is a decreasing
function of both $e$ and $e'$. Thus we may generally conclude that
$\beta_T$ must be a decreasing function of $e$ for an arbitrary
eccentricity distribution, with values in the same range 0.7--1.5.  In
particular Figure~\ref{f:beta} shows $\beta_T$ and $\beta_{\Omega}$
for a star cluster with a thermal eccentricity distribution $\D N =
2e\,\D e$ and number density proportional to $a^{-\gamma}$ where
$1.5<\gamma<2.5$.  Simple fitting formulae are\footnote{This result
  disagrees with the eccentricity dependence reported by
  \citet{2007MNRAS.379.1083G}, for reasons given in
  Section~\ref{s:comparison} below.}
\begin{align}\label{e:betafit}
 \beta_T(e) \simeq 1.05 - 0.3\, e\,,\quad
 \beta_{\Omega}(e) \simeq \frac{1.05 - 0.3\, e}{(1-e^2)^{1/2}}\,.
\end{align}
Thus we find that the angular-momentum re-orientation timescale is
approximately independent of the semimajor axis distribution
  (i.e., the exponent $\gamma$), and is also independent of the eccentricity for $0\leq e
\leq 0.75$, to within $20\%$ accuracy.  The
angular momenta of highly eccentric stars are re-oriented much more
rapidly. RMS-averaging over both $e$ and $e'$ for a thermal
distribution yields
$\langle\beta_T^2\rangle^{1/2}=0.85$.\footnote{The RMS average of
$\beta_{\Omega}$ over both $e$ and $e'$ in a thermal eccentricity distribution
is logarithmically divergent, $\langle \beta_{\Omega}^2\rangle^{1/2}\propto \ln(1-e_{\max})$
for $e_{\max}\rightarrow 1$.}

Pursuing the analogy to the random walk on the sphere, the decoherence
time $t_\phi$ takes the place of the timestep and
$\Omega_{\rms}t_\phi$, which we call the angular coherence length,
takes the place of the RMS angular displacement per timestep
$\langle\alpha^2\rangle$.  On timescales short compared to the
decoherence time, the orbit normals move in the mean field of the
cluster at a rate $d\Ln/dt$ which is approximately constant\footnote{
  As long as the mean-field potential is constant in time, ${\Ln}$
  moves with angular velocity $\partial H_{\E}/\partial \L$ along a
  closed path on the unit sphere that is a contour of constant
  $H_{\E}$, see Eq.~(\ref{e:EOM2}).}, and the angular variance is
\begin{align}
V_{\ell}(t) &= -\frac{4}{\ell(\ell+1)}\ln
\left|\left\langle P_{\ell}\hspace{-1pt}\big[\Ln_i(t+t_0)\cdot\Ln_i(t_0)\big] \right\rangle \right|,
\quad
t\,\lesssim\, t_{\phi, \ell}
\nonumber\\ &= \Omega_{\rms}^2t^2\,,
\label{e:V-coherent}
\end{align}
where the quadratic approximation in the second
line holds so long as the angular displacement is small
($V_\ell(t)\ll1$).  On timescales long compared to the decoherence
time, the orbital vectors execute a random walk, where
\begin{align}
V_{\ell}(t) = \frac{t}{t_{{\vrr},\ell}}, \quad t\,\gtrsim\, t_{\phi, \ell}\,,
\label{e:V-incoherent}
\end{align}
which defines the VRR time for the $\ell^{\rm th}$ harmonic
$t_{{\vrr},\ell}$. We identify the decoherence time
with the transition from quadratic to linear growth
of $V_{\ell}(t)$, that is,
\begin{equation}\label{e:tvrr-def}
t_{\phi,\ell} = \frac{1}{\Omega_{\rms}^2 t_{\vrr,\ell}}\,.
\end{equation}

\begin{figure}
\centering
\mbox{
\includegraphics{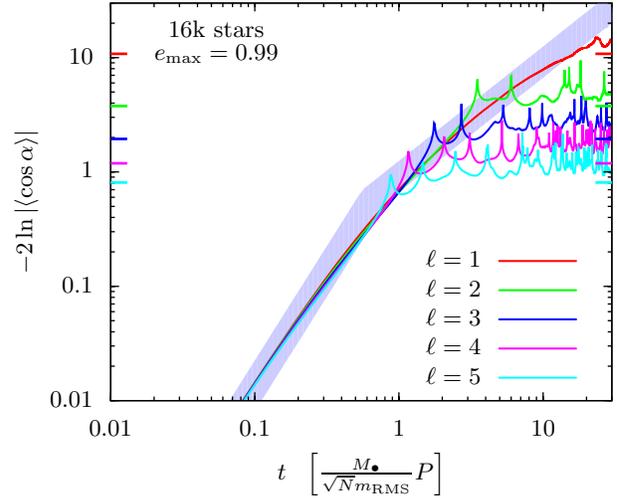}
}
\caption{\label{f:Vell} The evolution of the angular variance
  $V_{\ell}=-4\ell^{-1}(\ell+1)^{-1}\ln \big|{(NT)}^{-1}\int_0^{T} \D
  t_0\sum_{i=1}^N P_{\ell}[\cos \alpha_i(t,t_0)]\big|$ in a simulation
  with 16,384 stars.  Here $\alpha_i(t,t_0)$ is the angular distance
  between the angular-momentum vector of star $i$ at time $t_0$ and
  time $t+t_0$.  The stars are initially spherically distributed with
  nearly the same semimajor axis and a uniform distribution in the
  square of the eccentricity for $e\lesssim 0.99$ (i.e., uniform
  distribution on the energy surface in phase space). The angular
  variance is expected to grow quadratically at early times (coherent
  torques) and linearly at later times (random walk on a sphere) until
  the mode is fully mixed, as marked by short coloured lines on
    the vertical axis (Eq.\ \ref{e:Vell-sat}). The shaded region
  shows $\min[ (\beta_{\Omega} t f_{\vrr}/t_{\vrr})^2, t/t_{\vrr} ]$
  for reference where $t_{\vrr} = f_{\vrr}
  [M_{\bullet}/(\sqrt{N}m_{\rms})] P$, $0.9\leq \beta_{\Omega}\leq
  1.5$, and $0.8\leq f_{\vrr} \leq 1.5$.}
\end{figure}

For a single-component spherical cluster of stars, the torques are
comparable for different stars and constant for a characteristic time
$\sim\Omega_{\rms}^{-1}$, and therefore one might expect
$t_{\phi}\sim\Omega_{\rms}^{-1}$, so the angular coherence length is
$\Omega_{\rms}t_{\phi} \sim 1$.  In this case the formulae above yield
$t_{\vrr}\sim \Omega_{\rms}^{-1}$ so we write
\begin{equation}
t_{\vrr} = f_{\vrr}\frac{M_\bullet}{\sqrt{N}m_{\rms}}P\,,
\label{e:vrr}
\end{equation}
where $f_{\vrr}$ is a dimensionless constant of order unity. With
these definitions the decoherence time is\footnote{
  Using the notation of \citet{2009ApJ...698..641E}, the decoherence
  time is parameterized by the dimensionless constant $A_{\phi}$ as
  $t_{\phi} = A_{\phi} [M_{\bullet}/(\sqrt{N}m_{\rms})] P$.}
\begin{equation}\label{e:tphi}
t_\phi=\frac{1}{f_{\vrr}\beta_\Omega^2}\frac{M_\bullet}{\sqrt{N}m_{\rms}}P\,.
\end{equation}

\begin{figure*}
\centering
\mbox{
\includegraphics{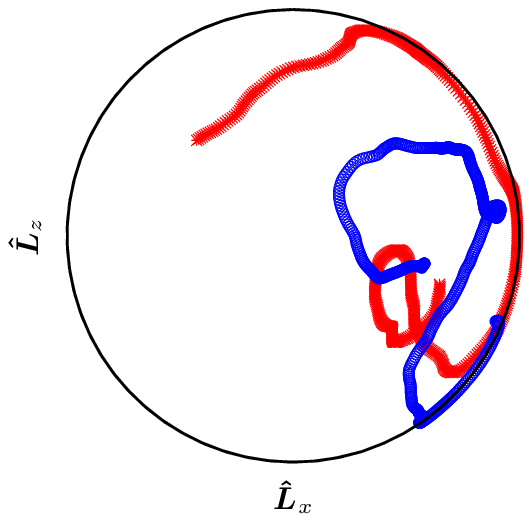}\quad
\includegraphics{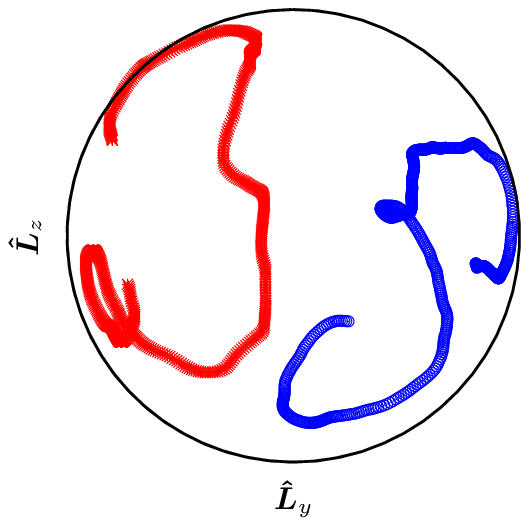}\quad
\includegraphics{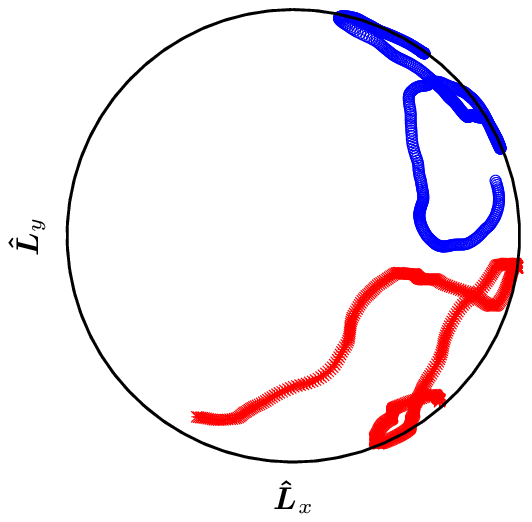}
}\\[2ex]
\mbox{
\includegraphics{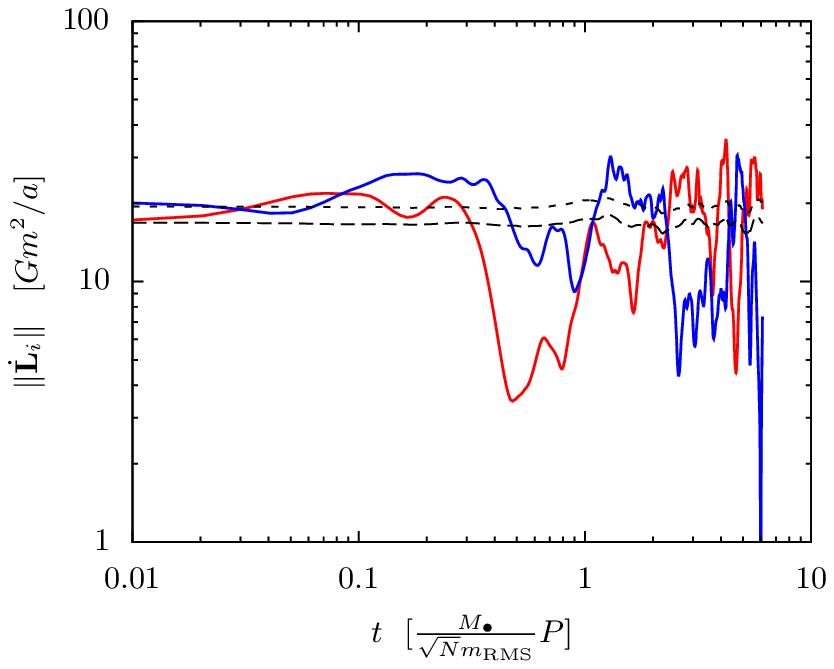}
\includegraphics{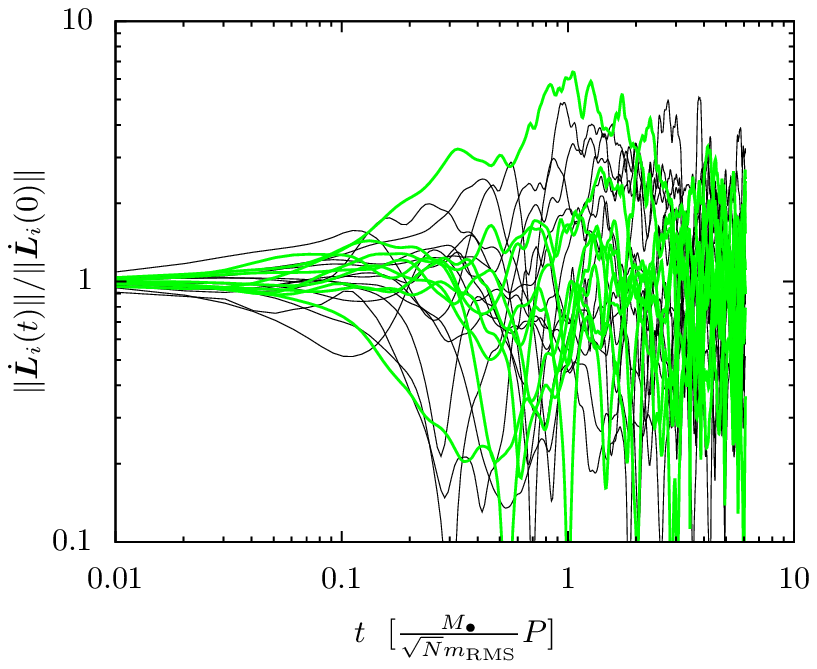}
}
\caption{\label{f:Lmapandtorque} {\it Top panels:} The evolution of
  the normalized angular-momentum vectors for two representative stars
  from a simulation similar to Figure~\ref{f:Vell}. The three panels
  show orthogonal projections on the $x$--$z$, $y$--$z$, and $x$--$y$
  planes.  The two stars were randomly selected from a subset of stars
  with roughly the mean eccentricity of the cluster $\langle e \rangle
  =0.58$.  The motion is shown for a time interval $t=7
  \,\Omega_{\rms}^{-1}$.  {\it Bottom panels:} The evolution of the
    torque as a function of time. The bottom left panel shows the two
    stars for which the trajectories are shown in the top panels.  The
    long-dashed and short-dashed lines show the mean of $\|\bm{T}_i\|$
    and $T_{\rms,i}$ of the cluster, respectively. In the bottom right
    panel, the solid green and black curves in the right panel show
    stars with nearly the mean eccentricity, and stars from the whole eccentricity
    range. The decoherence time is approximately independent of
    eccentricity.  }
\end{figure*}

\begin{figure*}
\mbox{\hspace{-20pt}\includegraphics{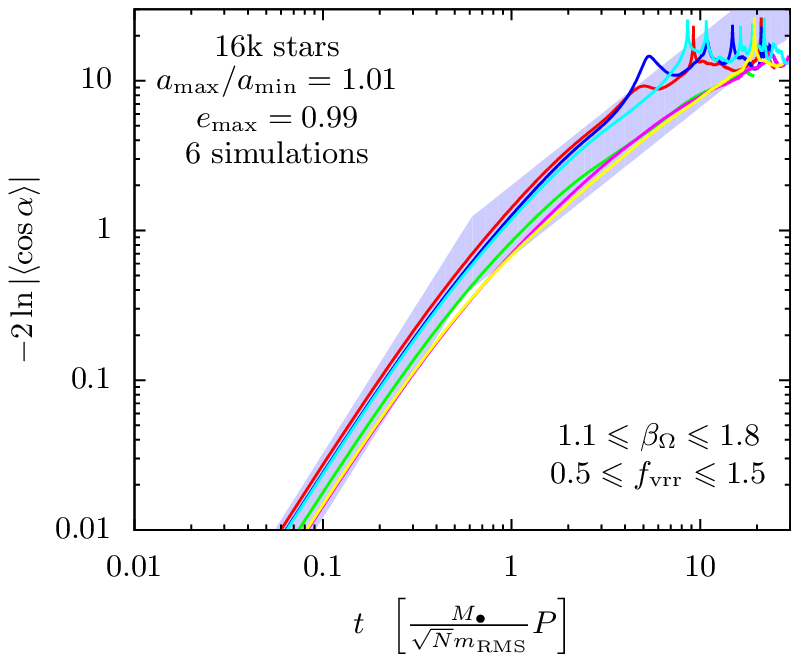}\qquad
\includegraphics{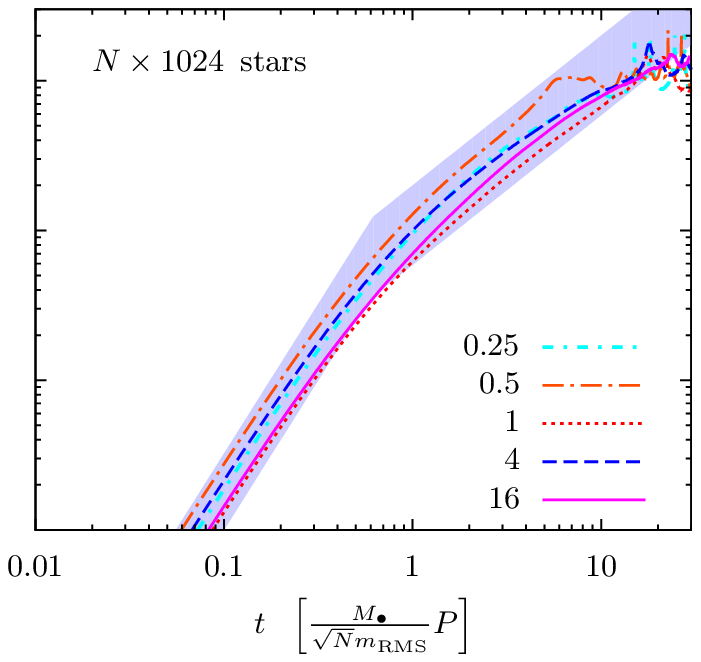}}\\
\mbox{
\vspace{2pt}\hspace{10pt}\includegraphics{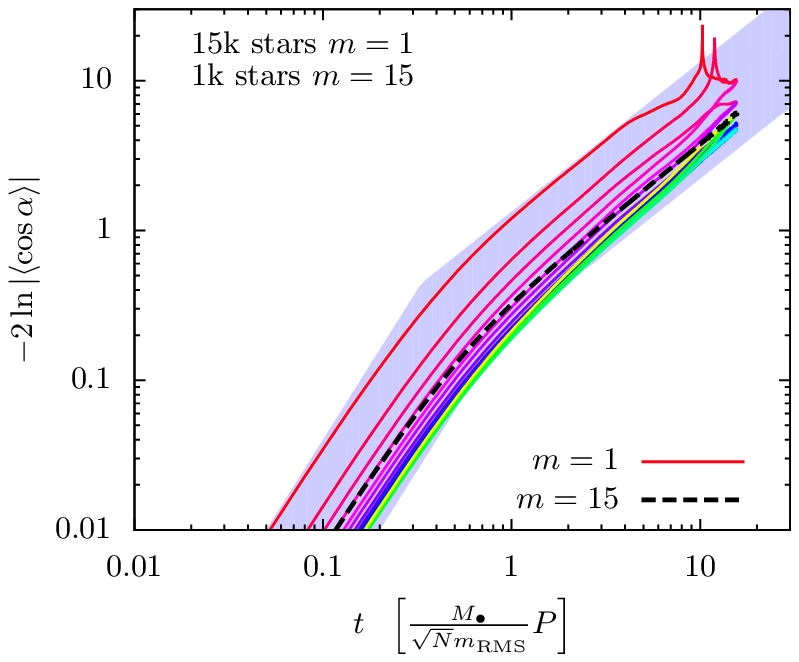}\qquad
\includegraphics{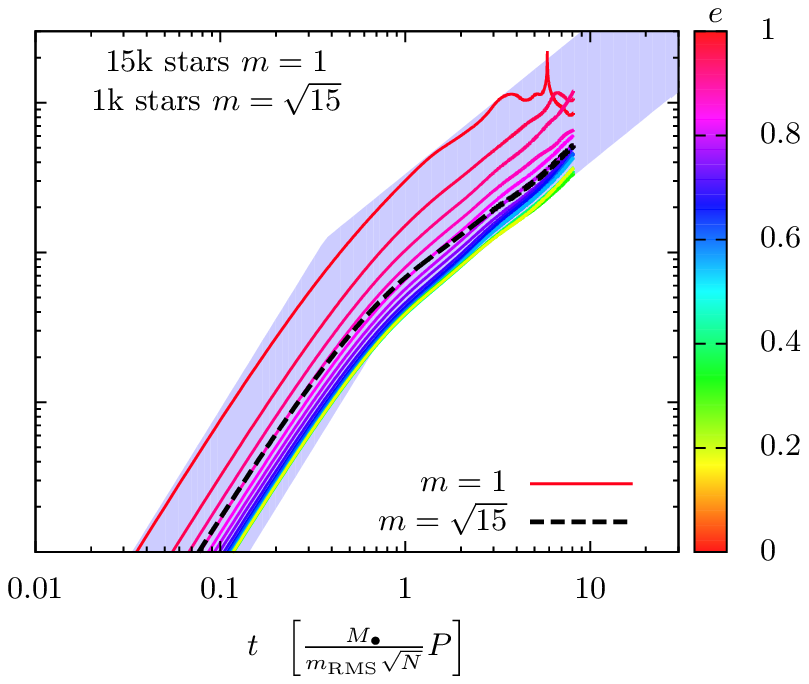}}
\centering
\caption{\label{f:correlation1} The angular variance $V_{\ell}(t)$ as in
  Figure~\ref{f:Vell}, but for $\ell=1$ and different initial
  conditions (top left), different numbers of stars (top right) and
  different RMS masses (bottom panels).  {\it Top left:} Different
  curves show the range spanned by six simulations with different
  initial conditions.  {\it Top right:} The number of stars $N$ is
  varied between 256 and 16384, the legend shows $N/1024$.  {\it
    Bottom left:} The stellar cluster is comprised of 15k low-mass and
  1k high-mass stars (left) so the total mass $Nm$ is the same for
  both groups.  The curves show $V_1$ for stars grouped in subsets
  containing 1k members, sorted by mass and eccentricity
    (curves are colored by eccentricity as shown on the right); solid
    and dashed lines have different stellar masses as labeled. {\it
    Bottom right:} Similar to bottom left, but with heavy stars
  $\sqrt{15}\times$ more massive than light stars.  The shaded regions show $1.1 \leq
  \beta_{\Omega}\leq 1.8$ and $0.5\leq f_{\vrr}\leq 1.5$ (top panels),
  $0.5 \leq \beta_{\Omega}\leq 2$ and $0.75\leq f_{\vrr}\leq 4.5$
  (bottom left), and $0.7 \leq \beta_{\Omega}\leq 3.0$ and $0.3\leq
  f_{\vrr}\leq 2.5$ (bottom right).  }
\end{figure*}

For a range of semimajor axes, the relaxation time Eq.~(\ref{e:vrr}) becomes
\begin{equation}
t_{\vrr}(a) = f_{\vrr}\frac{M_\bullet}{\sqrt{4\pi a^3n(a)}m_{\rms}}P(a)\,.
\label{e:vrr1}
\end{equation}
We measure the dimensionless parameters $\beta_\Omega$
and $f_{\vrr}$ using numerical simulations, from the behavior of
$V_\ell(t)$ at small and large times.

\begin{figure*}
\centering
\mbox{
\includegraphics[height=7.1cm]{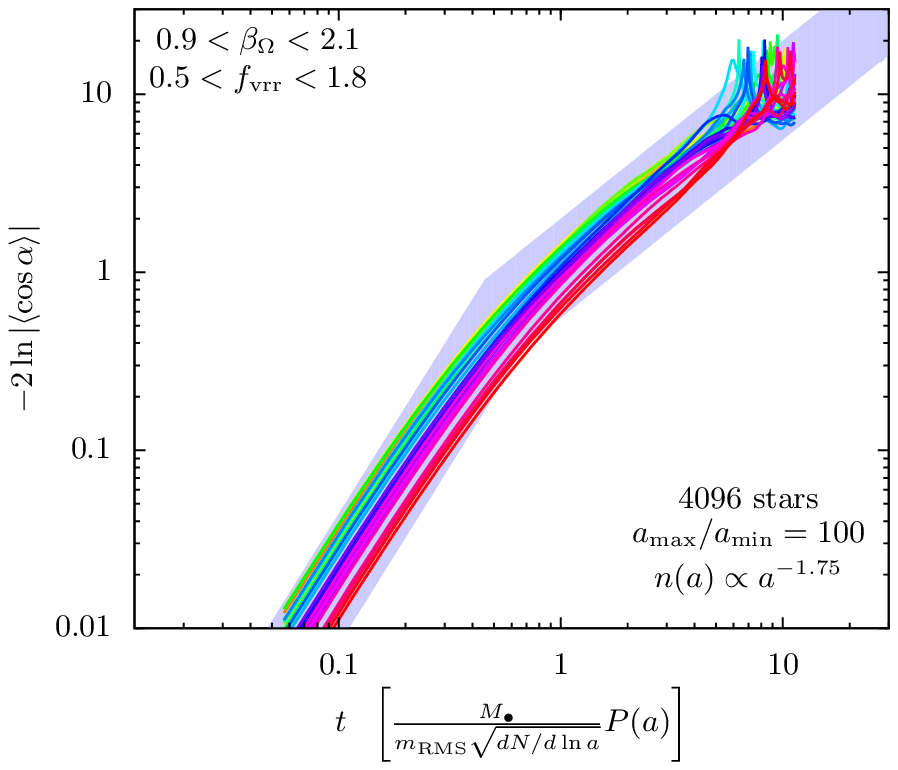}
\includegraphics[height=7.57cm]{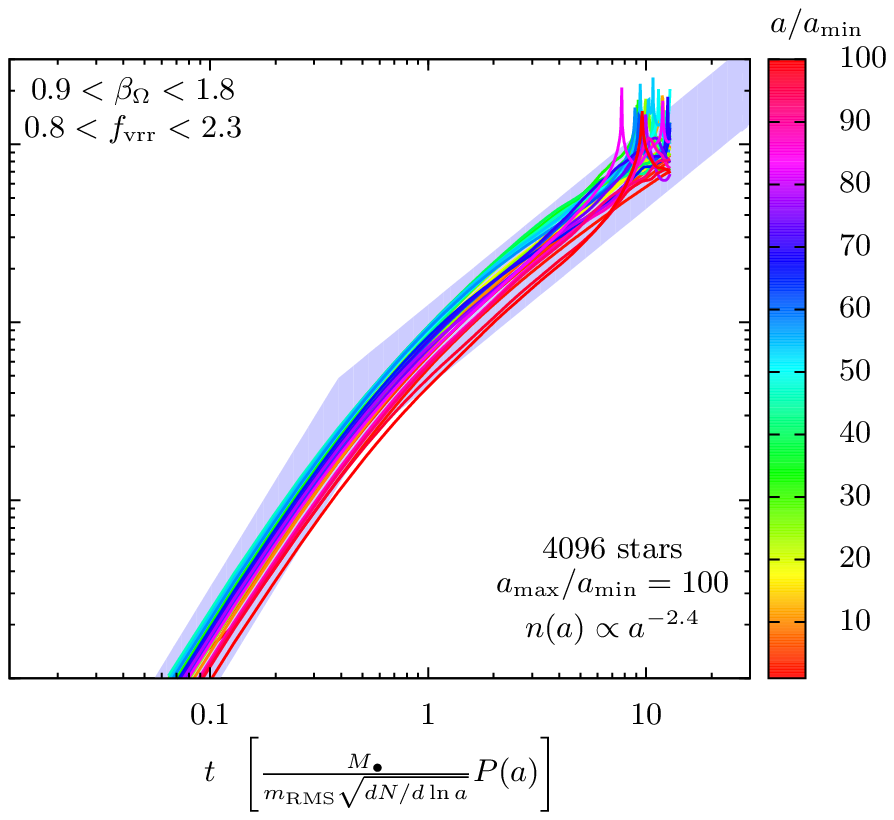}}\\
\mbox{
\includegraphics[height=7.1cm]{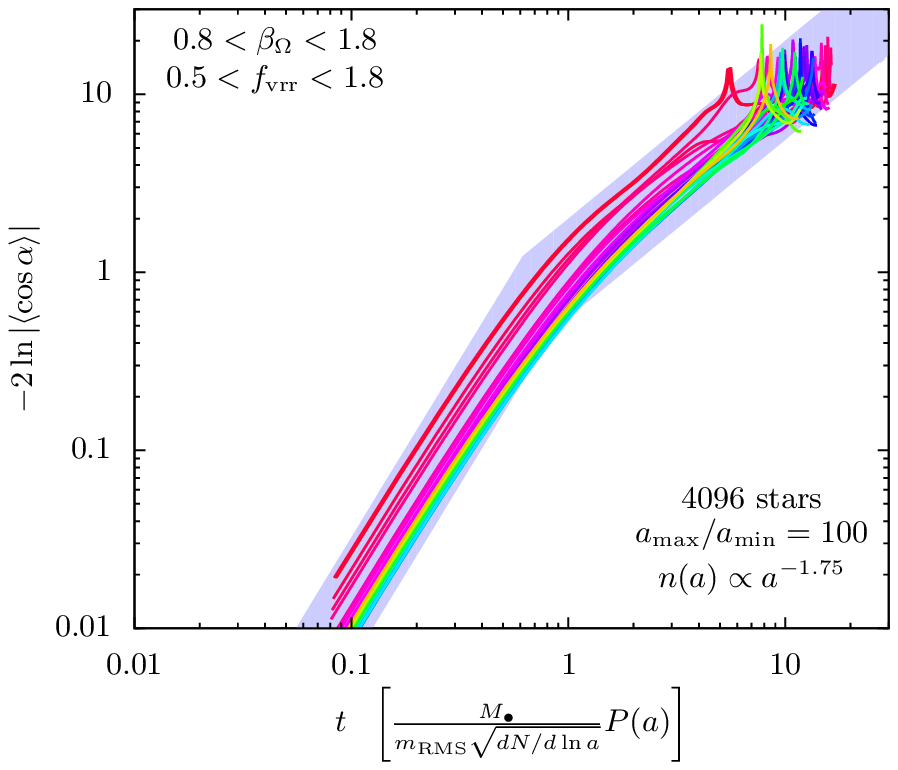}
\includegraphics[height=7.5cm]{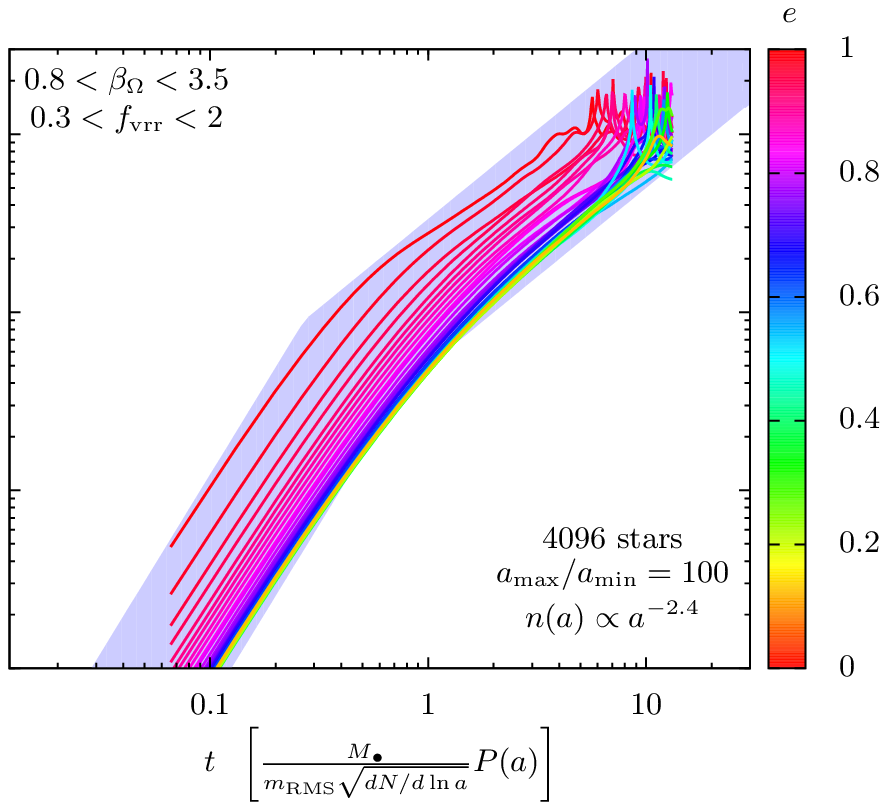}}
\caption{\label{f:correlation_a}
  VRR in a stellar cluster with a range of eccentricity $0\leq e
  \leq0.99$ ($\D N = 2e\D e$) and semimajor axis
  $a_{\max}/a_{\min}=100$ with number density $n(a)\propto a^{-1.75}$
  (left panels) and $r^{-2.4}$ (right panels).  We sort the stars with
  respect to their semimajor axis (top panels) and eccentricity
  (bottom panels) and group them into 32 bins containing 128 stars
  each.  The 32 curves in each panel shows $V_1=-2\ln|\langle \cos
  \alpha_i\rangle|$ as in Figure~\ref{f:correlation1} for the stars in
  the corresponding bins where $\alpha_i$ is the angular distance
  traversed by star $i$ in dimensionless time $\tau =
  t/[M_{\bullet}m_{\rms}^{-1}(\D N/\D \ln a)^{-1/2}P(a)]$ from some
  reference time $t_0$. We average over $i$ and $t_0$ for each $\tau$.
  The evolution of this quantity is quadratic in the initial coherent
  phase and linear during incoherent random mixing.  The curves are
  colored according to the semimajor axis (top) or eccentricity
  (bottom panels) as shown on the right.  The main systematic effect
  with semimajor axis is well captured by the relaxation time formula,
  the curves nearly overlap in these units despite a range of a factor
  of 56 (left panels) or 250 (right panels) in $t_{\vrr}(a)$. Residual
  variations are probably due to edge effects: stars near $a_{\min}$
  and $a_{\max}$ relax slower.  Since the curves nearly overlap for
  $e<0.7$, stars with small to moderately large eccentricities relax
  at nearly the same rate given by $t_{\vrr}(a)$.  However highly
  eccentric orbits $e>0.8$ relax by up to a factor 4--8 faster.  The
  eccentricity dependence in the coherent phase of the simulation is
  in perfect agreement with the direct calculation shown in
  Figure~\ref{f:beta}.}
\end{figure*}

\begin{figure*}
\centering
\mbox{\includegraphics[height=7.12cm]{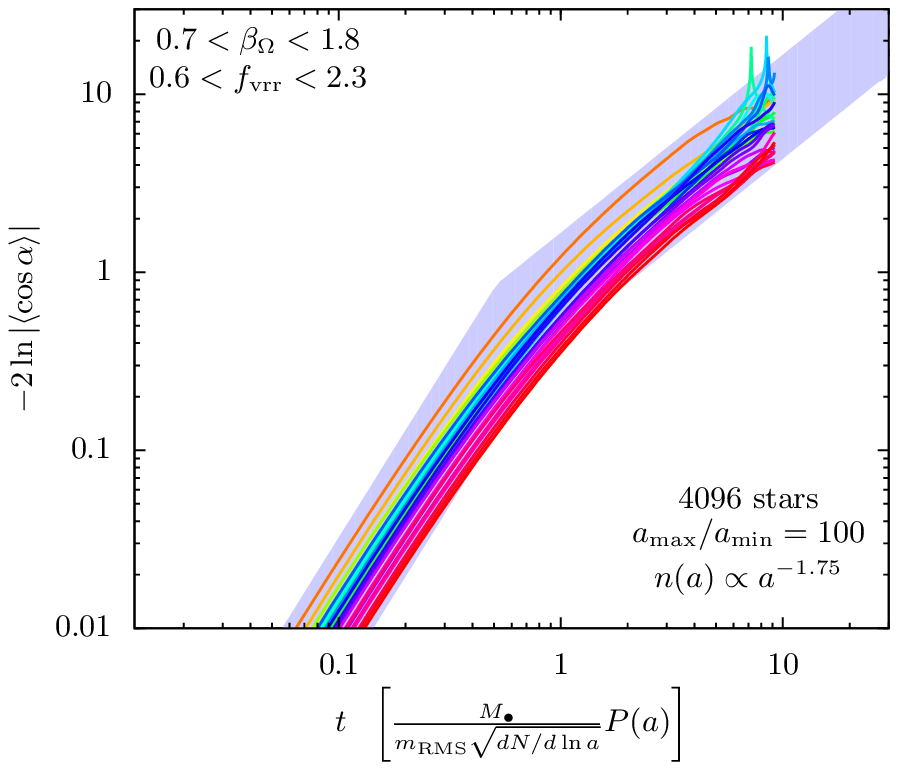}
\includegraphics[height=7.58cm]{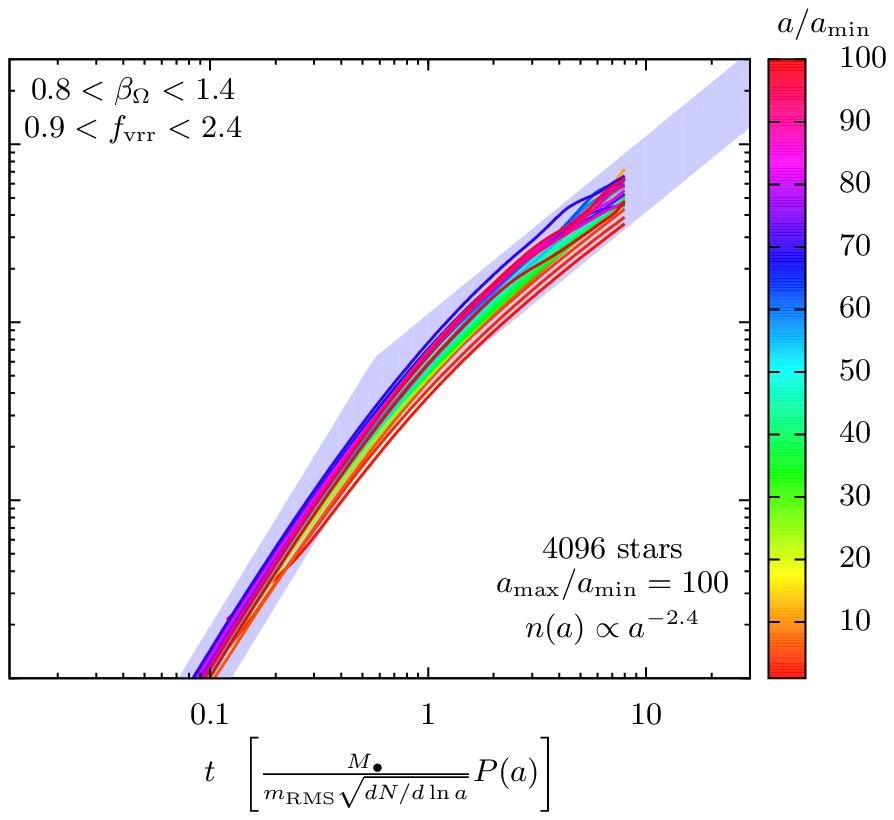}}\\
\mbox{
\includegraphics[height=7.12cm]{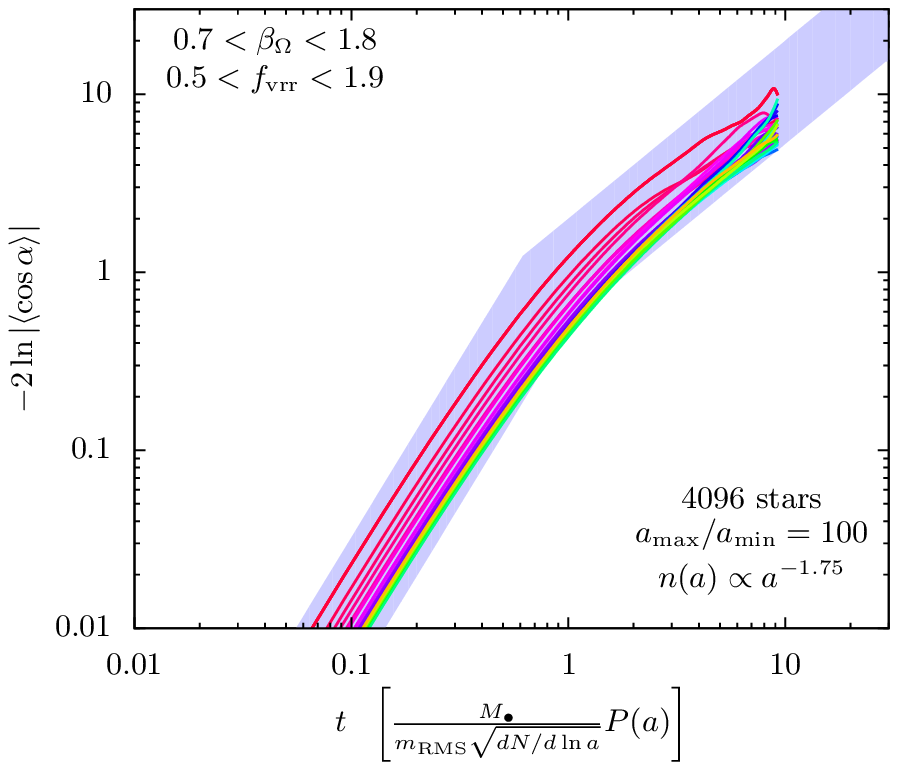}
\includegraphics[height=7.58cm]{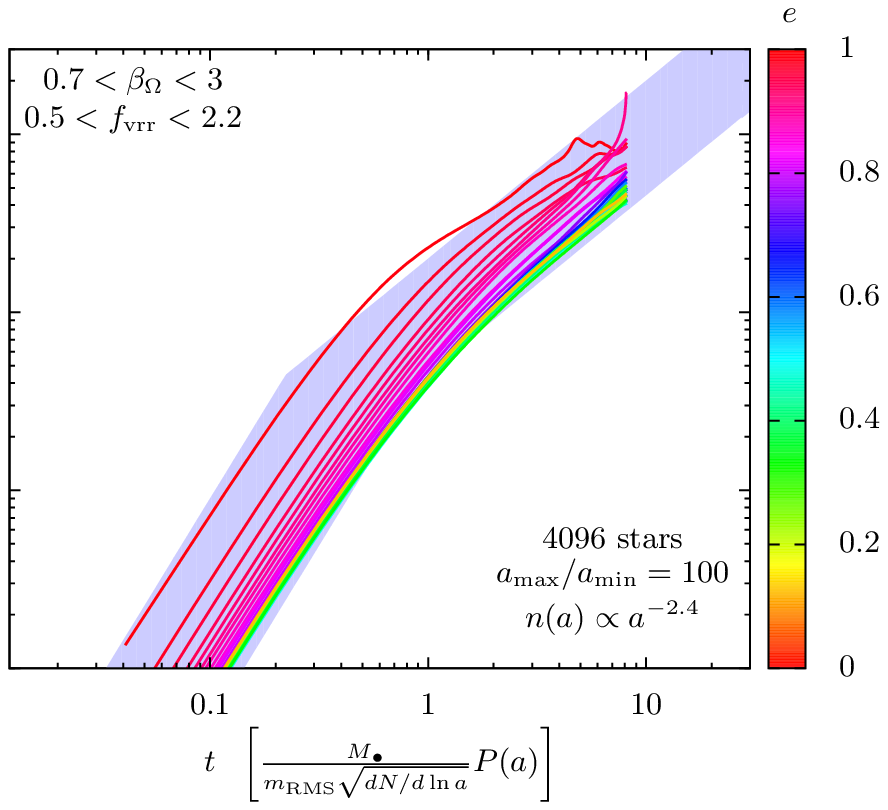}}
\caption{\label{f:correlation_a2}
Same as Figure~\ref{f:correlation_a} but with an eccentricity
distribution that is thermal below $e=0.4$ and flat at higher
  eccentricities, $dN\propto e de$ for $e<0.4$ and $dN\propto de$ otherwise.
The trends are very similar. The interaction calculation $\J_{ij\ell}$ was
truncated above multipole harmonic index $\ell_{\max}=20$ in this figure,
and above $50$ in Figure~\ref{f:correlation_a}.
}
\end{figure*}

Figure~\ref{f:Vell} shows $V_{\ell}(t)$ measured in a simulation of a
spherical cluster with 16,384 stars with nearly the same semimajor
axes and masses. The figure shows that indeed all $V_{\ell}$ grow
quadratically at first (coherent torques) and then linearly (random
walk), until eventually they saturate and thereafter execute random
variations. The dipole ($\ell=1$) mixes most slowly, higher harmonics
mix sooner. The curves with different $\ell$ approximately overlap
before they saturate; this behavior is in agreement with
Eq.~(\ref{e:Vell-theory1}) for Brownian motion even though the angular
coherence length is of order unity so the Brownian approximation is
questionable. The shaded region shows the model described by
Eqs.~(\ref{e:V-coherent})--(\ref{e:vrr}) with
$0.9\leq \beta_{\Omega}\leq 1.5$ and $0.8\leq f_{\vrr}\leq 1.5$,
the best-fit dimensionless torque and VRR factors are
$\beta_\Omega\approx 1.2$ and $f_{\vrr}\approx 1.2$.  The linear
evolution corresponding to a random walk starts where
$V_{\ell}(t_{\phi})=\beta_{\Omega}^{-2}f_{\vrr}^{-2} \approx 0.5 $ for
$1\leq \ell\leq 5$.  The angular coherence length is $\langle
\alpha^2\rangle^{1/2} = \Omega_{\rms}t_{\phi} \approx
\beta_{\Omega}^{-1}f_{\vrr}^{-1} \approx 0.7 \approx 39\,\rm deg$.
The horizontal lines show the expected saturated level of $V_{\ell}$
based on Eq.~(\ref{e:Vell-sat}), which is consistent with the curves.
Thus, our approximate treatment of the stochastic motion as a random
walk appears to provide a consistent model of the evolution shown in
Figure \ref{f:Vell}.

To show an example of the actual motion of angular-momentum vectors,
the top panel of Figure~\ref{f:Lmapandtorque} shows a time interval
$\sim 7\,\Omega_{\rms}^{-1}$ of the $\Ln_i$ trajectory for two stars
in a simulation similar to Figure~\ref{f:Vell}. The two stars are
chosen to have close to the mean eccentricity of the cluster. In this
case, our model approximates their motion as $\sim 10$ steps of a
random walk with an average step size of $30^\circ$. The time interval
shown corresponds to $\sim 3$ relaxation timescales, and $\sim 0.25$
of the complete mixing timescale for $\ell=1$.  The bottom left panel
of Figure~\ref{f:Lmapandtorque} shows the torque as a function of time
in units of $M_{\bullet} P
/(\sqrt{N}m_{\rms})=\beta_\Omega\Omega_{\rms}^{-1}\approx\Omega_{\rms}^{-1}$
for the same stars. The bottom right panel of
Figure~\ref{f:Lmapandtorque} shows the torque as a function of time
for a larger sample of stars: green curves show stars with nearly the
median eccentricity, black curves show stars from the full range of
eccentricities ($0\leq e<0.99$).  The torques vary substantially from
their initial values after a decoherence time $t_{\phi} \sim
(0.3$--$0.7)\,\Omega_{\rms}^{-1}$, which is consistent with our
earlier estimate from the angular variance.  The decoherence time is
similar for stars of all eccentricities.

Figure~\ref{f:correlation1} shows $V_\ell(t)$ in simulations with
different initial conditions, numbers of stars, and distributions of
stellar masses. In these simulations we continue to assume that all
stars have nearly the same semimajor axis, a spherical distribution in
angular-momentum space, and a thermal distribution of eccentricities
as in Figure~\ref{f:Vell}.
We find that Eq.~(\ref{e:vrr}) describes well the dependence of the
relaxation timescale $t_{\vrr}$ on the number of stars, although
 the fitted value of
$f_{\vrr}$ can vary by 30--40\% for different initial
conditions. In particular, in the upper right
panel we vary the number of stars by a factor $64$ but the variation
in scaled time at a fixed value of $V_1$ is less than a factor of two,
and shows no systematic trend with $N$.  Complete mixing occurs when
the angular variance saturates, which in these simulations occurs at
$t_{\sat}\sim 10$--$30\, t_{\rm vrr}$.
Note however that some of the curves do not display a perfectly linear growth
during incoherent evolution; in various runs $V_{\ell}(t)$
exhibits time dependence both shallower and steeper than linear.
Similar anomalous diffusion is often observed
in chaotic systems near phase transitions, in random walks where the
probability distribution of step size is top-heavy, and in systems with long-term memory
 (\citealt{PhysRevLett.83.2104,PhysRevE.82.021101,Gottwald21052013}; see also
 \citealt{2011MNRAS.411L..56G} and \citealt{2013ApJ...764...52B} for related findings in scalar resonant 
and two-body relaxation, respectively).

 The simulations in the bottom panels contain two groups of stars with
 the same total mass $Nm$ (bottom left panel) and the same value of
 $\sqrt{N}m$ (bottom right panel); the RMS masses in the clusters are
 $m_{\rms}=3.87$ and $1.40$ respectively.  Here $V_{1}(t)$ is shown
 for 1k element bins sorted by mass and eccentricity, with solid
 curves showing the low-mass stars, colors representing eccentricity
 as shown on the right, and dashed black curves showing the high-mass
 stars.  We find that the predicted scaling with $m_{\rms}$ captures
 the mass dependence well.  The relaxation is approximately eccentricity 
 independent for $e\lesssim 0.8$, and it is systematically faster
 for more eccentric orbits, but the decoherence time is roughly
 independent of eccentricity even for very eccentric orbits.  
 This is consistent with the observation
 that the mean field of the cluster is dominated by stars with
 $e\lesssim 0.8$ ($64\%$ of stars have $e<0.8)$ and the torque
 decreases weakly with eccentricity; thus the torque is approximately
 constant until the stars with $e\lesssim 0.8$ are re-oriented.

 Next let us relax the assumption of a fixed semimajor axis.  We
 distribute the orbits between $a_{\max}/a_{\min}=100$, and integrate
 for $\sim 10$ relaxation times at the outer edge of the cluster or
 $\sim10^3$ relaxation times at the inner edge of the cluster. To
 maintain numerical accuracy for such a large dynamic range, we reduce
 the number of stars to 4 thousand. Each star has the same mass and a
 thermal distribution of eccentricity ($dN = 2 e \,\D e$). We bin the
 stars according to semimajor axis or eccentricity to look for
 systematic effects in the relaxation time.
 Figure~\ref{f:correlation_a} shows the result of two simulations with
 number density profiles $n(r)\propto r^{-1.75}$ and $r^{-2.4}$
 respectively, which correspond to the observed distribution of
 B-stars and Wolf--Rayet/O stars, respectively
 \citep{2010ApJ...708..834B}. We find that the dependence of the
 relaxation time $t_{\vrr}(a)$ on semimajor axis $a$ is given
 approximately by Eq.~(\ref{e:vrr1}). Indeed, despite a range of a
 factor of $(0.3$--$6)\times10^{4}$ in the number density as a
 function of semimajor axis, there is less than a factor $\sim 3$
 variation in the angular variance $V_1(t)$ when time is measured in
 units of $t_{\vrr}(a)$ as given by Eq.~(\ref{e:vrr}). This is only a
 little larger than the factor $\sim 2$ variation seen for different
 realizations of the initial conditions (cf. top left panel of
 Figure~\ref{f:correlation1}).  Moreover most of this variation is
 seen for stars with semimajor axes near the cutoffs at $a_{\max}$ and
 $a_{\min}$, and so are probably due to ``edge effects''.

 The bottom panels of Figure~\ref{f:correlation_a} show the dependence
 of the relaxation rate on eccentricity: the rate is nearly
 independent of eccentricity for $e\lesssim 0.7$, but orbits with
 $e\,\gtrsim \,0.8$ relax faster on average, by as much as a factor of
 4--8.  This behavior is in good agreement with the direct calculation
 of $\beta_T$ and $\beta_\Omega$ shown in Figure~\ref{f:beta}.
   Highly eccentric orbits have a much larger $\beta_{\Omega}$;
   therefore they are re-oriented more rapidly and have a larger
   angular coherence length. However, the decoherence time is roughly
   independent of eccentricity since the torques are dominated by
   stars with $e\lesssim 0.7$ which have similar $\beta_\Omega$ and
   therefore are re-oriented at similar rates.
   Figure~\ref{f:correlation_a2} shows the evolution in the case where
   the number of high-eccentricity stars is smaller, but the results
   are very similar.
The figures show that $\beta_{\Omega}=0.95\pm 0.1$ 
and $f_{\rm vrr}=1.9\pm0.2$ for $e<0.7$, while for orbits with $e>0.8$
$\beta_{\Omega}$ and $f_{\rm vrr}$ are larger and smaller by up to factors of 
$\sim 3.5$ and $5$, respectively.

\subsection{Comparison with previous results}\label{s:comparison}

In this paper, we have explored an idealized model of how orbits in a
spherical stellar system undergo re-orientation due to torques from
other orbits.  Our model is based on the approximation that the rate
of apsidal precession is much faster than the rate at which the
orbital planes change their orientation. This approximation is valid
because the ratio of the re-orientation time to the apsidal precession
time in a cluster of $N\gg 1$ stars scales as $\sqrt{N}$ (see Section
\ref{s:intro}) which is likely to be valid for most stars in the
Galactic center with semimajor axes between $\sim0.003\pc$ and $\sim
1\pc$ (see Section~\ref{s:Hamiltonian} and Figure 1 of KT11).

There are many previous studies of resonant relaxation
\citep{1996NewA....1..149R, 2006ApJ...645.1152H,
  2007MNRAS.379.1083G,2009ApJ...698..641E,2009ApJ...702..884P,
  2009ApJ...705..361G,2010PhRvD..81f2002M,
  2011MNRAS.411L..56G,2011ApJ...738...99M,2011PhRvD..84d4024M,
  2011ApJ...726...61M,2012A&A...545A..70S,2013ApJ...763L..10A}.  Some
of these studies only computed the torques between fixed Kepler
ellipses, which are not relevant in the regime considered here where
the apsidal precession is faster than the re-orientation of the
ellipse. Some employed direct N-body simulations, which in principle
are more accurate than the approximations used here. However, due to
the computational cost of studying slow processes such as VRR with
direct N-body simulations, earlier studies were restricted to either
(i) small-$N$ systems, in which the vector and scalar resonant
relaxation timescales are not well-separated, or (ii) following the
N-body system for less than the apsidal precession period, so the
torque parameters $\beta_T$ and $\beta_\Omega$ (Eq.\
\ref{e:torque-coherent-powerlaw}) were measured for Keplerian ellipses
rather than annuli. Thus, either they did not measure the long-term average
values of $\beta_T$ and $\beta_\Omega$ that are relevant for VRR, or
they did not measure the coefficient $f_{\vrr}$ (Eq.\ \ref{e:vrr}) that
parametrizes incoherent VRR. We believe that the simulations in this paper
provide the first detailed study of VRR that represents
both the coherent and incoherent evolution for systems with a large
number of stars.

When comparing with earlier studies, we must account for definitions
of $\beta$ in these papers that are slightly different from ours:

\begin{itemize}

\item \citet{1996NewA....1..149R} defined $\beta^{\rm RT}$ using
  $\langle\|\T_i\|\rangle=\beta^{\rm RT} (2\pi)^{-1} \sqrt{N}Gm^2/a_i$
  where $N$ denotes the total number of stars. They carried out N-body
  simulations with $64\le N\le 8192$ and a range of semimajor axes
  $a_{\max}/a_{\min}=10$, with $n(a)\propto a^{-\gamma}$, $\gamma=2$,
  and a thermal distribution of eccentricities for $e\le
  e_{\max}=0.8$. Only 64 ``active'' stars interacted
  self-consistently; the rest exerted torques on the active stars but
  followed fixed orbits in either a point-mass or an isochrone
  potential (the isochrone was used to experiment with the effect of
  rapid apsidal precession).  They measured $\beta^{\rm RT}$ as
  $\beta^{\rm RT} \equiv \langle 2\pi a_i \|\T_i\|\rangle/(\sqrt{N}
  Gm^2 ) =(M_{\bullet}/\sqrt{N}m) \langle \|\T_i \|P_i / L_{{\rm
      c},i}\rangle_i$, where $L_{{\rm c},i}=m \sqrt{G M_{\bullet}
    a_i}$ is the angular momentum of a circular orbit. To compare this
  to our $\beta_T$ we must make two corrections. First, for a
  two-dimensional Gaussian distribution ($\T$ perpendicular to $\L$)
  $\langle \|\T\|\rangle = \half\pi^{1/2} T_{\rms} $.  Second, we
  measure $\beta_T$ using $dN/d\ln a$ whereas they use $N$; to make
  the conversion we note from Appendix~\ref{s:beta} that in a
  power-law density distribution $\beta_T$ and hence $ a
  \|\T\| /\sqrt{\D N /\D \ln a}$ is independent of $a$, so we have
  $\langle a_i\|\T_i\|\rangle/\sqrt{N} = (a_i \|\T_i\|/\sqrt{N})\langle
    \sqrt{\D N/\D\ln a}\,\rangle /\sqrt{\D N/\D\ln a_i}$. Then for the
  assumed number density profile ($\gamma=2$, $a_{\max}/a_{\min}=10$), we have $\beta^{\rm RT}
  = \half f_e\pi^{1/2} \beta_T {\int_{a_{\min}}^{a_{\max}} \D N\,(\D N /\D \ln
  a)^{1/2} /(N^{1/2}\int_{a_{\min}}^{a_{\max}} \D N)} =
  0.670f_e\, \beta_T$, where $f_e\simeq 1.2$ is a correction arising
  because Rauch \& Tremaine did not have any stars with
  $e>e_{\max}=0.8$ (cf.\ Fig.~\ref{f:betaT}).  They measured
  $\beta^{\rm RT} = 1.8\pm 0.1$ in the Kepler case where the
  background stars had no apsidal precession due to the unperturbed
  potential, and $\beta^{\rm RT} = 0.7\pm 0.1$ in the isochrone case
  with rapid apsidal precession, corresponding to $\beta_T=2.2\pm0.1$
  and $\beta_T=0.9\pm0.1$, respectively.

\item \citet{2007MNRAS.379.1083G} defined $\beta^{\rm GH}$ using
  $\langle\|\T_i\|\rangle=\beta^{\rm GH} \sqrt{N(<2a_i)}Gm^2/a_i$,
  where $N(<2a_i)$ denotes the number of stars with semimajor axis
  less than $2a_i$.  They calculated the orbit-averaged torques for
  fixed Keplerian wires using $N=10,000$ stars with density
  $n(a)\propto a^{-\gamma}$ and $\gamma=1.4$.  They found that the
  mean absolute torque along the minor axis of the orbit increased
  with eccentricity and conversely along the major axis, such that the
  total torque increased with eccentricity as $\beta^{\rm GH} = 1.76
  (e^2 + 0.5)/2\pi$ with an average over the eccentricity distribution
  ($\D N/\D e\propto 2e$) $\langle \beta^{\rm GH}\rangle = 1.76/2\pi$
  and an RMS $\langle (\beta^{\rm GH})^2\rangle^{1/2}=1.83/2\pi$. For
  a power-law density distribution our definition of the torque
  parameter is related to theirs as
$2\pi\beta^{\rm GH}= \pi^{1/2} (3-\gamma)^{1/2} 2^{(\gamma-5)/2}
\beta_T$. For $\gamma=1.4$ this yields $2\pi\beta^{\rm GH} =
0.64\,\beta_T$, so their result implies
$\beta_T=2.7(e^2+0.5)$. Averaging over a thermal distribution of
eccentricities yields $\langle \beta_T\rangle=2.6$
and $\langle \beta_T^2 \rangle^{1/2}=2.9$.

\item \citet{2009ApJ...698..641E} defined $\beta^{\rm EAK}$ using a
  similar definition as \citet{1996NewA....1..149R},
  $\langle\|\T_i\|\rangle=\beta^{\rm EKA} (2\pi)^{-1}
  \sqrt{N}Gm^2/a$. They conducted a number of N-body simulations with
  $N=200$ and a variety of semimajor axis distributions, number
  density $n\propto a^{-\gamma}$ with $1\leq \gamma\leq 1.75$.  They
  noted that the torque perpendicular to $\Ln$ was mostly along the
  instantaneous minor axis of the orbit, as in
  \citet{2007MNRAS.379.1083G}. Using the same arguments as for
  \citet{1996NewA....1..149R}, we get that $\beta^{\rm EKA} = 0.68\,
  \beta_T$ and $0.69\, \beta_T$ for $\gamma=1.75$ and $\gamma=1$,
  respectively.  Measuring the re-orientation correlation function for
  a few precession times, they found $\beta^{\rm EKA} = 1.83\pm 0.03$,
  which implies $\beta_T = 2.7$.
\end{itemize}

For comparison, our calculations yield $\beta_T\simeq 0.85\pm 0.1$
(Eq.\ \ref{e:betafit} and Fig.\ \ref{f:beta}), which is a factor 3
smaller than the results reported by \citet{1996NewA....1..149R},
\citet{2007MNRAS.379.1083G}, and \citet{2009ApJ...698..641E}.  The
systematically higher value of $\beta_T$ found in these investigations
arises because the torque on an orbit was averaged over a timescale
short compared to the apsidal precession period\footnote{The rate of
  re-orientation, as measured by $\beta_T$, can be even more rapid on
  timescales shorter than or comparable to the orbital period 
  \citep{2010PhRvD..81f2002M,2011CQGra..28v5029S,2012A&A...545A..70S}.}.
As shown by these studies, the largest component of the torque is
parallel to the minor axis of the Keplerian orbit; as the orbit
precesses, the direction of the largest torque precesses as well so
the mean torque averaged over a precession period is smaller than the
mean torque averaged over the orbital period. The use of torques
averaged over the apsidal precession period rather than the orbital
period is necessary to estimate the rate of VRR on timescales longer
than the apsidal precession period, so long as apsidal precession is
much faster than nodal precession.  This requirement is satisfied for stars of
small to moderate eccentricity at all radii in the Galactic centre
(see Fig.\ 1 of KT11), but can fail for nearly radial orbits at large
or small radii (see footnote \ref{foot:prec}).

An observation that supports this argument is that our estimate
$\beta_T\simeq 0.85\pm 0.1$ matches the estimate $\beta_T=0.9\pm0.1$
reported by \citet{1996NewA....1..149R} for the isochrone potential, in
which the stars are subject to rapid apsidal precession. Furthermore,
a similar rate of vector resonant relaxation was found using direct
N-body simulations, which
looked at the long term behavior of orbits close to the SMBH including
relativistic corrections (Kupi \& Alexander, private communication\footnote{talk presented at Stars and Singularities, Benoziyo Center for Astrophysics Workshop Series, Rehovot, Israel, \url{http://www.weizmann.ac.il/home/tal/Workshop09/talk_files/Kupi.pdf}}). Their rate
of re-orientation may be fitted by $\|\Delta \L\|/L = c_0 (\beta_{T,0} +
\beta_{T,1}) t/t_{\omega}$ if $t\,\lesssim\, t_{\omega}$, and $c_0
[\beta_{T,0} (t/t_{\omega})^{1/2} + \beta_{T,1} (t/t_{\omega})]$ if
$t_{\phi} \,\gtrsim\, t \,\gtrsim\, t_{\omega}$, where
$\beta_{T,0}+\beta_{T,1}\simeq 2.7$, $\beta_{T,1}\simeq 0.9$,
$t_\omega$ is the apsidal precession time, and $c_0$ is a
constant. Thus, part of the initial coherent torque becomes incoherent
over timescales longer than the apsidal precession time $t_{\omega}$,
leaving a much smaller coherent component thereafter.

An additional limitation of earlier studies is that they could 
not accurately characterize the properties of the random walk for the direction
$\Ln_i$ during the incoherent phase of VRR (i.e., the parameter
$f_{\vrr}$ of Eq.\ \ref{e:vrr}), mainly due to the computational cost
of long N-body integrations. Furthermore, previous simulations were
restricted to a small number of self-consistently interacting stars
(between $50$ and $200$) in which complete mixing sets in much earlier
(Eqs.~\ref{e:Vell-theory1})--(\ref{e:Vell-theory2}) which makes the
measurement of the parameters of the incoherent phase more difficult.\footnote{
\citet{2009ApJ...698..641E} defined $f_{\vrr}=1/(A_{\phi} \beta_{\Omega}^2)$, 
where $A_{\phi}$, set by the decoherence time in Eq.~(\ref{e:tphi}), 
was not determined.}
Finally, previous analyses used the simplified model
$\langle\|\L(t+t_0) - \L(t_0)\|/L \rangle \propto (t/t_{\rm
  vrr})^{1/2}$ to characterize VRR, which is
not appropriate if the angular coherence length is of order unity; it
is for this reason that we developed the analysis in
Section~\ref{s:random-theory} based on the random walk on the sphere.

\section{Summary}

We have introduced a new integrator, \textsc{n-ring}, to simulate
vector resonant relaxation in stellar clusters around supermassive
black holes.  \textsc{n-ring} integrates Hamilton's equations for $N$
stars, averaged over the orbital period and apsidal precession. The
code uses a multipole expansion (up to $\ell_{\max}=50$ in our
experiments) of the averaged inter-particle potential.  The code
decomposes the evolution into pairwise interactions, integrates the
averaged Hamiltonian exactly for each pairwise interaction, and
iterates over all $\half N(N-1)$ such interactions, thereby conserving
the total angular momentum exactly.  The coupling coefficients for
different multipole moments are generally complicated functions of the
semimajor axis and eccentricity, but can be calculated once and for
all at the start of the integration.

We have shown how to make the algorithm time-reversible and $n^{\rm
  th}$ order accurate (up to $n=8$ in our experiments). We constructed
a parallelization scheme, and increased the efficiency using a
time-block refinement and operator ordering.  Using a small computer
cluster of 32 cores, this integrator can accurately integrate the
evolution of a cluster of $\sim 10^4$ stars with a large range of
radii for $\sim 10$ relaxation times within 7 days.

The major challenges that limit the speed of the code include the
following.

\begin{enumerate}[leftmargin=0.5cm]

 \item The coupling coefficients driving resonant relaxation
can be strongly enhanced for orbits with nearly coincident periapsides
or apoapsides (see bottom panels of Figure \ref{f:energy-alpha}).

 \item
For radially overlapping orbits the coupling coefficients decline
relatively slowly, as $\ell^{-2}$, implying that all multipoles up to
$\ell\sim 1/I$ contribute equally to the motion for orbital
inclination $I$.

 \item The precession frequency between two radially
overlapping orbits diverges as their mutual inclination approaches
zero.

 \item Gravitational N-body integrations of star clusters,
galaxies, or large-scale structure benefit from the fact that most
stars are at large distances ($N\sim r^3$) so their collective
gravitational potential can be approximated by a few multipole
moments; in contrast, in the averaged problem investigated by
\textsc{n-ring} each star can interact strongly with all stars having
radially overlapping orbits. Thus there are no simple ways to reduce
the number of calculations per timestep below O$(N^2)$. However,
parallel execution on $N$ processors can reduce the computation time
to O$(N)$.

\end{enumerate}

We derived a stochastic model to describe a random walk with
an arbitrary distribution of step sizes on the unit sphere.  Expanding
the probability distribution in spherical harmonics shows that the
amplitudes of the spherical harmonics with $\ell>0$ decay exponentially during a
spherical random walk.  The angular variance $V_{\ell}\equiv
-2\ell^{-1}(\ell+1)^{-1}\ln|\langle P_{\ell}(\cos\alpha) \rangle|$
grows linearly in time where $\alpha$ is the angular distance
traversed by an orbit normal in time $t$ and $P_{\ell}(\cdot)$ are
Legendre polynomials.

We have investigated the long-term evolution of spherical stellar
systems with up to 16k stars, spanning a factor of up to 100 in
semimajor axis.  The simulations confirm that the
orbital orientation vectors initially evolve coherently
($V_{\ell}\propto t^2$) and then undergo a spherical random walk
($V_{\ell}\propto t$) until the system becomes fully mixed. The RMS
step size of the random walk in our simulations is
$\alpha_{\rms}\simeq 0.5$--1 radians and full mixing requires $(\ln
3N)/\alpha_{\rms}^2$ timesteps where $N$ is the number of stars.

In the initial coherent phase of vector resonant relaxation, the RMS
torques can be calculated exactly (Appendix~\ref{s:beta} and Figures
\ref{f:betaT} and \ref{f:beta}).  This confirmed the analytical
scaling relations with semimajor axis, number density, and component
mass, and showed perfect agreement with the simulations.  In
particular, the torque parameter is $\beta_T=0.8$--$1.5$ (see
Eq.~\ref{e:torque-coherent-powerlaw} and Figure~\ref{f:betaT}) for
different eccentricities. The rate of re-orientation of the orbital
plane follows a similar scaling with
$\beta_\Omega=\beta_T/(1-e^2)^{1/2}$ (Eq.~\ref{e:betafit}).  We found
that the torques are generally weakly decreasing functions of the
eccentricity in spherical clusters during vector resonant relaxation,
and in particular for a thermal eccentricity distribution
$\beta_T\simeq 1.05-0.3\,e$.  The rate of re-orientation of the orbit
axis is approximately independent of eccentricity for $e\lesssim 0.7$,
and much faster only for $e \,\gtrsim\, 0.8$. The rate of
re-orientation is smaller than has been observed in most\footnote{Except for the
isochrone simulations of \citet{1996NewA....1..149R} and Kupi \&
Alexander, as described in Section~\ref{s:comparison}.} N-body simulations
by a factor $\sim 3$, and most of this difference arises
because the torque perpendicular to the angular-momentum vector is
smaller when apsidal precession is rapid.

Our simulations confirm the formula for the vector resonant relaxation
timescale derived from a model of the relaxation as a random walk on
the sphere (Eq.~\ref{e:vrr}) and imply that the parameter
$f_{\vrr}\simeq 0.5$--2.1 depending mainly on eccentricity (Figures
\ref{f:correlation_a} and \ref{f:correlation_a2}).  In a thermal
distribution of eccentricities ($\D N \propto 2\,e\,\D e$), we find
that highly eccentric orbits $e\,\gtrsim\, 0.8$ relax faster by up to
a factor $5$; however, the vector resonant relaxation time for low-
and moderate-eccentricity orbits with $e\lesssim 0.7$ is practically
independent of eccentricity with $f_{\vrr}\simeq 1.9\pm 0.2$. The
simulations also show that the decoherence time of vector resonant
relaxation is roughly independent of eccentricity in the full eccentricity range.  The
angular-momentum vectors in the inner regions of our simulated cluster
undergo a stochastic random walk already when the vectors in the outer
parts of the cluster are still experiencing a coherent torque.  For a
cluster with a given number of stars, the relaxation rate is
proportional to the RMS stellar mass of the stellar cluster. Thus the
primary uncertainty in estimating the vector resonant relaxation near
the Galactic centre is the mass function of stars, stellar remnants,
gas clouds, etc.: the RMS stellar mass diverges even for a Salpeter
mass function unless a maximum-mass cutoff is imposed, and the mass
function in the Galactic centre is believed to be more top-heavy than
in the solar neighbourhood (see KT11 and references therein).

We found that the Markovian random walk on a sphere gives a good approximate
description of the long-term evolution under vector
resonant relaxation.  However, in some cases the temporal correlation
function displays deviations from this model even after averaging over
several mixing timescales (Figure~\ref{f:correlation1}), which
possibly indicates some level of persistent long-term memory in these
stellar systems.  In the future we will use \textsc{n-ring} to
examine resonant dynamical friction and vector resonant relaxation in
anisotropic systems.

The purpose of this paper has been twofold: first, to develop an
efficient and general numerical algorithm for simulating vector
resonant relaxation, and second, to relate the simple analytic
description of vector resonant relaxation to quantitative results from
our simulations of model star clusters surrounding central
black holes.

\section*{Acknowledgments}
BK was supported in part by the W.M. Keck Foundation Fund of the
Institute for Advanced Study and NASA grants NNX11AF29G and
NNX14AM24G.  Simulations were run on the Harvard Odyssey, CfA/ITC, 
and IAS clusters.

\bibliography {ms}

\begin{thebibliography}{}

\bibitem[\protect\citeauthoryear{{Alexander}, {Armitage} \&
  {Cuadra}}{{Alexander} et~al.}{2008}]{2008MNRAS.389.1655A}
{Alexander} R.~D.,  {Armitage} P.~J.,    {Cuadra} J.,  2008, \mnras, 389, 1655

\bibitem[\protect\citeauthoryear{{Antonini} \& {Merritt}}{{Antonini} \&
  {Merritt}}{2013}]{2013ApJ...763L..10A}
{Antonini} F.,  {Merritt} D.,  2013, \apjl, 763, L10

\bibitem[\protect\citeauthoryear{{Bar-Or} \& {Alexander}}{{Bar-Or} \&
  {Alexander}}{2014}]{2014CQGra..31x4003B}
{Bar-Or} B.,  {Alexander} T.,  2014, Classical and Quantum Gravity, 31, 244003

\bibitem[\protect\citeauthoryear{{Bar-Or}, {Kupi} \& {Alexander}}{{Bar-Or}
  et~al.}{2013}]{2013ApJ...764...52B}
{Bar-Or} B.,  {Kupi} G.,    {Alexander} T.,  2013, \apj, 764, 52

\bibitem[\protect\citeauthoryear{{Bartko}, {Martins}, {Fritz}, {Genzel},
  {Levin}, {Perets}, {Paumard}, {Nayakshin}, {Gerhard}, {Alexander},
  {Dodds-Eden}, {Eisenhauer}, {Gillessen}, {Mascetti}, {Ott}, {Perrin}, {Pfuhl}
  \& {Reid}}{{Bartko} et~al.}{2009}]{2009ApJ...697.1741B}
{Bartko} H.,  {Martins} F.,  {Fritz} T.,  {Genzel} R.,  {Levin} Y.,  {Perets}
  H.,  {Paumard} T.,  {Nayakshin} S.,  {Gerhard} O.,  {Alexander} T.,
  {Dodds-Eden} K.,  {Eisenhauer} F.,  {Gillessen} S.,  {Mascetti} L.,  {Ott}
  T.,  {Perrin} G.,  {Pfuhl} O.,    {Reid} M.,  2009, \apj, 697, 1741

\bibitem[\protect\citeauthoryear{{Bartko}, {Martins}, {Trippe}, {Fritz},
  {Genzel}, {Ott}, {Eisenhauer}, {Gillessen}, {Paumard}, {Alexander},
  {Dodds-Eden}, {Gerhard}, {Levin}, {Mascetti}, {Nayakshin}, {Perets}, {Perrin}
  \& {Pfuhl}}{{Bartko} et~al.}{2010}]{2010ApJ...708..834B}
{Bartko} H.,  {Martins} F.,  {Trippe} S.,  {Fritz} T.,  {Genzel} R.,  {Ott} T.,
   {Eisenhauer} F.,  {Gillessen} S.,  {Paumard} T.,  {Alexander} T.,
  {Dodds-Eden} K.,  {Gerhard} O.,  {Levin} Y.,  {Mascetti} L.,  {Nayakshin} S.,
   {Perets} H.,  {Perrin} G.,    {Pfuhl} O.,  2010, \apj, 708, 834

\bibitem[\protect\citeauthoryear{{Binney} \& {Tremaine}}{{Binney} \&
  {Tremaine}}{2008}]{bt08}
{Binney} J.,  {Tremaine} S.,  2008, {Galactic Dynamics, 2nd Edition}.
Princeton University Press, Princeton NJ

\bibitem[\protect\citeauthoryear{{Byrd} \& {Friedman}}{{Byrd} \&
  {Friedman}}{1971}]{ByrdFriedman}
{Byrd} P.,  {Friedman} M.,  1971, {Handbook of Elliptic Integrals for Engineers
  and Scientists}.
Springer-Verlag, Berlin

\bibitem[\protect\citeauthoryear{{Casas}, {Murua} \& {Nadinic}}{{Casas}
  et~al.}{2012}]{2012CoPhC.183.2386C}
{Casas} F.,  {Murua} A.,    {Nadinic} M.,  2012, Computer Physics
  Communications, 183, 2386

\bibitem[\protect\citeauthoryear{{Coffey} \& {Kalmykov}}{{Coffey} \&
  {Kalmykov}}{2012}]{2012leas.book.....C}
{Coffey} W.~T.,  {Kalmykov} Y.~P.,  2012, {The Langevin Equation: With
  Applications to Stochastic Problems in Physics, Chemistry and Electrical
  Engineering (3rd Edition)}.
World Scientific, Singapore

\bibitem[\protect\citeauthoryear{{Debye}}{{Debye}}{1929}]{1929pomo.book.....D}
{Debye} P.,  1929, {Polare Molekeln}.
Hirzel, Leipzig

\bibitem[\protect\citeauthoryear{{Eilon}, {Kupi} \& {Alexander}}{{Eilon}
  et~al.}{2009}]{2009ApJ...698..641E}
{Eilon} E.,  {Kupi} G.,    {Alexander} T.,  2009, \apj, 698, 641

\bibitem[\protect\citeauthoryear{Gottwald \& Melbourne}{Gottwald \&
  Melbourne}{2013}]{Gottwald21052013}
Gottwald G.~A.,  Melbourne I.,  2013, Proc.\ Nat.\ Acad.\ Sci., 110, 8411

\bibitem[\protect\citeauthoryear{{Gualandris} \& {Merritt}}{{Gualandris} \&
  {Merritt}}{2009}]{2009ApJ...705..361G}
{Gualandris} A.,  {Merritt} D.,  2009, \apj, 705, 361

\bibitem[\protect\citeauthoryear{{G{\"u}rkan}}{{G{\"u}rkan}}{2011}]{2011MNRAS.%
411L..56G}
{G{\"u}rkan} M.,  2011, \mnras, 411, L56

\bibitem[\protect\citeauthoryear{{G{\"u}rkan} \& {Hopman}}{{G{\"u}rkan} \&
  {Hopman}}{2007}]{2007MNRAS.379.1083G}
{G{\"u}rkan} M.,  {Hopman} C.,  2007, \mnras, 379, 1083

\bibitem[\protect\citeauthoryear{{Hopman}}{{Hopman}}{2009}]{2009ApJ...700.1933%
H}
{Hopman} C.,  2009, \apj, 700, 1933

\bibitem[\protect\citeauthoryear{{Hopman} \& {Alexander}}{{Hopman} \&
  {Alexander}}{2006}]{2006ApJ...645.1152H}
{Hopman} C.,  {Alexander} T.,  2006, \apj, 645, 1152

\bibitem[\protect\citeauthoryear{{Ivanov}, {Polnarev} \& {Saha}}{{Ivanov}
  et~al.}{2005}]{2005MNRAS.358.1361I}
{Ivanov} P.~B.,  {Polnarev} A.~G.,    {Saha} P.,  2005, \mnras, 358, 1361

\bibitem[\protect\citeauthoryear{{Jackson}}{{Jackson}}{1998}]{1998clel.book...%
..J}
{Jackson} J.~D.,  1998, {Classical Electrodynamics, 3rd Edition}.
Wiley-VCH, New York

\bibitem[\protect\citeauthoryear{{Kocsis} \& {Tremaine}}{{Kocsis} \&
  {Tremaine}}{2011}]{2011MNRAS.412..187K}
{Kocsis} B.,  {Tremaine} S.,  2011, \mnras, 412, 187 (KT11)

\bibitem[\protect\citeauthoryear{Kumar, Harbola \& Lindenberg}{Kumar
  et~al.}{2010}]{PhysRevE.82.021101}
Kumar N.,  Harbola U.,    Lindenberg K.,  2010, Phys. Rev. E, 82, 021101

\bibitem[\protect\citeauthoryear{Latora, Rapisarda \& Ruffo}{Latora
  et~al.}{1999}]{PhysRevLett.83.2104}
Latora V.,  Rapisarda A.,    Ruffo S.,  1999, Phys. Rev. Lett., 83, 2104

\bibitem[\protect\citeauthoryear{{L{\"o}ckmann}, {Baumgardt} \&
  {Kroupa}}{{L{\"o}ckmann} et~al.}{2009}]{2009MNRAS.398..429L}
{L{\"o}ckmann} U.,  {Baumgardt} H.,    {Kroupa} P.,  2009, \mnras, 398, 429

\bibitem[\protect\citeauthoryear{{Madigan}, {Hopman} \& {Levin}}{{Madigan}
  et~al.}{2011}]{2011ApJ...738...99M}
{Madigan} A.-M.,  {Hopman} C.,    {Levin} Y.,  2011, \apj, 738, 99

\bibitem[\protect\citeauthoryear{{Merritt}}{{Merritt}}{2013}]{2013degn.book...%
..M}
{Merritt} D.,  2013, {Dynamics and Evolution of Galactic Nuclei}.
Princeton University Press, Princeton NJ

\bibitem[\protect\citeauthoryear{{Merritt}, {Alexander}, {Mikkola} \&
  {Will}}{{Merritt} et~al.}{2010}]{2010PhRvD..81f2002M}
{Merritt} D.,  {Alexander} T.,  {Mikkola} S.,    {Will} C.~M.,  2010, \prd, 81,
  062002

\bibitem[\protect\citeauthoryear{{Merritt}, {Alexander}, {Mikkola} \&
  {Will}}{{Merritt} et~al.}{2011}]{2011PhRvD..84d4024M}
{Merritt} D.,  {Alexander} T.,  {Mikkola} S.,    {Will} C.~M.,  2011, \prd, 84,
  044024

\bibitem[\protect\citeauthoryear{{Merritt} \& {Vasiliev}}{{Merritt} \&
  {Vasiliev}}{2011}]{2011ApJ...726...61M}
{Merritt} D.,  {Vasiliev} E.,  2011, \apj, 726, 61

\bibitem[\protect\citeauthoryear{{Merritt} \& {Vasiliev}}{{Merritt} \&
  {Vasiliev}}{2012}]{2012PhRvD..86j2002M}
{Merritt} D.,  {Vasiliev} E.,  2012, \prd, 86, 102002

\bibitem[\protect\citeauthoryear{{Perets}, {Gualandris}, {Kupi}, {Merritt} \&
  {Alexander}}{{Perets} et~al.}{2009}]{2009ApJ...702..884P}
{Perets} H.~B.,  {Gualandris} A.,  {Kupi} G.,  {Merritt} D.,    {Alexander} T.,
   2009, \apj, 702, 884

\bibitem[\protect\citeauthoryear{{Pfuhl}, {Alexander}, {Gillessen}, {Martins},
  {Genzel}, {Eisenhauer}, {Fritz} \& {Ott}}{{Pfuhl}
  et~al.}{2014}]{2014ApJ...782..101P}
{Pfuhl} O.,  {Alexander} T.,  {Gillessen} S.,  {Martins} F.,  {Genzel} R.,
  {Eisenhauer} F.,  {Fritz} T.~K.,    {Ott} T.,  2014, \apj, 782, 101

\bibitem[\protect\citeauthoryear{{Rauch} \& {Tremaine}}{{Rauch} \&
  {Tremaine}}{1996}]{1996NewA....1..149R}
{Rauch} K.,  {Tremaine} S.,  1996, New Astr., 1, 149

\bibitem[\protect\citeauthoryear{{Roberts} \& {Ursell}}{{Roberts} \&
  {Ursell}}{1960}]{Roberts_Ursell60}
{Roberts} P.,  {Ursell} H.,  1960, Phil. Trans. R. Soc. Lond. A, 252, 317

\bibitem[\protect\citeauthoryear{{Sabha}, {Eckart}, {Merritt}, {Zamaninasab},
  {Witzel}, {Garc{\'{\i}}a-Mar{\'{\i}}n}, {Jalali}, {Valencia-S.}, {Yazici},
  {Buchholz}, {Shahzamanian}, {Rauch}, {Horrobin} \& {Straubmeier}}{{Sabha}
  et~al.}{2012}]{2012A&A...545A..70S}
{Sabha} N.,  {Eckart} A.,  {Merritt} D.,  {Zamaninasab} M.,  {Witzel} G.,
  {Garc{\'{\i}}a-Mar{\'{\i}}n} M.,  {Jalali} B.,  {Valencia-S.} M.,  {Yazici}
  S.,  {Buchholz} R.,  {Shahzamanian} B.,  {Rauch} C.,  {Horrobin} M.,
  {Straubmeier} C.,  2012, \aap, 545, A70

\bibitem[\protect\citeauthoryear{{Sadeghian} \& {Will}}{{Sadeghian} \&
  {Will}}{2011}]{2011CQGra..28v5029S}
{Sadeghian} L.,  {Will} C.~M.,  2011, Classical and Quantum Gravity, 28, 225029

\bibitem[\protect\citeauthoryear{{Saha} \& {Tremaine}}{{Saha} \&
  {Tremaine}}{1994}]{1994AJ....108.1962S}
{Saha} P.,  {Tremaine} S.,  1994, \aj, 108, 1962

\bibitem[\protect\citeauthoryear{{Sch{\"o}del}, {Eckart}, {Alexander},
  {Merritt}, {Genzel}, {Sternberg}, {Meyer}, {Kul}, {Moultaka}, {Ott} \&
  {Straubmeier}}{{Sch{\"o}del} et~al.}{2007}]{2007A&A...469..125S}
{Sch{\"o}del} R.,  {Eckart} A.,  {Alexander} T.,  {Merritt} D.,  {Genzel} R.,
  {Sternberg} A.,  {Meyer} L.,  {Kul} F.,  {Moultaka} J.,  {Ott} T.,
  {Straubmeier} C.,  2007, \aap, 469, 125

\bibitem[\protect\citeauthoryear{{Suzuki}}{{Suzuki}}{1990}]{1990PhLA..146..319%
S}
{Suzuki} M.,  1990, Physics Letters A, 146, 319

\bibitem[\protect\citeauthoryear{{Suzuki}}{{Suzuki}}{1994}]{1994PhyA..205...65%
S}
{Suzuki} M.,  1994, Physica A, 205, 65

\bibitem[\protect\citeauthoryear{{Touma}, {Tremaine} \& {Kazandjian}}{{Touma}
  et~al.}{2009}]{2009MNRAS.394.1085T}
{Touma} J.,  {Tremaine} S.,    {Kazandjian} M.,  2009, \mnras, 394, 1085

\bibitem[\protect\citeauthoryear{{Tremaine}}{{Tremaine}}{2005}]{2005ApJ...625.%
.143T}
{Tremaine} S.,  2005, \apj, 625, 143

\bibitem[\protect\citeauthoryear{{Trotter}}{{Trotter}}{1959}]{Trotter}
{Trotter} H.,  1959, Proc. Amer. Math. Soc., 10, 545

\bibitem[\protect\citeauthoryear{{Tuckerman}, {Berne} \& {Martyna}}{{Tuckerman}
  et~al.}{1992}]{1992JChPh..97.1990T}
{Tuckerman} M.,  {Berne} B.,    {Martyna} G.,  1992, Jour. Chem. Phys., 97,
  1990

\bibitem[\protect\citeauthoryear{{Yoshida}}{{Yoshida}}{1990}]{1990PhLA..150..2%
62Y}
{Yoshida} H.,  1990, Physics Letters A, 150, 262

\end{thebibliography}
\appendix
\onecolumn
\section{Apsidal precession}\label{app:precession}

We calculate the apsidal precession rate $\Omega_{\rm prec}$ of stellar orbits due to the
gravitational field from a spherical near-Keplerian stellar system.
For a stellar system with enclosed mass $M_*(r) \ll M_{\bullet}$ we
have
\citep{2005ApJ...625..143T}
\begin{equation}
 \Omega_{\rm prec} = \frac{\Omega}{\pi M_{\bullet} e} \int_0^{\pi} \D \psi\; M_*[r(\psi)] \cos \psi
\end{equation}
where $\Omega = (G M_{\bullet})^{1/2} a^{-3/2}$ is the average orbital
angular frequency, $\psi$ is the true anomaly, and the radius
 is given by $r(\psi) = p / ( 1 + e \cos\psi)$, where $e$ is the
eccentricity, $p=a(1-e^2)$ is the semi-latus rectum, and $a$ is the
semimajor axis. The precession is retrograde for any positive-definite
spherical mass distribution \citep{2005ApJ...625..143T}.

The integral can be simplified for power-law mass distributions of the
form\footnote{For the Galactic centre $ s = 1.8$ and 1.25 for
  $r\lesssim 0.2$\,pc and $r\gtrsim 0.2$\,pc, respectively
\citep{2007A&A...469..125S,2009MNRAS.398..429L}.}
 $M_*(r) = M_0 (r/r_0)^{ s} $. In this case \citep{2005MNRAS.358.1361I}
\begin{equation}
 \Omega_{\rm prec} =
\frac{\Omega }{\pi  e} \frac{M_0}{M_{\bullet}} \left(\frac{p}{r_0}\right)^{ s}
\int_0^{\pi} \frac{\cos\psi\,\D \psi}{(1+e\cos \psi)^s}
=\Omega\frac{M_0}{M_{\bullet}}
\left(\frac{a}{r_0}\right)^s\frac{(1-e^2)^{(s+1)/2}}{e^2}\left[P_{s-2}(\chi)-\chi
  P_{s-1}(\chi)\right]\quad
  \mbox{where} \quad \chi\equiv \frac{1}{\sqrt{1-e^2}}.
\end{equation}
Here $P_n$ denotes the Legendre function of order $n$. In terms of the
density $\rho(r)=(4\pi r^2)^{-1}\D M(r)/\D r$, we have
\begin{equation}
 \Omega_{\rm prec} = \frac{4\pi G\rho(a)}{\Omega
   s}\frac{(1-e^2)^{(s+1)/2}}{e^2}
\left[P_{s-2}(\chi)-\chi P_{s-1}(\chi)\right]\quad
  \mbox{where} \quad \chi\equiv \frac{1}{\sqrt{1-e^2}}.
\end{equation}
For $e\rightarrow 0$ and arbitrary $ s>0$,
\begin{equation}
  \Omega_{\rm prec} = -\frac{2\pi G \rho(a) }{\Omega }  \left[ 1
    +\left( \frac{1}{4}-\frac{5s}{8}+\frac{s^2}{8}\right)e^2 +\mathcal{O}(e^4) \right].
\end{equation}
For some values of $s$ there are analytic expressions valid for all
eccentricities \citep{2013degn.book.....M}:
\begin{align}\label{e:appprec}
  \Omega_{\rm prec} &= - \frac{2\pi G \rho(a) }{  \Omega } \sqrt{1-e^2}
\times\left\{
\begin{array}{ll}
1 				& \text{if~} \rho(r)\propto r^{-1}, \quad  s =2 \\
2/(1+\sqrt{1-e^2}) 	& \text{if~} \rho(r)\propto r^{-2}, \quad  s =1.
\end{array}
\right.
\end{align}

\section{Interaction energy}\label{app:interactionenergy}

Here we simplify the orbit- and precession-averaged interaction energy
between two stars (Eq.\ \ref{e:Hannuli}), which is a four-dimensional
integral over the two annular surfaces.  The evaluation of this integral depends on
the radial geometry of the two annuli. In particular let $R_1$ and
$R_2$ be the set of all radii occupied by the annuli of the two orbits
(e.g., $R_1=\{r\, |\, r_{p1}\le r\le r_{a1}\}$ where $r_{p1}$ and $r_{a1}$
are the periapsis and apoapsis of orbit 1). We call the orbits
``non-overlapping'' if they occupy disjoint ranges of radius, $R_1\cap
R_2=\emptyset$; we say that orbit 1 is ``embedded'' in orbit 2 if
$R_1\subset R_2$; we call the orbits ``identical'' if $R_1=R_2$
(even if the orbits are mutually inclined); and we say the orbits are
``overlapping'' if $R_1\cap R_2\not=\emptyset$ but $R_1\not\subset
R_2$ and $R_2\not\subset R_1$.

We show that the interaction energy can be reduced to a sum over a
series of one-dimensional integrals in the general case, and to a sum
over a series of closed analytic expressions for non-overlapping or
identical orbits.

We need first to find the gravitational potential energy between two
circular rings of radius $r$ and $r'$, inclined by an angle $I$. We
expand the inverse distance in spherical harmonics\footnote{We use
  the orthonormal definition for spherical harmonics \citep{1998clel.book.....J}
\begin{equation}\label{e:Y}
 Y_{\ell m}(\theta,\varphi) = \sqrt{\frac{2\ell + 1}{4\pi} \frac{(\ell - m)!}{(\ell + m)!} } P_{\ell}^{m}(\cos\theta) e^{i m \varphi} \,,
\end{equation}
where $P_{\ell}^m(x)$ are associated Legendre polynomials, defined by
\begin{equation}
 P_{\ell}^{m}(x) = \frac{(-1)^m}{2^\ell\, \ell!} (1 - x^2)^{m/2} \frac{\D^{\ell + m}}{\D x^{\ell + m}} (x^2 - 1)^{\ell}\,.
\end{equation}
In particular for $m=0$, $P_{\ell}^{0}(x) = P_{\ell}(x)$ are Legendre polynomials.
},
\begin{equation}
  \frac{1}{\|\r  -
    \r'\|}=\sum_{\ell=0}^\infty\frac{4\pi}{2\ell+1}\frac{\min(r,r')^\ell}{\max(r,r')^{\ell+1}}Y_{\ell m}^{*}(\theta,\varphi)Y_{\ell m}(\theta',\varphi').
\end{equation}
We orient the coordinate
systems such that the unprimed ring lies in the equator. Then
averaging the inverse distance over this ring is equivalent to
averaging over $\varphi$, and in this average all terms except $m=0$
disappear. Thus
\begin{equation}
  \left\langle \frac{1}{\|\r  -       \r'\|}\right\rangle_\phi
  =\sum_{\ell=0}^\infty \frac{\min(r,r')^\ell}{\max(r,r')^{\ell+1}}
    P_\ell(0)P_\ell(\cos\theta').
\end{equation}
Now $\cos\theta'=\sin I\sin\psi$ where $\psi$ is the azimuthal angle
in the primed ring, measured from the line of nodes with the unprimed
ring. Then
\begin{equation}
\frac{1}{2\pi}\int_0^{2\pi}d\psi\,P_\ell(\sin
I\sin\psi)=P_\ell(0)P_\ell(\cos I),
\end{equation}
where
for integer $\ell\ge0$ $P_\ell(0)$ is given by Eq.\ (\ref{e:pnzero}).
With this result the average
becomes
\begin{equation}
  \left\langle \frac{1}{\|\r  -       r'\|}\right\rangle_{\phi,\psi}
  =\sum_{\ell=0}^\infty \frac{\min(r,r')^\ell}{\max(r,r')^{\ell+1}}
    |P_\ell(0)|^2P_\ell(\cos I).
\end{equation}

Now the dependence of the interaction energy (\ref{e:Hannuli}) on the
radial and angular variables separates,
\begin{equation}\label{e:Hintdefinion}
  H^{( i  j )}_{\E} = - G  \sum_{\ell=0}^{\infty} R_\ell \Phi_\ell
\end{equation}
where
\begin{equation}\label{e:R_ell}
R_\ell \equiv R_\ell(a_i ,a_j,e_i, e_j) =
\int_{r_{pi}}^{r_{ai}}\D r
\int_{r_{pj}}^{r_{aj}} \D r'  \sigma_i(r)
\sigma_j(r') rr' \frac{\min(r,r')^\ell}{\max(r,r')^{\ell+1}}
\end{equation}
and
\begin{equation}
\Phi_\ell= 4\pi^2 [P_{\ell}(0)]^2\,P_{\ell}(\cos I)
 \label{e:Phi_ell}
\end{equation}
which vanishes for odd $\ell$.

The radial integral $R_\ell$ is evaluated using Eq.~(\ref{e:sigma}) for the surface density:
\begin{equation}\label{e:Rdef}
R_\ell =\frac{m_im_j}{4\pi^4 a_i a_j} S_\ell
\end{equation}
where
\begin{equation}\label{e:S_ldef}
S_\ell =
\int\limits_{r_{pi}}^{r_{ai}}\D r
\int\limits_{r_{pj}}^{r_{aj}} \D r' \,
\frac{r}{\sqrt{r - r_{pi}}\sqrt{r_{ai}- r}}
\frac{r'}{\sqrt{r' - r_{pj}}\sqrt{r_{aj} - r'}} \frac{\min(r,r')^\ell}{\max(r,r')^{\ell+1}}\,.
\end{equation}
In the following three subsections the calculation of $S_\ell$ is done separately for orbits that are
non-overlapping, identical, and overlapping or embedded in radius.

The quantity $S_\ell$ is related to the dimensionless parameter
$s_\ell$ defined in Eq.~(\ref{e:s_ijl}) by
\begin{equation}\label{e:sijl-def}
 s_{\ell}=\frac{S_{\ell}}{\pi^2 \alpha^{\ell} a_{\In}}\,,
\end{equation}
where $\alpha=a_{\In}/a_{\Out}$, $a_{\In}=\min(a,a')$, and $a_{\Out}=\max(a,a')$.

\subsection{Non-overlapping orbits}

As usual, in this subsection the subscripts ``in'' and ``out''
denote the orbits with the smaller and larger semimajor axis.  If
there is no radial overlap then $r_{a,\In}<r_{p,\Out}$, and we may assume
$r_{\In} = r'$ and $r_{\Out} = r$ throughout the integration domain in
Eq.~(\ref{e:S_ldef}).  Thus the integrals can be evaluated
independently.
\begin{equation}
S_\ell =
\int\limits_{r_{p,\Out}}^{r_{a,\Out}}\D r
\frac{r^{-\ell}}{\sqrt{r - r_{p,\Out}}\sqrt{r_{a \Out} - r}}
\int\limits_{r_{p,\In}}^{r_{a,\In}} \D r'  \,
\frac{{r'}^{\ell+1}}{\sqrt{r' - r_{p,\In}}\sqrt{r_{a,\In} - r'}}\,.
\label{e:radial integrals}
\end{equation}
We can transform the first integral to the same form as the second by
introducing the variable $u=1/r$:
\begin{equation}
\int\limits_{r_{p,\Out}}^{r_{a,\Out}}\D r \frac{r^{-\ell}}{\sqrt{r - r_{p,\Out}}\sqrt{r_{a,\Out} - r}}
=\frac{1}{\sqrt{r_{p,\Out} r_{a,\Out}}}
\int\limits_{u_{a,\Out}}^{u_{p,\Out}}\D u
\frac{u^{\ell-1}}{\sqrt{u - u_{a,\Out}}\sqrt{u_{p,\Out} - u}}\label{e:inverse}
\end{equation}
After this change of variables both integrals in Eq.~(\ref{e:radial
  integrals}) have the same algebraic form with a
  non-negative integer
exponent in the numerator for $\ell>0$:
\begin{align}
\int\limits_{x_{\min}}^{x_{\max}}\D x
\frac{x^{n}}{\sqrt{x - x_{\min}}\sqrt{x_{\max} - x}}
= \pi (x_{\rm max}x_{\rm min})^{n/2}P_n\left(\frac{x_{\rm max}+x_{\rm
      min}}{2\sqrt{x_{\rm max}x_{\rm min}}}\right).\label{e:radiallegendre}
\end{align}
Now we set $x_{\max,\min}=a(1\pm e)$ (for $x=r$) or
$x_{\max,\min}=1/[a(1\mp e)]$ (for $x=u$), and we obtain
\begin{equation}\label{e:S-nonoverlapping}
S_\ell =\pi^2
\frac{a_{\In}^{\ell+1}(1-e_{\In}^2)^{(\ell+1)/2}}{a_{\Out}^\ell(1-e_{\Out}^2)^{\ell/2}}
P_{\ell+1}\left(\chi_{\In}\right) P_{\ell-1}\left(\chi_{\Out}\right)
\ \quad\mbox{where}\quad\ \chi\equiv\frac{1}{\sqrt{1-e^2}}\quad(\ell>0).
\end{equation}
For $\ell=0$ we can directly use Eq.~(\ref{e:radial integrals}), which
can be evaluated using Eq.~(\ref{e:radiallegendre}) with $n=0$ and $1$
to yield $S_0 = \pi^2 a_{\In}$.

\subsection{Identical orbits}\label{s:Overlap-complete}

Next we discuss the special case where $r_{p,\In}=r_{p,\Out}$ and $r_{a,\In}=r_{a,\Out}$, which also admits a
closed-form solution. The technique introduced here may be generalized for the overlapping or embedded cases
as we show in the following subsection.

In this case the integrals over $r<r'$ and $r>r'$ are identical. We calculate the contribution from $r<r'$.
Change integration variables in Eq.~(\ref{e:S_ldef}) $(r,r')\rightarrow(\phi,\phi')$ such that
  $r=a(1+e\cos \phi)$
\begin{equation}\label{e:Sidentical}
S_{\ell} =
2 a \int_{0}^{\pi} \D \phi
\int_{0}^{\phi} \D \phi'
\frac{(1 + e \cos\phi)^{\ell+1}}{(1 + e \cos\phi')^{\ell}}
\end{equation}
In this section we use the following shorthand notation to simplify the expressions
\begin{equation}
 h\equiv \frac{1}{e}\,, \quad
 s\equiv \frac{1 - e}{2e}\,.
\end{equation}

First we evaluate the $\phi'$ integral.
We may eliminate the $\ell$ dependence in the denominator by realizing that it is
the $(\ell-1)^{\rm th}$ complete derivative with respect to $h$,
\begin{equation}
\int_{0}^{\phi} \D \phi'
\frac{1}{(1 + e \cos\phi')^{\ell}}
=
\frac{(-1)^{\ell -1}}{(\ell - 1 )!}h^{\ell}\frac{d^{\ell-1}}{dh^{\ell-1}}
\int_{0}^{\phi}
\frac{\D \phi'}{h + \cos\phi'}\,.
\end{equation}
This integral can be evaluated with a half-angle substitution
\begin{equation}
\int_{0}^{\phi}
\frac{\D \phi'}{h + \cos\phi'}
= \frac{2}{\sqrt{h^2 - 1 }} \arctan \left[ \sqrt{\frac{h-1}{h+1}} \tan \left(\frac{\phi}{2}\right)\right]\,.
\end{equation}
Substitute in Eq.~(\ref{e:Sidentical}) and change to half angles $\phi
\rightarrow \phi/2$, which gives
\begin{equation}\label{e:Sidenticalh}
S_{\ell} =
\frac{2^{\ell+4} a }{h}
\frac{(-1)^{\ell -1}}{(\ell - 1 )!}
\frac{\partial^{\ell-1}}{\partial h^{\ell-1}}
\int_{0}^{\pi/2} \D \phi \;
\frac{(s +  \cos^2\phi)^{\ell+1}}{\sqrt{h^2 - 1 }}
 \arctan \left( \sqrt{\frac{h-1}{h+1}} \tan \phi \right)\,.
\end{equation}

Next, expand $(s+\cos^2\phi)^{\ell +1}$ with the binomial identity
\begin{equation}\label{e:Sidentical-nstep}
\int_{0}^{\pi/2} \D \phi \;
(s +  \cos^2\phi)^{\ell+1}
 \arctan \left( \sqrt{\frac{h-1}{h+1}} \tan \phi \right)
=
\sum_{n=0}^{\ell+1} \binom{\ell+1}{n}s^{\ell + 1 - n}
\int_{0}^{\pi/2} \D \phi \;
\cos^{2 n} \phi
 \, \arctan \left( \sqrt{\frac{h-1}{h+1}} \tan \phi \right)
\end{equation}
where $\binom{\ell+1}{n} =(\ell+1)!/[n!\, (\ell +1 - n)!]$.
Switch variables to $x=\tan\phi$. The integral is then
\begin{equation}
\int_{0}^{\pi/2} \D \phi \;
\cos^{2 n} \phi
 \, \arctan \left( \sqrt{\frac{h-1}{h+1}} \tan \phi \right)
=
\int_{0}^{\infty}
\frac{\D x }{(1 + x^2)^{n+1}}\arctan \left( \sqrt{\frac{h-1}{h+1}}\,x \right)\,.
\end{equation}
The $n+1$ exponent in the denominator may be eliminated by expressing the integrand as the $n^{\rm th}$
derivative as follows:
\begin{equation}
=
\lim_{\gamma\rightarrow 1}
\frac{(-1)^n}{n!}\frac{\partial^{n}}{\partial \gamma^{n}}
\int_{0}^{\infty} \frac{\D x}{\gamma+x^2} \;\arctan \left( \sqrt{\frac{h-1}{h+1}}x \right)
=\lim_{\gamma\rightarrow 1}\frac{(-1)^n}{n!}
\frac{\partial^{n}}{\partial \gamma^{n}}
\left[ \frac{1}{\sqrt{\gamma}} \int_{0}^{\pi/2} \D \theta \;
\arctan \left( \sqrt{\gamma \frac{h-1}{h+1}} \tan \theta \right) \right]
\end{equation}
where in the second step we changed integration variables to $x=\sqrt{\gamma}\tan \theta$.
We can now define
\begin{equation}\label{e:Legendre-chi-integral}
 F(h,\gamma) =
\frac{1}{\sqrt{\gamma}}\int_{0}^{\pi/2} \arctan ( q \tan \theta ) \;\D \theta=
\frac{\chi_{L}(q) - {\rm arctanh}(q) \ln(q)}{\sqrt{\gamma}}\,
\quad {\rm where}~~q\equiv q(h,\gamma) =  \sqrt{\gamma \frac{h-1}{h+1}}\,.
\end{equation}
where we have evaluated the integral using the Legendre-$\chi$ function
\begin{equation}
 \chi_{L}(z) = \sum_{n=0}^{\infty} \frac{z^{2n+1}}{(2n+1)^2}.
\end{equation}
Substituting back in Eq.~(\ref{e:Sidentical-nstep}),
\begin{equation}\label{e:Sidentical-nstep-final}
\int_{0}^{\pi/2} \D \phi \;
(s +  \cos^2\phi)^{\ell+1}
 \arctan \left( \sqrt{\frac{h-1}{h+1}} \tan \phi \right)
=
\lim_{\gamma\rightarrow 1} \sum_{n=0}^{\ell+1} \binom{\ell+1}{n} \frac{(-1)^n}{n!} s^{\ell + 1 - n}
\frac{\partial^n}{\partial\gamma^n}F(h,\gamma).
\end{equation}
The sum may be simplified using the following result, valid for any
function $F$:
\begin{equation}
\sum_{n=0}^{\ell+1} \binom{\ell+1}{n} \frac{(-1)^n}{n!} s^{\ell + 1 - n}
\frac{\partial^n}{\partial\gamma^n}F(h,\gamma)=
\lim_{z\rightarrow 0}\frac{(-1)^{\ell+1}}{(\ell+1)!} \frac{\partial^{\ell+1}}{\partial z^{\ell+1}}
\left[\frac{1}{1+sz} F\left(h, \gamma+\frac{z}{1+sz}\right)\right].
\end{equation}
Now we substitute back into Eq.~(\ref{e:Sidenticalh}) and take the
limit $\gamma\rightarrow 1$.  The result may be simplified using the
substitution $z=2x$ and replacing $h$ with $h+y$ where $y\rightarrow
0$, and then using $2s=h-1$. We get
\begin{align}\label{e:Sidentical-final}
S_{\ell} &=
\lim_{\substack{x\rightarrow 0\\y\rightarrow 0}}
\frac{2 a e }{(\ell-1)!(\ell+1)!}
\frac{\partial^{\ell+1}}{\partial x^{\ell+1}}\frac{\partial^{\ell-1}}{\partial y^{\ell-1}}
G(h,x,y)
\end{align}
where $G(h,x,y)$ is
\begin{equation}\label{e:generatingfunction-def}
G(h,x,y)=
 \frac{4}{[(h-1)x+1](y+h+1)}
 \left[
    \frac{\chi_{L}(Q)}{Q} - \frac{{\rm arctanh}(Q)}{Q} \ln(Q)
 \right]
\quad{\rm where}\quad Q\equiv\sqrt{\left(\frac{(h+1)x+1}{(h-1)x+1}\right)\left(\frac{y+h-1}{y+h+1}\right)}\,.
\end{equation}
$G(h,x,y)$ is a bivariate generating function
of $S_{\ell}$.
This expression may be further manipulated to arrive at a more compact
and symmetric form, using the substitutions $x\to (h_{\In}-h)/(h^2-1)$,
$y\to h_{\Out}-h$:
\begin{equation}\label{e:Sidentical-h}
  S_{\ell} =
\lim_{\substack{h_{\In}\rightarrow h\\h_{\Out}\rightarrow h}}
\frac{8a}{h}\frac{(h^2-1)^{\ell+\frac{3}{2}}}{(\ell-1)!(\ell+1)!}
\frac{\partial^{\ell+1}}{\partial h_{\In}^{\ell+1}}\frac{\partial^{\ell-1}}{\partial h_{\Out}^{\ell-1}}
 \frac{\chi_{L}(Q) - {\rm arctanh}(Q) \ln Q}{\sqrt{(h_{\In}^2-1)(h_{\Out}^2-1)}}
\quad{\rm where}\quad
Q\equiv \sqrt{\frac{(h+1)(h_{\In}-1)(h_{\Out}-1)}{(h-1)(h_{\In}+1)(h_{\Out}+1)}}.
\end{equation}

Using
\begin{equation}
\frac{\D \chi_{L}(q)}{\D q} = \frac{{\rm arctanh\,}q}{q}\,,\quad{\rm and}\quad
\frac{\D({\rm arctanh\,}q)}{\D q} = \frac{1}{2q}\left( \frac{1}{1-q} - \frac{1}{1+q} \right)\,\,,
\end{equation}
the result for $\ell\geq 3$ is
\begin{align}\label{e:Sell-identical-result}
  S_{\ell} = 4ae\left\{ 4 A_{\ell} D_{\ell}
\left[\chi_{\Out}\left(\sqrt{\frac{1-e}{1+e}}\right) + \frac{1}{2}{\rm arcsech}(e)\, {\rm arctanh}(e) \right]
-\left[\left(\frac{1}{\ell} + \frac{1}{\ell+1}\right)\frac{1}{e} + B_{\ell} D_{\ell}\right]{\rm arctanh}(e)
+  C_{\ell} D_{\ell} + E_{\ell}\right\}
\end{align}
where $A_{\ell}$, $B_{\ell}$, $C_{\ell}$, $D_{\ell}$, and $E_{\ell}$ are
\begin{align}
 A_{\ell}&= (-1)^{\ell+1}\sum_{i=0}^{\ell-1} P_{2i} P_{2\ell-2-2i}\,\left(\frac{1-e}{1+e}\right)^{i + \frac{1}{2}}\,,\\
 B_{\ell}&=
(-1)^{\ell}\sum_{i=0}^{\ell-2}\sum_{n=0}^{i}\sum_{m=i}^{\ell-2}
\frac{2\, P_{2n}\, P_{2i-2n} P_{2m-2i} P_{2\ell - 4 - 2m}}{m-n+1}
\left(\frac{1-e}{1+e}\right)^{i + 1}\,,\\
C_{\ell} &=
\sum_{i=0}^{\ell-3}\sum_{j=i}^{\ell-3}\sum_{n=0}^{i}\sum_{m=j}^{\ell-3}
\frac{(-1)^{\ell+i-j}P_{2n}P_{2i-2n}P_{2m-2j}P_{2\ell-6-2m}}{(1+j-i)(\ell-1 -m +n +j -i)}
\left(\frac{1-e}{1+e}\right)^{i + 1}\,,\nonumber\\&\quad
-\sum_{i=0}^{\ell-3}\sum_{j=0}^{i}\sum_{n=0}^{j}\sum_{m=i}^{\ell-3}
\frac{(-1)^{\ell+j-i}P_{2n}P_{2j-2n}P_{2m-2i}P_{2\ell-6-2m}}{(1+i-j)(\ell-1 -m +n + i- j)}
\left(\frac{1-e}{1+e}\right)^{i + 2}\,,\\
D_{\ell} &= \sum_{n=0}^{\ell+1}\binom{\ell + 1}{n}(-1)^n P_{2n}(0)\, \left(\frac{2e}{1-e}\right)^{n-1} \\
E_{\ell} &= \sum_{i=0}^{\ell-2}\sum_{j=0}^{i}\sum_{n=i+1}^{\ell+1}\sum_{m=i+1}^{n}
\binom{i}{j}\binom{\ell+1}{n}
\frac{(-1)^{n+j} P_{2n-2m}P_{2m - 2 - 2i} }{(\ell -1 -j)\,m}
\left[1 - \left(\frac{1-e}{1+e}\right)^{\ell - 1 -j} \right]
 \left(\frac{2e}{1-e}\right)^{n-i-2}
\nonumber\\&\quad
- \sum_{j=1}^{\ell - 1} \binom{\ell-1}{j}\frac{(-1)^j}{j}
\left[1 - \left(\frac{1-e}{1+e}\right)^{j} \right]\left[\frac{1}{2}\left(\frac{1}{\ell} + \frac{1}{\ell+1}\right)
+\left(\frac{\ell+1}{\ell} - \frac{\ell}{(\ell+1)(j+1)} \right) \frac{1-e}{2e}
\right]-\frac{1}{\ell+1}\label{e:Sell-identical-result-end}
\end{align}
where $P_{2n}\equiv P_{2n}(0)$ (see Eq.\ \ref{e:pnzero}).
However Eq.~(\ref{e:Sell-identical-result}) is numerically ill-behaved for
$e>0.5$ and $\ell\geq 35$, since in this case $D_{\ell}>10^{15}$ and $E_{\ell}< -10^{14}$ in a way that
the transcendental functions $A_{\ell}D_{\ell}[\chi_{L}(q)+ \frac{1}{2}{\rm arcsech}(e)\, {\rm arctanh}(e) ]$
cancel out the algebraic terms $C_{\ell}D_{\ell} + E_{\ell}$ to at least 14 significant digits.

Numerically we find that $S_{\ell} \propto \ell^{-1} \ln \ell $ as $\ell\rightarrow\infty$.

\subsection{Overlapping or embedded orbits}\label{s:Overlap-partial}

Finally we consider the most general case, in which the orbits overlap
in radius. The derivation is similar to that of the previous subsection.

We start by changing the integration variables in
Eq.~(\ref{e:S_ldef}),
$(r,r')\rightarrow(\phi_{\In},\phi_{\Out})$ such that
  $r_{\In}=a_{\In}(1+e_{\In}\cos\phi_{\In})$ with a similar definition
  for $\phi_{\Out}$ and $a_{\In}\le a_{\Out}$:
\begin{equation}
S_\ell =
\int_0^{\pi} \D \phi_{\In} \int_0^{\pi} \D \phi_{\Out} \, \frac{
\min(a_{\In}(1 + e_{\In} \cos\phi_{\In}), a_{\Out}(1 + e_{\Out} \cos\phi_{\Out}))^{\ell+1}
}{\max(a_{\In}(1 + e_{\In} \cos\phi_{\In}), a_{\Out}(1 + e_{\Out}
  \cos\phi_{\Out}))^{\ell}}.
\label{e:www}
\end{equation}
We may take a factor $(a_{\In}/a_{\Out})^{\ell}$ outside of the integral
as defined in Eq.~(\ref{e:sijl-def}) to arrive at Eqs.~(\ref{e:Hresult})--(\ref{e:s_ijl}) in the main
text.

The integration domain can be separated into two parts depending on which $a_i(1+e_i\cos\phi_i)$ is larger:
\begin{equation}
S_\ell = S_\ell^{-} + S_\ell^{+}\,,\quad{\rm where}\quad
S_\ell^{+} =
\frac{a_{\In}^{\ell+1}}{a_{\Out}^{\ell}}\hspace{5pt}\int \hspace{-90pt}
\int\limits_{\substack{0\leq \phi_{\In},\phi_{\Out}<\pi\\\hspace{75pt} a_{\In}(1+e_{\In}\cos\phi_{\In})<a_{\Out}(1+e_{\Out}\cos\phi_{\Out})}}
\hspace{-70pt}\D \phi_{\In}\,  \D \phi_{\Out} \;
\frac{(1 + e_{\In} \cos\phi_{\In})^{\ell+1}}{(1 + e_{\Out} \cos\phi_{\Out})^{\ell}},\label{e:S+1}
\end{equation}
and $S_\ell^-$ is obtained similarly, by switching the stellar indices
``$\In$'' $\leftrightarrow$ ``$\Out$'' in $S_\ell^+$.  The quantity $S_\ell^+$ gives the contribution
to the interaction energy from the regions where the orbit with the
larger semimajor axis has larger radius than the orbit with the smaller semimajor axis, and vice versa for $S_\ell^-$. For non-overlapping
orbits $S_{\ell}^-$ vanishes.

We follow the analysis of the previous subsection to convert
$S_{\ell}^+$ to a generating function. To this end we introduce a
similar notation
\begin{equation}
 h_{\Out}\equiv \frac{1}{e_{\Out}}\,, \quad
 s_{\In}\equiv \frac{1 - e_{\In}}{2e_{\In}}\,.
\end{equation}
First we simplify the denominator by differentiating with respect to
$h_{\Out}$,
\begin{equation}
 S_{\ell}^+ = \frac{a_{\In}^{\ell+1}}{a_{\Out}^{\ell}}\frac{(-1)^{\ell-1}}{(\ell-1)!}h_{\Out}^{\ell}\frac{\partial^{\ell-1}}{\partial h_{\Out}^{\ell-1}}
\int_{0}^{\pi}\D\phi_{\In} \;(1+e_{\In}\cos\phi_{\In})^{\ell+1}
\hspace{-100pt}\int\limits_{\substack{0\leq \phi_{\Out}<\pi\\\hspace{80pt} a_{\In}(1+e_{\In}\cos\phi_{\In})<a_{\Out}(1+e_{\Out}\cos\phi_{\Out})}}\hspace{-85pt}
 \frac{\D \phi_{\Out}}{h_{\Out}+\cos \phi_{\Out}}\,.
\end{equation}
Now make the substitutions $y=\sqrt{(h_{\Out}-1)/(h_{\Out}+1)} \tan(\phi_{\Out}/2)$ and $\phi_{\In} \rightarrow \phi_{\In} /2$
\begin{equation}
 S_{\ell}^+ = \frac{(2e_{\In} a_{\In})^{\ell+1}}{a_{\Out}^{\ell}} \frac{(-1)^{\ell-1}}{(\ell-1)!}
h_{\Out}^{\ell}\frac{\partial^{\ell-1}}{\partial h_{\Out}^{\ell-1}}
\int_{0}^{\pi/2}\D\phi_{\In} \;\frac{(s_{\In}+ \cos^2\phi_{\In})^{\ell+1}}{\sqrt{h_{\Out}^2 - 1}}
\int\limits_{\substack{0\leq y<\infty\\
D(y)}}
 \frac{\D y}{1 + y^2}
\,.
\end{equation}
where the domain $D(y)$ is defined such that
\begin{equation}\label{e:D(y)}
\left[(r_{p,\In} - r_{a,\Out})\tan^2\phi_{\In} + (r_{a,\In}-r_{a,\Out}) \right]\frac{h_{\Out}-1}{h_{\Out}+1}
\leq
 \left[(r_{p,\Out} - r_{p,\In})\tan^2 \phi_{\In} +
(r_{p,\Out} - r_{a,\In})\right] y^2\,;
\end{equation}
the ``$\In$'' and ``$\Out$'' indices in $r_p$ and $r_a$ continue to
refer to the orbits with the smaller and larger semimajor axes.  To
carry out the integral we must express the integration bound
explicitly for $y$. We introduce angles where the sign of the
left-hand and right-hand sides changes in Eq.~(\ref{e:D(y)}):
\begin{align}
 \phi_l &= {\rm arctan}\sqrt{\frac{r_{a,\In}-r_{a,\Out}}{r_{a,\Out} - r_{p,\In}}}~~{\rm if }~~
r_{p,\In}< r_{a,\Out}\leq r_{a,\In}\,,\quad
\phi_l=0~~{\rm if }~~ r_{a,\In}\leq r_{a,\Out}\,,
\\
 \phi_r &= {\rm arctan}\sqrt{\frac{r_{a,\In}-r_{p,\Out}}{r_{p,\Out} - r_{p,\In}}}~~{\rm if }~~
r_{p,\In}<r_{p,\Out}\leq r_{a,\In} \,,\quad
\phi_r=0~~{\rm if }~~ r_{a,\In}\leq r_{p,\Out}\,,\quad
\phi_r=\frac{\pi}{2}~~{\rm if }~~ r_{p,\Out}\leq r_{p,\In}\,.
\end{align}
Note that $\phi_l$ and $\phi_r$ are continuous across
$r_{p,\Out}=r_{a,\In}$ and $r_{a,\In}=r_{a,\Out}$. It is easy to show that $\phi_l\leq \phi_r$. We also
define the function
\begin{align}\label{e:Theta}
 \Theta(t) &=
\sqrt{\frac{(r_{p,\In} - r_{a,\Out})t^2 + (r_{a,\In}-r_{a,\Out})}{(r_{p,\Out} - r_{p,\In})t^2 + (r_{p,\Out} - r_{a,\In})}}\,.
\end{align}
With these definitions, the integral over $D(y)$ can be carried
separately over the individual regions
\begin{align}\label{e:Sgeneral-s}
 S_{\ell}^+ &= 4\frac{(2a_{\In}e_{\In})^{\ell+1}}{(a_{\Out} e_{\Out})^{\ell}} \frac{(-1)^{\ell-1}}{(\ell-1)!}
\frac{\partial^{\ell-1}}{\partial h_{\Out}^{\ell-1}}
\bigg\{
\int_{\phi_l}^{\phi_r}\D\phi_{\In} \;\frac{(s_{\In}+ \cos^2\phi_{\In})^{\ell+1}}{\sqrt{h_{\Out}^2 - 1}}
{\rm arctan}\left[
\sqrt{\frac{h_{\Out}-1}{h_{\Out}+1}} \Theta(\tan\phi_{\In})
\right] \nonumber \\
& \qquad + \int_{\phi_r}^{\pi/2}\D\phi_{\In} \;\frac{(s_{\In}+ \cos^2\phi_{\In})^{\ell+1}}{\sqrt{h_{\Out}^2 - 1}}
\frac{\pi}{2}\bigg\}
\end{align}

We may turn this into a generating function by manipulations analogous
to Eqs.~(\ref{e:Sidenticalh})--(\ref{e:Sidentical-final}). We find that
\begin{align}\label{e:Sgeneral-G}
S_{\ell}^+ &=
\lim_{\substack{x\rightarrow 0\\y\rightarrow 0}}
\frac{1 }{(\ell-1)!(\ell+1)!}
\frac{\partial^{\ell+1}}{\partial x^{\ell+1}}\frac{\partial^{\ell-1}}{\partial y^{\ell-1}}
G_+(x,y)
\end{align}
where $G_+(x,y)\equiv G_+(x,y; r_{p,\In},r_{a,\In},r_{p,\Out},r_{a,\Out})$ is a
bivariate generating function of $S_{\ell}^+$ given by
\begin{equation}\label{e:Sgeneral-G2}
G_+(x,y)=
 \frac{4}{\sqrt{(y + r_{a,\Out})(y+r_{p,\Out})}\sqrt{( 1+ r_{p,\In}x )( 1+ r_{a,\In}x )}}
\left\{
\int_{\theta_{l}(x)}^{\theta_{r}(x)}
{\rm arctan}
\left\{Q_{2}(y) \,\Theta\left[ Q_{1}(x) \tan\theta \right] \right\}\D\theta
+
\frac{\pi}{2}\left[\frac{\pi}{2} - \theta_{r}(x)\right]\right\}
\end{equation}
where
\begin{align}
Q_{1}(x)&=\sqrt{\frac{1+r_{a,\In}x}{1+r_{p,\In}x}}\,,
\qquad
Q_{2}(y)=\sqrt{\frac{y+r_{p,\Out}}{y+r_{a,\Out}}}\,,\\
 \theta_{l}(x) &=
\left\{
\begin{array}{ll}
0 & {\rm if~~} r_{a,\In}\leq r_{a,\Out}\,,
\\
{\rm arctan}
\left[
\left.
\sqrt{
  \frac{\textstyle r_{a,\In} - r_{a,\Out}}{\textstyle r_{a,\Out} - r_{p,\In}}
}
\right/
Q_{1}(x)
\right]
&
 {\rm if~~}
r_{p,\In}< r_{a,\Out}\leq r_{a,\In}\,,
\end{array}
\right.
\\
 \theta_{r}(x) &=
\left\{
\begin{array}{ll}
0 & {\rm if~~} r_{a,\In} \leq r_{p,\Out}\,,
\\
{\rm arctan}
\left[
\left.
\sqrt{
  \frac{\textstyle r_{a,\In}-r_{p,\Out}}{\textstyle r_{p,\Out}-r_{p,\In}}
}
\right/
Q_{1}(x)
\right]
&
 {\rm if~~}
r_{p,\In}< r_{p,\Out}\leq r_{a,\In}\,,
\\
\half\pi &
 {\rm if~~}
r_{p,\Out}\leq r_{p,\In}\,.
\end{array}
\right.
\end{align}
The analogous generating function $G_-(x,y)$ for $S_\ell^-$ is
obtained by switching the indices
``$\In$'' $\leftrightarrow$ ``$\Out$'' in $G_+(x,y)$.

Note that the generating function for $S_{\ell}^+$ in
Eq.~(\ref{e:Sgeneral-G2}) is not unique. In particular,
  $c^{\ell+1}d^{\ell-1}G_+(cx,dy)$
  is also a generating function of $S_{\ell}^+$ for
arbitrary constants $c$ and $d$.\footnote{ Another transformation that
  preserves $S_{\ell}$ is the one introduced in
  Eq.~(\ref{e:inverse}), which reverses the roles of the orbits, i.e.
\begin{equation}
 G(x,y;r_{p,\In},r_{a,\In},r_{p,\Out},r_{a,\Out})
\leftrightarrow
\sqrt{\frac{r_{p,\In}r_{a,\In}}{ r_{p,\Out} r_{a,\Out} }} G(x,y;r_{a,\Out}^{-1},r_{p,\Out}^{-1},r_{a,\In}^{-1},r_{p,\In}^{-1})
\end{equation}
is also a generating function of $S_{\ell}$ that satisfies
Eq.~(\ref{e:Sgeneral-G}).  The roles of $h_{\In}$ and $h_{\Out}$ are
reversed in the corresponding Eq.~(\ref{e:Sidentical-h}) for the
transformed orbits.  } We use this property to arrive at a more
compact and symmetric form analogous to Eq.~(\ref{e:Sidentical-h})
\begin{align}
S_{\ell}^{+} =&
\frac{a_{\In}^{\ell+1}}{a_{\Out}^{\ell}}(1-e_{\In}^2)^{\ell+\frac{3}{2}}
\frac{h_{\In}^{\ell+2}h_{\Out}^{\ell}}{(\ell+1)!(\ell-1)!}
\frac{\partial^{\ell+1}}{\partial h_{\In}^{\ell+1}}
\frac{\partial^{\ell-1}}{\partial h_{\Out}^{\ell-1}}\nonumber\\
&\qquad
\frac{4}{\sqrt{(h_{\In}^2-1)(h_{\Out}^{2}-1)}}
 \left\{
 \int_{\theta'_{l}(h_{\In})}^{\theta'_{r}(h_{\In})}
 {\rm arctan}
 \left\{Q(h_{\Out}) \,\Theta\left[ \frac{Q(h_{\In})}{Q(1/e_{\In})} \tan\theta' \right] \right\}\D\theta'
 +
 \frac{\pi}{2}\left(\frac{\pi}{2} - \theta'_{r}(h_{\In})\right)\right\}
\label{e:Sgeneral-h}
\end{align}
where $h_{\In}=1/e_{\In}$ and we have introduced
\begin{equation}
 Q(x)=\sqrt{\frac{x - 1}{x + 1}}\,,
\end{equation}
and
\begin{align}
 \theta'_l(h_{\In}) &=
\left\{
\begin{array}{ll}
0 & {\rm if~~} r_{a,\In}\leq r_{a,\Out}\,,
\\
{\rm arctan}
 \left[
\sqrt{
  \frac{\textstyle r_{a,\In} - r_{a,\Out}}{\textstyle r_{a,\Out} - r_{p,\In}}
}
 \frac{\textstyle Q(1/e_{\In})}{\textstyle Q(h_{\In})}
 \right]
&
 {\rm if~~}
r_{p,\In}\leq r_{a,\Out}\leq r_{a,\In}\,,
\end{array}
\right.
\\
 \theta'_{r}(h_{\In}) &=
\left\{
\begin{array}{ll}
0 & {\rm if~~} r_{a,\In} \leq r_{p,\Out}\,,
\\
{\rm arctan}
\left[
\sqrt{
  \frac{\textstyle r_{a,\In}-r_{p,\Out}}{\textstyle r_{p,\Out}-r_{p,\In}}
}
 \frac{\textstyle Q(1/e_{\In})}{\textstyle Q(h_{\In})}
\right]
&
 {\rm if~~}
r_{p,\In}\leq r_{p,\Out}\leq r_{a,\In}\,,
\\
\half\pi &
 {\rm if~~}
r_{p,\Out}\leq r_{p,\In}\,.
\end{array}
\right.
\end{align}
Note the distinction between $e_{\In}^{-1}$ and $h_{\In}$: while the two are equal, $\partial/\partial h_{\In}$ does not
act on $e_{\In}^{-1}$. Here $Q(1/e_{\In})=\sqrt{r_{p,\In}/r_{a,\In}}$.

Equations~(\ref{e:Sgeneral-s}),
(\ref{e:Sgeneral-G})--(\ref{e:Sgeneral-G2}), and (\ref{e:Sgeneral-h})
are valid for all eccentricities $0<e_i< 1$ ($i=\In$ or $\Out$) in both the
overlapping/embedded and non-overlapping cases.
We may recover
the special cases derived for non-overlapping and identical orbits as follows.  For identical orbits
$a_{\In}=a_{\Out}$, $e_{\In}=e_{\Out}$, so $h_{\In}=h_{\Out}=1/e$, $\theta'_{l}=0$,
$\theta'_{r}=\half\pi$, and $\Theta(\cdot)$ is the identity function
(see Eq.~\ref{e:Theta}), and we recover
Eqs.~(\ref{e:Legendre-chi-integral}) and (\ref{e:Sidentical-h})
given that $S_\ell=S^+_\ell+S^-_\ell=2S^+_\ell$ in this case.
For non-overlapping orbits $\theta'_{l}=\theta'_{r}=0$
and so the integration domain in Eq.~(\ref{e:Sgeneral-h})
is empty. Then the quantity in braces in Eq.~(\ref{e:Sgeneral-h})
is just $\pi^2/4$ and $S_\ell^-$ vanishes.
The evaluation of $S_\ell$ using Eq.~(\ref{e:Sgeneral-h})
reduces to finding the derivatives of $1/\sqrt{h^2-1}$. These generate the
Legendre polynomials,
\begin{equation}
 \frac{x^{\ell+1}}{\ell!}\frac{\partial^{\ell}}{\partial x^{\ell}}\frac{1}{\sqrt{x^2-1}} =
\frac{(-1)^{\ell}}{(1-x^{-2})^{(\ell+1)/2}}P_{\ell}\left(\frac{1}{\sqrt{1-x^{-2}}}\right)
\end{equation}
and we recover Eq.~(\ref{e:S-nonoverlapping}) for non-overlapping orbits.

\subsection{Classification of orbits}\label{s:app:classes}

The generating function (\ref{e:Sgeneral-G})--(\ref{e:Sgeneral-G2}) is
useful to understand the behavior of the interaction energy shown in
Figure~\ref{f:energy-alpha}.  This function generates the functions
$S_{\ell}^+$ and $S_{\ell}^-$ that determine the resonant relaxation
Hamiltonian $H_{\E}$; these are
piecewise smooth functions of the periapsis and apoapsis distances
$\{r_{pi},r_{ai},r_{pj},r_{aj}\}$ and have discontinuous derivatives
for special values of $\{r_{pi},r_{ai},r_{pj},r_{aj}\}$.  We may
classify the orbits accordingly as follows. For simplicity we
  assume that the labels are chosen so that orbit $i$ is the
  ``smaller'' orbit; here ``smaller'' means the smaller periapsis,
  $r_{pi}\le r_{pj}$, or if the periapsides are equal the smaller
  apoapsis.

There are 14 topologically different
radial configurations where the interaction energy behaves
differently, with distinct large--$\ell$ asymptotics.  These are
defined by the relative radial locations of the singularities in the
radial density function $\sigma(r)$ in Eq.~(\ref{e:sigma}),
i.e., $r_{pi}$, $r_{ai}$, $r_{pj}$, and $r_{aj}$.  Three of the 14
configurations have a nonzero measure, i.e.
\begin{enumerate}[leftmargin=0.5cm]
 \item $r_{pi}< r_{ai} < r_{pj} < r_{aj}$:  non-overlapping
   orbits, $r_{i}<r_{j}$ everywhere, with $S_\ell^-=0$ and $S_{\ell}=S_{\ell}^+$,\label{i:nonoverlap1}
 \item $r_{pi}< r_{pj} < r_{ai} < r_{aj}$: overlapping orbits,\label{i:overlap-alter1}
  \item $r_{pi}< r_{pj} < r_{aj} < r_{ai}$: embedded orbits
   with $r_{j}\subset r_{i}$,\label{i:overlap-embed2}
\end{enumerate}
There are 11 pathological configurations of zero measure when at least two of
$\{r_{pi},r_{ai},r_{pj},r_{aj}\}$ coincide---six configurations where exactly
two coincide, two configurations where two distinct pairs coincide (i.e.,
$r_{pi}=r_{pj} < r_{ai}=r_{aj}$, $r_{pi}=r_{ai} < r_{pj}=r_{aj}$), two configurations where three coincide,
and one configuration where all four coincide. Six of the 11 pathological
configurations involve circular orbits.  In particular, 4 have one circular
and one eccentric orbit, 1 has two distinct circular orbits, and 1 has
two circular orbits with the same radius.

The configurations \ref{i:nonoverlap1}--\ref{i:overlap-embed2}
with non-zero measure are the most important.  The behavior of
$S_{\ell}$ is different in these three regions as shown in
Figure~\ref{f:energy-alpha} in the main text.  As a function of the
semimajor axis ratio $\alpha<1$, $S_{\ell}$ has a plateau for
overlapping/embedded orbits and local maxima at the edges of the overlapping/embedded
regions where two of the radial turning points coincide.  Once the
orbits are non-overlapping, $S_{\ell}$ decays quickly as $\alpha$
decreases, i.e., $S_{\ell}\propto \alpha^{\ell}$.  The figure shows
that $S_{\ell}$ varies continuously as a function of $\alpha$, but at
the transition between overlapping and non-overlapping orbits its
first derivatives with respect to $\alpha$ are (approximately)
discontinuous, especially for large $\ell$.

This classification scheme does not distinguish cases where the
semimajor axes coincide ($\alpha=1$); however the interaction energy is
typically a smooth function of $a_i/a_j$ across $a_{i}=a_{j}$ for
eccentric orbits.

\subsection{Convergence}
\label{app:convergence}

How many terms of the infinite sum must one account for to accurately calculate the interaction Hamiltonian?
We use the following asymptotic properties of Legendre polynomials:
\begin{align}\label{e:Pasymptotic}
 P_{\ell}(\cos \theta)
&=\left(\frac{\theta}{\sin\theta}\right)^{1/2}J_0[(\ell+\half)\theta][1+
\mathcal{O}(\ell^{-1})], \quad 0\le\theta\le\half\pi\\
P_{\ell}\left(\frac{1}{\sqrt{1-e^2}}\right) &=
\left(\frac{\xi}{\sinh\xi}\right)^{1/2}I_0\big[(\ell+\half)\xi\big][1+\mathcal{O}(\ell^{-1})],\quad
\xi\equiv \tanh^{-1}e.
\label{e:Pasymptotic2}
\end{align}
Here $J_0$ and $Y_0$ are Bessel functions and $I_0$ is a modified Bessel
functions. In evaluating these expressions the following properties of
Bessel functions are useful:
\begin{align}
J_0(x)&=\left(\frac{2}{\pi x}\right)^{1/2}\big[\cos(x-\ffrac{1}{4}\pi)+\mathcal{O}(x^{-1})\big] \label{eq:bessone} \\
I_0(x)&=\frac{e^x}{\sqrt{2\pi
    x}}\big[1+\mathcal{O}(x^{-1})\big]. \label{eq:besstwo}
\end{align}
From these results, or from Eq.\ (\ref{e:pnzero}) and Stirling's
formula, it is straightforward to show that $P_{2\ell}(0)^2\rightarrow 1/(\pi\ell)$  for
large $\ell$;
Substituting in Eqs.~(\ref{e:xxxyyy}) and (\ref{e:Jijell}) for
non-overlapping or marginally overlapping orbits, we find that the coupling coefficients in
the Hamiltonian asymptotically satisfy
\begin{align}\label{e:Jasymptotic-nonoverlap}
 \J_{\ell}^{\rm asymp} &=
\frac{G m_{\In} m_{\Out}}{\pi^2 \ell^2} \frac{r_{a,\In}^\ell}{r_{p,\Out}^{\ell +1}} \frac{[(1+e_{\In})(1-e_{\Out})]^{3/2}}{(e_{\In}e_{\Out})^{1/2}}[1
+ \mathcal{O}(\ell^{-1})]~~{\rm if}~~ \ell \,\gtrsim \,\max\left(\frac{5}{e_{\In}},\frac{5}{e_{\Out}}\right)\,,\;
r_{p,\Out} \geq r_{a,\In}\,,\;
\text{and $\ell\in{\rm even}$}.
\end{align}
If one or both orbits are circular, the
asymptotic decay of $\J_{\ell}$ is slower by factors of $\ell^{1/2}$
 and $\ell$, respectively.
Note that $\J_{\ell}\propto \ell^{-2}$ for marginally overlapping orbits where $r_{p,\Out}=r_{a,\In}$.

For overlapping or embedded orbits, we may derive the asymptotic form of the
coupling coefficients using the stationary phase
approximation. For large $\ell$ the double integral in Eq.~(\ref{e:S+1}) is dominated by
the region where $a_{\In} (1+e_{\In} \cos\phi_{\In}) \approx a_{\Out}
( 1 + e_{\Out} \cos \phi_{\Out})$.  We define $\phi_*\equiv
\phi_*(\phi_{\Out})$ to satisfy $a_{\In} (1+e_{\In} \cos\phi_{*}) =
a_{\Out} ( 1 + e_{\Out} \cos \phi_{\Out})$, and replace the
integration variable $\phi_{\In}$ with $\phi_* + \Delta$.  After
substituting in Eq.~(\ref{e:S+1}) and expanding $\cos(\phi_* +
\Delta)$ to first order in $\Delta$ we get
\begin{equation}\label{e:S+approx1}
S_{\ell}^+ = a_{\Out} \int_{\phi_{l}}^{\phi_{r}} \D\phi_{\Out}
\int_0^{\Delta_{\max}}\D \Delta\;
(1+e_{\Out}\cos\phi_{\Out})
\left[ 1 - \frac{a_{\In} e_{\In} \sin \phi_*(\phi_{\Out})}{a_{\Out}(1+e_{\Out}\cos\phi_{\Out})}\Delta \right]^{\ell+1}\,,
\end{equation}
where
\begin{equation}
\phi_l = \arccos\left( \frac{r_{a,\In} - a_{\Out}}{a_{\Out} e_{\Out}} \right)\quad{\rm and}\quad
\phi_r = \arccos\left( \frac{r_{p,\In} - a_{\Out}}{a_{\Out} e_{\Out}} \right)
\end{equation}
if both are real, and $\phi_l =0$ and/or $\phi_r = \pi$ otherwise.
For large $\ell$, the integrand decays exponentially as a function of
$\Delta$, so we can extend the integration domain to
$0\leq\Delta<\infty$.  Approximate the bracket in
Eq.~(\ref{e:S+approx1}) using $\lim_{n\to\infty} (1+x/n)^n = e^x$,
carry out the $\Delta$ integral, and change the integration variable
to simplify the result:
\begin{equation}\label{e:S+approx2}
S_{\ell}^+ \approx \frac{1}{\ell}\frac{a_{\Out}^2}{a_{\In}}
\int_{\phi_{l}}^{\phi_{r}} \D\phi_{\Out}
\frac{(1+e_{\Out}\cos\phi_{\Out})^2}{e_{\In} \sin\phi_*(\phi_{\Out})} =
\frac{1}{\ell} \int_{\max(r_{p,\In},r_{p,\Out})}^{\min(r_{a,\In},r_{a,\Out})}
\frac{r^2\; \D r}{\sqrt{(r-r_{p,\In})(r-r_{p,\Out})(r_{a,\In}-r)(r_{a,\Out}-r)}}\,.
\end{equation}
Note that the integral in Eq.~(\ref{e:S+approx2}) is independent of $\ell$.
It can be evaluated in a closed form using a M\"obius transform\footnote{
\url{http://math.stackexchange.com/questions/669301/closed-form-integral-int-bc-fracx2-sqrtx-ax-bc-xd-x-dx}}
\citep{ByrdFriedman}:
\begin{align}\label{e:I2}
  I^{(2)}(a,b,c,d) = \int_{b}^{c}
\frac{r^2\; \D r}{\sqrt{(r-a)(r-b)(c-r)(d-r)}}
= (c-b)\sqrt{\frac{k^2-\lambda^2}{1-\lambda^2}}
\left\{
\begin{array}{ll}
 & K(k) \left(\frac{c+b}{c-b}\right)^2\\
+ & \frac{2}{\lambda}\left[K(k) - (1-\lambda^2)\Pi(\lambda^2,k)\right] \left(\frac{c+b}{c-b}\right)\\
+ & K(k) + \frac{1-\lambda^2}{\lambda^2-k^2}\left[ E(k) - (1-k^2)\Pi(\lambda^2,k) \right]
\end{array}
\right\}
\end{align}
for $a<b<c<d$, where
\begin{equation}
\lambda = \frac{\Lambda}{1+\sqrt{1-\Lambda^2}}, \quad \Lambda =
\frac{\tilde{a}+\tilde{d}}{1+\tilde{a}\tilde{d}}, \quad
\tilde{a} = \frac{2a - (b+c)}{c-b}, \quad \tilde{d} =
\frac{2d-(b+c)}{c-b}, \quad k = \frac{1 - \lambda \tilde{d}}{\tilde{d}
  - \lambda},
\end{equation}
and $K(k)$, $E(k)$, and $\Pi(k)$ are complete elliptic integrals\footnote{
We use the definitions
\begin{equation}
K(k) = \int_0^1 \frac{dz}{\sqrt{(1-z^2)(1-k^2z^2)}}\,,\quad
E(k)  = \int_0^1 \sqrt{\frac{1-k^2 z^2}{1-z^2}} dz\,,\quad{\rm and}\quad
\Pi(\eta,k) = \int_0^1 \frac{dz}{(1 - \eta z^2)\sqrt{(1-z^2)(1-k^2 z^2)}}\,.
\end{equation}
}.
Similarly, it may be shown that
$S_{\ell}^-$ and $S_{\ell}^+$ are asymptotically equal for overlapping or embedded orbits with
distinct periapsides and apoapsides.
After substituting in Eqs.~(\ref{e:Hintdefinion}) and (\ref{e:HRR})--(\ref{e:Jijell}) we arrive at
the asymptotic form for overlapping or embedded orbits
\begin{equation}\label{e:Jasymptotic-overlap}
\J_{\ell}^{\rm asymp} =
\frac{4}{\pi^3 \ell^2}\frac{G m_{\In} m_{\Out}}{a_{\In} a_{\Out}}
I^{(2)}(r_{p<},r_{p>},r_{a<},r_{a>})\quad{\rm if}\quad
r_{p>} < r_{a<}\,,\; r_{p<}\neq r_{p>}\,,\;r_{a<}\neq r_{a>}\,,\,  \ell\;\gtrsim\;\frac{2r_{p>}}{r_{a<}-r_{p>}}
\text{and $\ell\in{\rm even}$},
\end{equation}
where $r_{p<} = \min(r_{p,\In}, r_{p,\Out})$, $r_{p>} = \max(r_{p,\In}, r_{p,\Out})$,
and similarly for $r_{a<}$ and $r_{a>}$.

Using a combination of analytic arguments and numerical experiments,
we find that the terms in the sum over $\ell$ comprising the
Hamiltonian (Eq.~\ref{e:HRR}) decrease asymptotically for the
different configurations defined in Appendix~\ref{s:app:classes} as
follows\footnote{In all of these equations $\cos \ell I$ is shifted
  by a phase of order $-\pi/4$ not shown for simplicity, see
  Eqs.~(\ref{e:Pasymptotic}) and (\ref{eq:bessone}).}.

\begin{figure*}
\centering
\mbox{\includegraphics{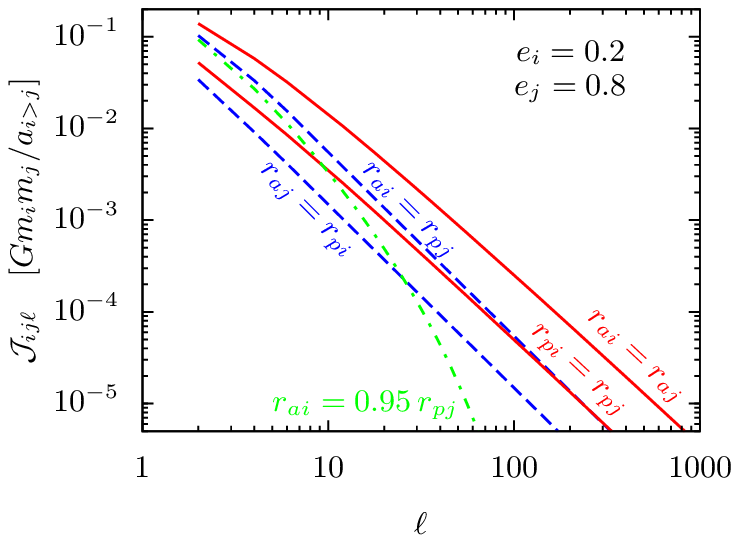}
\includegraphics{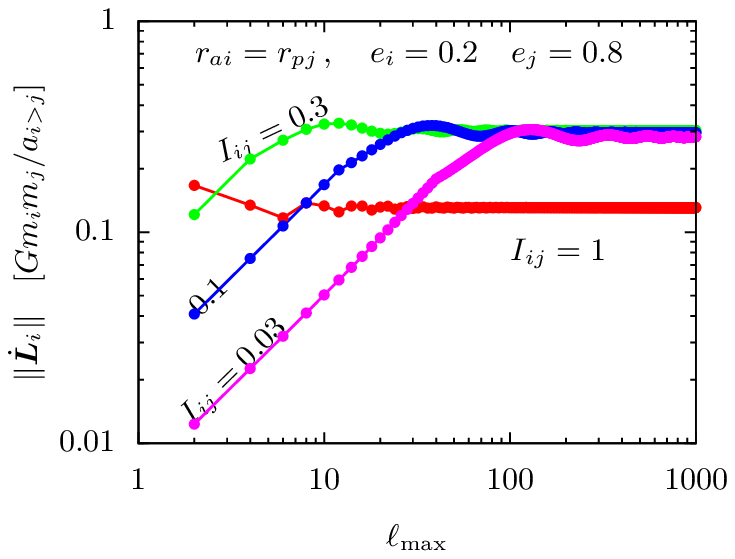}
}
\caption{\label{f:convergence} {\it Left:} Asymptotic behavior of the
  coupling coefficients of the Hamiltonian as a function of the (even)
  multipole order $\ell$, for orbit pairs with $e_i=0.2$ and
  $e_j=0.8$.  Red solid curves show marginally embedded orbits,
  $r_{pi}=r_{pj}$ and $r_{ai}=r_{aj}$ respectively, which scale
  asymptotically as $\ell^{-2} \ln \ell$.  Blue dashed curves show
  marginally overlapping orbits, $r_{pi}=r_{aj}$ and $r_{pj}=r_{ai}$
  respectively, which scale asymptotically as $\ell^{-2}$.  The green
  dash-dotted curve shows a non-overlapping orbit,
  $r_{aj}=0.95\,r_{pi}$, for which the coupling coefficient declines
  exponentially at high $\ell$.  The coupling coefficients are
  continuous functions of the orbital parameters, so the coefficients
  of all overlapping orbits with these eccentricities lie in between
  the red and blue curves shown, and all non-overlapping orbits lie
  below the blue curves.  {\it Right:}  The torque on star $i$ due to
  star $j$ as a function of $\ell_{\max}$, for different inclinations
  as marked in radians. The orbits are marginally overlapping,
  $r_{ai}=r_{pj}$. As the inclination tends to zero, an accurate evaluation
  of the torque requires more and more $\ell$ multipoles.
    }
\end{figure*}

\begin{itemize}[leftmargin=0.5cm,itemsep=1ex]

\item $\ell^{-2.5}\alpha^{\ell}\cos(\ell I)/\sqrt{\sin I}$ for non-coplanar,  non-overlapping
or marginally overlapping, eccentric orbits,
where $\alpha\equiv r_{a,\In}/r_{p,\Out}<1$;

 \item $\ell^{-2}\alpha^{\ell}$ for coplanar, non-overlapping or marginally overlapping,
   eccentric orbits;

 \item $\ell^{-2}\alpha^{\ell}\cos(\ell I)/\sqrt{\sin I}$ for
   non-coplanar,  non-overlapping orbits, one circular and one
   eccentric;

 \item $\ell^{-1.5}\alpha^{\ell}$ for coplanar,
   non-overlapping orbits, one circular and one eccentric;

 \item $\ell^{-1.5}\alpha^{\ell}\cos(\ell I)/\sqrt{\sin I}$ for
   non-coplanar circular orbits with different radii;

 \item $\ell^{-1}\alpha^{\ell}$ for coplanar circular orbits with
   different radii;

 \item $\ell^{-2.5}\cos(\ell I)/\sqrt{\sin I}$ for non-coplanar
   overlapping or embedded orbits;

 \item $\ell^{-2}$ for coplanar overlapping or embedded orbits;

 \item $\ell^{-2.5}\ln \ell/\sqrt{\sin I}$ for non-coplanar embedded
   orbits where the periapsides or the apoapsides coincide
   ($r_{p,\In}=r_{p,\Out}$ or $r_{a,\In}=r_{a,\Out}$);

 \item $\ell^{-2}\ln \ell$ for coplanar embedded orbits where the
   periapsides or the apoapsides coincide;

 \item $\ell^{-2}\cos(\ell I)/\sqrt{\sin I}$ for non-coplanar orbits,
   one circular and one eccentric, with the same peri- or apoapsides
   ($r_{a,\In}=r_{p,\Out}=r_{a,\Out}$);

 \item $\ell^{-1.5}$ for coplanar orbits, one circular and one
   eccentric, with the same peri- or apoapsides
   ($r_{p,\In}=r_{a,\In}=r_{p,\Out}$ or
   $r_{a,\In}=r_{p,\Out}=r_{a,\Out}$);

 \item $\ell^{-1.5}\cos(\ell I)/\sqrt{\sin I}$ for non-coplanar
   circular orbits with the same radii
   ($r_{p,\In}=r_{a,\In}=r_{p,\Out}=r_{a,\Out}$);

 \item $\ell^{-1}$ for coplanar circular orbits with the same radii
   ($r_{p,\In}=r_{a,\In}=r_{p,\Out}=r_{a,\Out}$).

\end{itemize}
The interaction energy sum in Eq.~(\ref{e:Hamiltonian-result})
converges for all but the last of these cases, in which the
interaction energy has a logarithmic singularity in
$a_{\Out}-a_{\In}$.  Figure~\ref{f:convergence} shows examples of
$\J_{\ell}$ for orbits with eccentricities $0.2$ and 0.8. The
asymptotic relations for non-overlapping or marginally overlapping orbits
(Eq.~\ref{e:Jasymptotic-nonoverlap}) approximate $\J_{\ell}$ to within $50\%$ already at
$\ell=2$.

The rate of convergence for an asymptotic scaling $\ell^{-k}$ is
related to the Riemann $\zeta$ function of order $k$.  The absolute
error when neglecting $\ell\geq \ell_0$ is then typically proportional
to
\begin{equation}
\sum_{\ell=\ell_0}^{\infty} \frac{1}{\ell^{k}} =
 \zeta\left(k,\ell_0\right)\,.
\end{equation}
For all overlapping or embedded orbits other than a set of measure
zero, we have $k=2$ for coplanar orbits and $k=\frac{5}{2}$ for
non-coplanar orbits, so the relative error from neglecting $\ell_0\geq
10$ is of order $\zeta(2,10)/\zeta(2)=0.064$ for coplanar orbits and
$\zeta(\frac{5}{2},10)/\zeta(\frac{5}{2})=0.017$ for non-coplanar
orbits.  Thus, the error in calculating the Hamiltonian should be only of order a few percent if we
account for at least the first four non-zero multipoles in the
interaction.  Similarly, multipoles up to and including $\ell=12$ and
60 must be accounted for in order to reach $1\%$ accuracy for the
non-coplanar and coplanar cases, respectively, and $\ell=60$ and $600$
for $0.1\%$ accuracy.  The convergence rate is exponentially faster
for non-overlapping orbits; for example, if $r_{a,\In}/r_{p,\Out} \le
0.3$ then by including all multipoles up to $\ell=10$ we expect to
achieve an accuracy of $10^{-7}$--$10^{-8}$.

The equations of motion converge more slowly. In Eq.~(\ref{e:EOM2}) we found that
\begin{align}\label{e:EOM2app}
\dot{\L }_i =   \bm{\Omega}_i \times \L _i\,,\quad{\rm where}\quad
\bm{\Omega}_i =   -\sum_{j\ell} \frac{\J_{ij\ell}}{L_i L_j}
P'_{\ell}\big(\Ln_i\cdot \Ln_j\big)\, \L _j \,.
\end{align}
From Eqs.~(\ref{e:Pasymptotic}) and (\ref{eq:bessone}) we get
\begin{align}\label{e:P'cosI1}
P'_{\ell}(\cos\theta)
&=\frac{\ell\,\theta^{1/2}}{\sin^{3/2}\theta}\big\{J_1[(\ell-\half)\theta]+\mathcal{O}(\ell^{-1})\big\} \\
&=\sqrt{\frac{2\, \ell}{\pi \sin^3 \theta}}\big\{\cos[(\ell + \half)\theta - \ffrac{3}{4}\pi]
+ \mathcal{O}(\ell^{-1})\big\} \quad{\rm if}\quad \ell \,\gtrsim\, \frac{2}{\theta}.
\label{e:P'cosI2}
\end{align}
This shows that non-coplanar orbits precess around their total angular
momentum vector, with an angular velocity that is convergent if
$\J_{ij\ell} < C_{ij} \ell^{-1.5}$ for large $\ell$ for some $C_{ij}$
constant.  This condition is generally met by all non-overlapping
orbits and also by overlapping or embedded eccentric orbits.  However,
for nearly coplanar overlapping or embedded eccentric orbits, the sum over the
multipoles converges more slowly.  The right panel of Figure
\ref{f:convergence} shows the convergence of the precession rate by
truncating the torque sum at different $\ell_{\max}$ for different
inclinations, when the orbits have $e_i=0.2$ and $e_j=0.8$ and
$r_{ai}=r_{pj}$.
For overlapping or embedded orbits truncating the sum
at some $\ell_{\max}$ leads to an accurate evaluation of the torque
unless the orbits are nearly parallel or antiparallel, with mutual
inclination $I<\half\pi/\ell_{\max}$ or $I>\pi -
\half\pi/\ell_{\max}$. For non-overlapping orbits with
$r_{a,\In}/r_{p,\Out}<0.3$, $\ell_{\max}=10$ is sufficient for a
tolerance of $10^{-6}$ at arbitrary inclinations.

\subsection{Extrapolating to $\ell\rightarrow \infty$}\label{s:asymptotics}

Neglecting the contribution of terms with $\ell>\ell_{\max}$ in the
equations of motion is equivalent to an effective gravitational
softening.  Alternatively, the asymptotic relations we have derived
may be used to extrapolate the contribution of terms in the equations
of motion with $\ell \le \ell_{\max}$ to $\ell\to\infty$. We start by
rewriting the second of Eqs.~(\ref{e:EOM2app}) as
\begin{align}\label{e:Omegaasymptotics2}
 \bm{\Omega}_i =\bm{\Omega}^{\rm asymp}_i + (\bm{\Omega}_i-\bm{\Omega}^{\rm asymp}_i)
 =\bm{\Omega}^{\rm asymp}_i - \sum_{j\ell} \frac{\tilde{\J}_{ij\ell}}{L_iL_j} P'_{\ell}(\Ln_i\cdot\Ln_j)\L_j
\end{align}
where
\begin{align}
 \tilde{\J}_{ij\ell} &= \J_{ij\ell} - \frac{\ell^2\J_{ij\ell}^{\rm asymp}}{(\ell+1)(\ell+2)}\\
 \bm{\Omega}^{\rm asymp}_i &= -\sum_{j\ell}\frac{\ell^2\J_{ij\ell}^{\rm asymp}}{(\ell+1)(\ell+2)L_iL_j} P'_{\ell}(\Ln_i\cdot\Ln_j)\L_j\,
\label{e:Omegaasymptotics0}
 \end{align}
 and $\J_{ij\ell}^{\rm asymp}$ is defined in Eq.\
 (\ref{e:Jasymptotic-overlap});
note that the numerator $\ell^2\J_{ij\ell}^{\rm asymp}$ is
 independent of $\ell$ for overlapping or embedded orbits (Eq.\
 \ref{e:Jasymptotic-overlap}) or proportional to $\alpha^\ell$ for
 non-overlapping orbits (Eq.~\ref{e:Jasymptotic-nonoverlap}).  We now use
 the generating function of the Legendre polynomial $(1-2\alpha
 z+\alpha^2)^{-1/2}=\sum_{\ell=0}^{\infty} \alpha^\ell P_{\ell}(z)$.
 Integrating this expression twice with respect to $\alpha$ and taking
 the even part in $z$ gives the identity
\begin{equation}\label{e:asymptotic-Legendresum}
\sum_{\ell>0,\rm even} \frac{\alpha^{\ell}}{(\ell+1)(\ell+2)}P_{\ell}(z)
= \frac{g(\alpha,z) + g(\alpha,-z)}{2\alpha^2}\,,
\end{equation}
where we have used the fact that $P_\ell(z)$ is an even function of
$z$ if $\ell$ is even, and odd if $\ell$ is odd. We have introduced the function
\begin{equation}\label{e:asymptotic-Legendresum2}
 g(\alpha,z) = 1-\frac{\alpha^2}{2}-\sqrt{1+\alpha^2-2\alpha z}+(\alpha-z)\ln\left(\frac{\alpha-z
 + \sqrt{1 + \alpha^2 - 2 \alpha z}}{1-z}\right)\,.
\end{equation}
We differentiate Eq.~(\ref{e:asymptotic-Legendresum}) with respect to
$z$ and then substitute in Eq.~(\ref{e:Omegaasymptotics0}) with
$z=\Ln_i\cdot \Ln_j$. Next we replace $\J^{\rm asymp}_{ij\ell}$ with
the expressions from Eqs.~(\ref{e:Jasymptotic-nonoverlap}) and
(\ref{e:Jasymptotic-overlap}) for non-overlapping and overlapping/embedded orbits. We obtain
\begin{align}\label{e:Omegaasymptotics1}
 \bm{\Omega}_{i}^{\rm asymp}
 &= -\sum_{\substack{j \in\rm overlapping \\ \rm /embedded}}
 \frac{8}{\pi^2 P_i}\frac{m_j}{M_{\bullet}}\frac{I^{(2)}(r_{p<},r_{p>},r_{a<},r_{a>})}{a_j(1-e_{i}^2)^{1/2}}
 g_2(1,\Ln_i\cdot \Ln_j)\Ln_j
 \nonumber\\&\quad-\!\!\!
\sum_{j\in\rm non-overlapping}
\frac{2}{\pi P_{i}}\frac{m_j}{M_{\bullet}}\frac{[(1+e_{\In})(1-e_{\Out})]^{3/2}}{[e_{i}e_{j}(1-e_{i}^2)]^{1/2}}
\frac{a_i r_{p,\Out}}{r_{a,\In}^2}\,g_2\hspace{-2pt}\left(\frac{r_{a,\In}}{r_{p,\Out}},\Ln_i\cdot \Ln_j\right)
\Ln_j
\end{align}
where $g_2(\alpha,z)\equiv
\frac12\frac{d}{dz}[g(\alpha,z)+g(\alpha,-z)]$ is a closed-form
combination of
elementary analytic functions\footnote{For overlapping or embedded
  orbits $\alpha=1$ and
\begin{align}
 g_2(1,z) &= \frac14\left(\sqrt{\frac{2}{1-z}} - \sqrt{\frac{2}{1+z}}
 + \frac{1}{1+\sqrt{(1-z)/2}} - \frac{1}{1+\sqrt{(1+z)/2}}\right)
 +\frac12 \ln\left[\left( \frac{1+ \sqrt{(1+z)/2}}{1+ \sqrt{(1-z)/2}}\right)\sqrt\frac{1-z}{1+z} \right]\nonumber\\
 &=
 \frac14\left(\frac{1}{s} + \frac{1}{1+s} + 2 \ln\frac{s}{1+s}\right)
 -\frac14\left(\frac{1}{c} + \frac{1}{1+c} + 2 \ln\frac{c}{1+c}\right)\,,
 \end{align}
 where in the second line $s=\sin (I/2)$, $c=\cos (I/2)$, and $z=\cos
 I$.  In particular for $0<I\ll 1$, $s\approx I/2$, $c\approx 1$, and
 so $g_2(1,\cos I)\approx1/(2I)$.  Similar scaling relations apply if
 $I\approx \pi$.  }.  Since $\tilde{\J}_{ij\ell}$ decays to zero as
$\ell\to\infty$ much more quickly than ${\J}_{ij\ell}$, using
Eq.~(\ref{e:Omegaasymptotics2}) yields much more accurate results than
(\ref{e:EOM2app}) if the sum over $\ell$ in the second term is
truncated at $\ell_{\max}\gg 1$.

As shown in Section~\ref{s:pairwise}, the dynamical interaction of
each $i$--$j$ pair is a precession of $\bm{K}_{ij}=(\L_i-\L_j)/2$
around their total angular-momentum vector $2\bm{J}_{ij}=\L_i+\L_j$
with angular velocity $\bm{\Omega}_{ij}$, while $\Ln_i\cdot\Ln_j$,
$\|\bm{K}_{ij}\|$, and $\bm{J}_{ij}$ are fixed. The angular
  velocity $\bm{\Omega}_{ij}$ is obtained from $\bm{\Omega}_i$ by
  replacing $\L_j$ by $2\bm{J}_{ij}$ (Eqs.\ \ref{e:EOM2} and
  \ref{e:omjk}).  Summing over $\ell$ gives asymptotically
\begin{equation}\label{e:Omegaasymptotics3}
 \bm{\Omega}_{ij}^{\rm asymp} \!=\!
\left\{\!\!\!
\begin{array}{ll} \displaystyle
  -\frac{4}{\pi^3 M_{\bullet}}\frac{I^{(2)}(r_{p<},r_{p>},r_{a<},r_{a>})}{a_{\In}^{3/2} a_{\Out}^{3/2}
    [(1-e_{\In}^2)(1-e_{\Out}^2)]^{1/2}}
  \,g_2(1,\Ln_i\cdot \Ln_j)\,(\L_i + \L_j) &\!\!\text{overlapping,~} r_{a,\In} > r_{p,\Out}\,,
  \nonumber\\[1.2em] \displaystyle
  -\frac{1}{\pi^2 M_{\bullet}} \frac{r_{p,\Out}}{r_{a,\In}^2}
  \frac{[(1+e_{\In})(1-e_{\Out})]^{3/2}}{[a_{\In}a_{\Out}(1-e_{\In}^2)(1-e_{\Out}^2)e_{\In}e_{\Out}]^{1/2}}
  \,g_2\hspace{-2pt}\left(\frac{r_{a,\In}}{r_{p,\Out}},\Ln_i\cdot \Ln_j\right) (\L_i + \L_j)
  & \!\!\text{non-overlapping,~} r_{a,\In} \leq r_{p,\Out}.
\end{array}
\right.
\end{equation}

In practice, the high $\ell$ terms contribute most significantly if
the mutual inclination $I_{ij}$ is small or near $\pi$ and if the
orbits are overlapping or embedded\footnote{Note that
  $P'_{\ell}(1)=\ell(\ell+1)/2$ and $\J_{ij\ell}^{\rm asymp}\propto
  \ell^{-2}$ for overlapping or embedded orbits.}, since then $g_2(1,\cos
I)\simeq 1/(2I)$ or $1/[2(\pi-I)]$.  For nearly parallel overlapping or
embedded orbits, the instantaneous precession of $i$ due to $j$ in
Eq.~(\ref{e:Omegaasymptotics1}) simplifies to
\begin{equation}\label{e:Omegaasymptotics4}
 \bm{\Omega}_{i}^{\rm asymp} \approx -  \frac{4}{ \pi^2 P_i} \frac{m_j}{M_{\bullet}}
 \frac{I^{(2)}(r_{p<},r_{p>},r_{a<},r_{a>})}{a_{j}(1-e_i^2)^{1/2}}
 \frac{\Ln_j}{I_{ij}}\quad{\text{if $I_{ij}\ll 1$ and if $i$ and $j$ are overlapping or embedded}}\,.
\end{equation}

For nearly coplanar orbits orthogonal to the $z$-axis, we may
approximate $\hat{L}_{iz}\approx 1$ and so $ \Ln_{i} \approx
\hat{\bm{e}}_z + \hat{L}_{xi}\hat{\bm{e}}_x + \hat{L}_{yi}
\hat{\bm{e}}_y $.  The angular-momentum vectors are approximately
confined to a plane, and the mutual inclination is approximately the
Euclidean distance between the angular-momentum vectors in the plane,
$I_{ij}= \|\Ln_{i}-\Ln_{j}\| $. Thus,
\begin{equation}\label{e:thinoverlapping}
 \dot{\Ln}_i \approx   -\sum_j
 \frac{4}{\pi^2 P_i} \frac{m_j}{M_{\bullet}} \frac{I^{(2)}(r_{p<},r_{p>},r_{a<},r_{a>})}{a_j(1-e_i^2)^{1/2}}
 \frac{\Ln_j \times \Ln_i}{\|\Ln_{i}-\Ln_{j}\|}
\qquad \text{for a thin stellar disk of overlapping/embedded orbits\,.}
\end{equation}
These equations are similar to the equations of motion for a point vortex system on the sphere, where
the torque is proportional to $\sum_{j}{(\Ln_j \times \Ln_i)}/{\|\Ln_{i}-\Ln_{j}\|^2}$.

\subsection{Summary}

Now we can substitute the radial and the azimuthal integral (\ref{e:R_ell})
into the interaction energy~(\ref{e:Hintdefinion}).

For  non-overlapping orbits, $r_{p \Out}>r_{a \In}$,
\begin{equation}\label{e:Hamiltonian-result-nonoverlap}
H^{}_{\E} = -\frac{G m_{\In} m_{\Out}}{a_{\Out}} -\frac{ G m_{\In} m_{\Out}}{a_{\In} a_{\Out}}
 \sum_{\ell=2}^{\infty}
\frac{b_{\In}^{\ell+1}}{b_{\Out}^{\ell}}
P_{\ell}(0)^2
 P_{\ell+1}(\chi_{\In})P_{\ell-1}(\chi_{\Out})
\,P_{\ell}(\cos I).
\end{equation}
where $b_i=a_i \sqrt{1-e_i^2}$ is the semiminor axis,
$\chi_i=a_i/b_i=1/\sqrt{1-e_i^2}$ is the aspect ratio, and the sum is
over even $\ell$. For identical orbits, the generating function of the
radial integral for each multipole is a combination of transcendental
functions (Eq.~\ref{e:generatingfunction-def}), and the closed-form
formula is a lengthy expression given by
Eqs.~(\ref{e:Sell-identical-result})--(\ref{e:Sell-identical-result-end}). In
the general case of overlapping or embedded orbits, we have derived the generating function of the
radial integral in two parts $S_{\ell}^+$ and $S_{\ell}^-$. The
generating function is a one-dimensional integral
(\ref{e:Sgeneral-G})--(\ref{e:Sgeneral-G2}).  Equivalent expressions
for $S_{\ell}^+$ are Eqs.~(\ref{e:Sgeneral-s}) or
(\ref{e:Sgeneral-h}).  $S_{\ell}^{-}$ can be calculated with these
equations by reversing the indices $\In\leftrightarrow \Out$.
The interaction energy is then
\begin{equation}\label{e:Hamiltonian-result}
H_{\E} = -\frac{ G m_{\In} m_{\Out}}{a_{\Out}}
 \sum_{\ell=0}^{\infty} P_{\ell}(0)^2 s_{ \ell}\,\alpha^\ell P_{\ell}(\cos I).
\end{equation}
where $s_{\ell}$ is related to $S_{\ell}=S_{\ell}^+ + S_{\ell}^-$ by
Eq.\ (\ref{e:sijl-def}). Note that $s_{\ell}$ is dimensionless, and
hence independent of the overall dimensional scale, and $s_{\ell}=1$
for circular non-overlapping orbits.  The sum over $\ell$ converges
for all cases except for a set of measure zero.  The asymptotic form
of the multiplicative prefactor of $P_{\ell}(\cos I)$ is given in
closed form by Eqs.~(\ref{e:Jasymptotic-nonoverlap}) and
(\ref{e:Jasymptotic-overlap}) for non-overlapping and overlapping/embedded
orbits respectively. The corresponding asymptotic precession rate
is given by Eq.~(\ref{e:Omegaasymptotics3}).

\section{Random walk on a sphere}\label{s:app:randomwalk}

Here we derive the eigenfunctions and eigenvalues of the stochastic random walk
on a unit sphere, as used in Section \ref{s:random}.
Let us assume an initial probability $\rho_0(\r)$, and that $\r$ moves
an angle $\alpha$ in a random direction on the sphere at each step of
the walk.
Thus, the probability density after the $n^{\rm th}$ step is set by the probability density
of the preceding step as
\begin{align}\label{e:randomwalk-rho1}
\rho_n(\r) = \int_{S_2} \D \r' p_{\r, \r'}\,\rho_{n-1}(\r')\,.
\end{align}
where $p_{\r, \r'}$ is the transition probability between two points
$\r$ and $\r'$. The transition probability must vanish if
$\cos\gamma\equiv\r\cdot\r'$ differs from $\mu\equiv \cos\alpha$, and must
satisfy $\int_{S_2} d\r' \,p_{\r,\r'}=1$ for all $\r$ to conserve
probability.  These conditions require that
\begin{equation}
 p_{\r, \r'} = \frac{1}{2\pi}\delta(\cos\gamma-\mu).
\label{e:randomwalk-transitionprob}
\end{equation}
Next we will use the following identities of the Legendre polynomials,
\begin{align}
&\delta(\cos \gamma - \mu)= \sum_{\ell=0}^{\infty}
\frac{2\ell+1}{2}P_{\ell}(\cos\gamma)P_{\ell}(\mu)\,,\\
&P_{\ell}(\cos\gamma)
=
\frac{4\pi}{2\ell+1}\sum_{m=-\ell}^{\ell}
Y_{\ell m}(\r)Y_{\ell m}^*(\r')\,.
\end{align}
where $Y_{\ell m}(\r)$ are orthonormal spherical harmonics\footnote{
We use the definition in Eq.~(\ref{e:Y}) for $Y_{\ell m}(\r)$ which satisfies
\begin{equation}
 \int_{S_2}  Y_{\ell m}(\r)Y_{\ell' m'}^*(\r)\, \D \r =
 \delta_{\ell\, \ell'}\delta_{m\, m'}\quad{\rm if}\quad
 \ell\geq 0 \;\text{and}\; -\ell \leq m\leq \ell\,\;\text{and similarly for $\ell'$ and $m'$}.
\end{equation}
}.
Substituting into Eq.~(\ref{e:randomwalk-transitionprob}) gives
\begin{equation}
p_{\r, \r'}  =\sum_{\ell=0}^{\infty}\sum_{m=-\ell}^{\ell}
 P_{\ell}(\mu)\, Y_{\ell m}(\r)Y_{\ell m}^*(\r')\,.
\end{equation}
Next we substitute in Eq.~({\ref{e:randomwalk-rho1}}):
\begin{align}\label{e:randomwalk-rhoappendix}
 \rho_n(\r) = \sum_{\ell=0}^{\infty}\sum_{m=-\ell}^{\ell}
 P_{\ell}(\mu)\, Y_{\ell m}(\r) \int_{S_2} d\r'\,Y_{\ell
   m}^*(\r')\rho_{n-1}(\r').
\end{align}
In particular, if $\rho_{n-1}(\r')=Y_{LM}(\r')$ for some $L\geq 0$ and
$-L\leq M\leq L$, then $\rho_n(\r)=P_L(\mu)Y_{LM}(\r)$ by the
orthonormal property of the spherical harmonics. Thus the spherical
harmonics are eigenfunctions of the linear operator
(\ref{e:randomwalk-rho1}) or (\ref{e:randomwalkdef}) with eigenvalue
$P_L(\mu)$.

\section{Torque parameter}
\label{s:beta}

If each star in the cluster has the same semimajor axis and the cluster is
spherical, the dimensionless torque parameter for star $i$,
defined in Eq.~(\ref{e:betaT0}) simplifies to
\begin{equation}\label{e:betaT}
  \beta_T = \frac{2\pi a}{G m_i m_{\rms}} \left\langle\sum_\ell
    \frac{\ell(\ell+1)}{2\ell+1}\J_{ij\ell}^2\right\rangle_j^{1/2},
\end{equation}
where the average is over the distribution of stars $j$.  If there is
a distribution of semimajor axes, we replace $N\rightarrow \D N/\D \ln
a$---as we did in going from Eq.~(\ref{e:torque-coherent}) to
Eq.~(\ref{e:torque-coherent-powerlaw}) or from Eq.\
(\ref{e:vrr}) to (\ref{e:vrr1})---so Eq.~(\ref{e:betaT}) becomes
\begin{equation}\label{e:betaT5}
  \beta_T = \frac{2\pi a }{G m_i m_{\rms} }
\left(\frac{\D \ln N}{\D \ln a}\right)^{-1/2}
\left\langle\sum_\ell
      \frac{\ell(\ell+1)}{2\ell+1}\J_{ij\ell}^2\right\rangle_j^{1/2}.
\end{equation}
Here $a\equiv a_i$.
Now substitute $\J_{ij\ell}$ from Eq.~(\ref{e:Jijell}).  If the number
of stars in the range $[a,a+\D a]$, $[e,e+\D e]$, and $[m,m+\D m]$
is $\D N = 4\pi a^2 n(a,e,m)\,\D a\,\D e\,\D m$ then
\begin{align}
  \beta_T &= \frac{2\pi a}{m_{\rms}(\D N/\D \ln a)^{1/2}}
\bigg\{\sum_{\ell>0,\rm even} \frac{\ell(\ell+1)}{2\ell+1}[P_{\ell}(0)]^4
\int_{0}^{\infty}\D m'
\int_{0}^{1} \D e'
 \int_{0}^{\infty} \D a'\; 4\pi (a')^2 n(a',e',m') \frac{[\min(a,a')]^{2\ell}}{[\max(a,a')]^{2\ell+2}}
\nonumber\\&\quad\times
  (m')^2 s_{\ell}^2(\alpha,e_{\In},e_{\Out})\bigg\}^{1/2}
\end{align}
Here $s_{\ell}^2(\alpha,e_{\In},e_{\Out})$ is defined in Eq.~(\ref{e:s_ijl}), $\alpha = \min(a,a')/\max(a,a')$,
 $(e_{\In},e_{\Out})=(e',e)$ if $a'\leq a$, and $(e_{\In},e_{\Out})=(e,e')$ for $a'\geq a$.
Now let us assume that the $a$--$e$--$m$ distribution is separable as $n(a,e,m)=f(m)f(e)n(a)$,
where the distribution functions $f(m)$ and $f(e)$ have unit integrals.
Changing integration variable to $\alpha$ we get
\begin{align}
\beta_T= 2\pi \bigg\{ \frac{1}{n(a)} \sum_{\ell>0,\rm even} \frac{\ell(\ell+1)}{2\ell+1}[P_{\ell}(0)]^4
\int_{0}^{1} \D e' \,f(e')\int_{0}^{1} \D \alpha
 \left[ \alpha^2 n(a \alpha ) s_{\ell}^2(\alpha,e',e) + \alpha^{-2} n(a/\alpha) s_{\ell}^2(\alpha,e,e')\right]
\alpha^{2\ell}  \bigg\}^{1/2}
\label{e:betaT2}
\end{align}
For a power-law density profile $n(a)\propto a^{-\gamma}$,
\begin{align}\label{e:betaT3}
  \beta_T &= 2\pi
\left\{\sum_{\ell>0,\rm even} \frac{\ell(\ell+1)}{2\ell+1}[P_{\ell}(0)]^4
\int_{0}^{1} \D e' \,f(e')
  \int_{0}^{1} \D \alpha\, \left[\alpha^{2-\gamma} s_{\ell}^2(\alpha,e',e)
+\alpha^{\gamma-2} s_{\ell}^2(\alpha,e,e')\right]
\alpha^{2\ell} \right\}^{1/2}
\end{align}
Generally, the terms in the sum scale as $\ell^{-3}$ for overlapping
or embedded eccentric orbits, and $\ell^{-3}\alpha^{2\ell}$ for
eccentric non-overlapping orbits. The lowest order (i.e quadrupole)
terms dominate $\beta_T$ in a spherical cluster.  Note that $\beta_T$
is independent of the stellar mass $m$, the RMS stellar mass
$m_{\rms}$ and the semimajor axis $a$ (for a power-law density), but
it may depend on the exponent of the power law $\gamma$, the
eccentricity $e$ of the test star, and the distribution $f(e)$ of the
eccentricities of the cluster stars.  For circular orbits $e=e'=0$,
$s_{\ell}(\alpha,1,1)=1$ and we find that $\beta_T$ is between 1.507
and 1.526 if $1<\gamma<3$. We find a weak $\gamma$ dependence for
general eccentric orbits as well. For eccentric overlapping or embedded orbits we find
that $\beta_T$ is systematically smaller for all $1<\gamma<3$ and $f(e')$
as shown in Figure~\ref{f:betaT}.

\end{document}